\documentclass[letterpaper]{article} % DO NOT CHANGE THIS
\usepackage{aaai25}  % DO NOT CHANGE THIS
\usepackage{times}  % DO NOT CHANGE THIS
\usepackage{helvet}  % DO NOT CHANGE THIS
\usepackage{courier}  % DO NOT CHANGE THIS
\usepackage[hyphens]{url}  % DO NOT CHANGE THIS
\usepackage{graphicx} % DO NOT CHANGE THIS
\usepackage{natbib}  % DO NOT CHANGE THIS AND DO NOT ADD ANY OPTIONS TO IT
\usepackage{caption} % DO NOT CHANGE THIS AND DO NOT ADD ANY OPTIONS TO IT
\usepackage{algorithm}
\usepackage{algorithmic}

\usepackage{newfloat}
\usepackage{listings}
\DeclareCaptionStyle{ruled}{labelfont=normalfont,labelsep=colon,strut=off} % DO NOT CHANGE THIS
\floatstyle{ruled}
\newfloat{listing}{tb}{lst}{}
\floatname{listing}{Listing}
\usepackage[dvipsnames]{xcolor}
\usepackage{amsmath} 
\usepackage{subfigure}

\title{A Weighted Similarity Metric for Community Detection in Sparse Data}
\author {
    Yong Zhang\textsuperscript{\rm 1},
    Eric Herrison Gyamfi\textsuperscript{\rm 2}
}
\affiliations {
    \textsuperscript{\rm 1} Corporate Functions R\&D, Discovery\& Innovation Platforms, P\&G \thanks{This paper has been accepted for Workshop of AI for Social Impact at AAAI 2025.}\\
    \textsuperscript{\rm 2} Department of Mathematical Sciences, Division of Statistics and Data Science, University of Cincinnati \\
    
    zhang.y.13@pg.com, gyamfien@mail.uc.edu 
}
\begin{document}

\maketitle

\begin{abstract}
Many Natural Language Processing (NLP) related applications involves topics and sentiments derived from short documents such as consumer reviews and social media posts. Topics and sentiments of short documents are highly sparse because a short document generally covers a few topics among hundreds of candidates. Imputation of missing data is sometimes hard to justify and also often unpractical in highly sparse data. We developed a method for calculating a weighted similarity for highly sparse data without imputation. This weighted similarity is consist of three components to capture similarities based on both existence and lack of common properties and pattern of missing values. As a case study, we used a community detection algorithm and this weighted similarity to group different shampoo brands based on sparse topic sentiments derived from short consumer reviews. Compared with traditional imputation and similarity measures, the weighted similarity shows better performance in both general community structures and average community qualities. The performance is consistent and robust across metrics and community complexities.
\end{abstract}

\begin{links}
    \link{Code \& Dataset}{https://github.com/zhy5186612/Weighted_Similarity_Without_Imputation}
    % \link{Datasets}{https://aaai.org/example/datasets}
    % \link{Extended version}{https://aaai.org/example/extended-version}
\end{links}

%%%
\section{Introduction}

NLP and Large Language Model (LLM) have been increasingly used by Consumer Packaged Goods and other industries to extract insights from large amount of textual data. These textual data includes consumer reviews and complaints, open ended comments and verbatims in a survey and social media posts. While NLP and LLM can easily conduct summaries and extract topics and sentiments from these textual data, advanced analytics on topics and sentiments are generally needed to gain insights to enable better and faster consumer understanding and product innovation. The challenge of using advanced analytics and models arises when there are many missing data. The challenge is particularly prominent when using topic and sentiment data derived from short documents such as consumer reviews and social media posts. As an example, a review may only contain few topics out of hundreds or even thousands topics.          

Imputation is a natural solution for missing data. Mainstream imputation methods include replacing with mean, k-nearest neighbors (KNN) and Multivariate Imputation by Chained Equations (MICE). Different imputation techniques can significantly impact the effectiveness and qualities of downstream analytics. Sometimes imputation of missing data is hard to justify without knowing the missing mechanism. Furthermore, it is also unpractical to impute missing values if a dataset is highly sparse (e.g., 90\% values are missing).  

Similarity is an important metric for many advanced analytics and models. Community detection and/or clustering are good examples. Community detection in complex networks aims at identifying groups of nodes that are more densely connected internally than with the rest of the network \citep{Brzozowski2023CommunityDI, FORTUNATO201075}. Community detection and/or clustering rely heavily on similarities between data points. Both missing data and its imputation method can significantly affect the similarity quality and its related downstream analytics. A new similarity metric without imputation is needed and essential for meaningful downstream NLP applications involving sparse topics and sentiments from short documents.  

This paper is organized as follows. In next section, we provide a review of the traditional imputation techniques and similarity metrics utilized in this paper. We then formulate the theoretical framework of the weighted similarity. We proceed to describe a case study using the weighted similarity and imputed similarities based on an anonymized dataset. We conclude the paper after presenting results and discussion.

\section{Imputation and Similarity Methods}

Imputation is generally used to fill in missing values under the assumption of missing at random. Here we employed several different imputation methods for calculating different similarity metrics in the context of community detection. We compared the qualities of detected communities through different imputation and similarity measures to assess the merits of the new weighted similarity without imputation.      

\subsection{Imputation Methods}

Mean imputation is one of the simplest methods to handle missing data. It replaces missing values with the mean of the available values for that feature. This method is straightforward and computationally efficient \citep{Little2002}. However, it can distort the variance and covariance structures of the data, leading to potential biases in subsequent analyses. The K-Nearest Neighbor imputation is a more sophisticated method that replaces missing values with the average of the \( K \) nearest neighbors' values. This approach considers the local structure of the data, providing a more context-aware imputation \citep{Troyanskaya2001}. MICE is an advanced imputation technique that uses multiple iterations and models to predict missing values \citep{Buuren2011}.

\subsection{Similarity Measures}

Accurately measuring similarity between data points is crucial for tasks such as clustering and community detection. Traditional similarity measures offer different perspectives on how to quantify similarity, each with its strengths and applications. This paper considers imputed Cosine similarity, Euclidean distance, Canberra distance and Spearman correlation coefficient as benchmarks for the weighted similarity without imputation. 

Cosine similarity measures the cosine of the angle between two vectors, focusing on the orientation rather than magnitude. This measure is particularly effective for high-dimensional data where the magnitude of vectors can vary significantly \citep{Tan2005}. However, it does not handle missing values directly and is sensitive to the sparsity of the vectors, which can affect its accuracy in some cases. The Euclidean distance is one of the most intuitive measures of similarity, representing the straight-line distance between two points in multidimensional space \citep{Sneath1973}. Euclidean distance is sensitive to the scale of the data, requiring complete data and normalization for meaningful comparisons. 

Canberra distance is a weighted version of the Manhattan distance, giving higher weight to differences where values are small. This characteristic makes it particularly sensitive to small changes in values, which can be beneficial for datasets with a wide range of scales \citep{Lance1967}. Canberra distance is sensitive to zero values and requires complete data, which can limit its applicability in datasets with missing values. Spearman correlation measures the rank correlation between two variables, assessing how well the relationship between two variables can be described using a monotonic function \citep{Spearman1904}. It also requires complete data and does not capture linear relationships as effectively as Pearson correlation.

The appendices A has detailed descriptions on imputation methods and similarity metrics.

\section{Formulation of the Weighted Similarity} \label{sec:methods}

\begin{figure}[h!] % Options [h!] for placement
    \centering
        \includegraphics[scale=0.53]{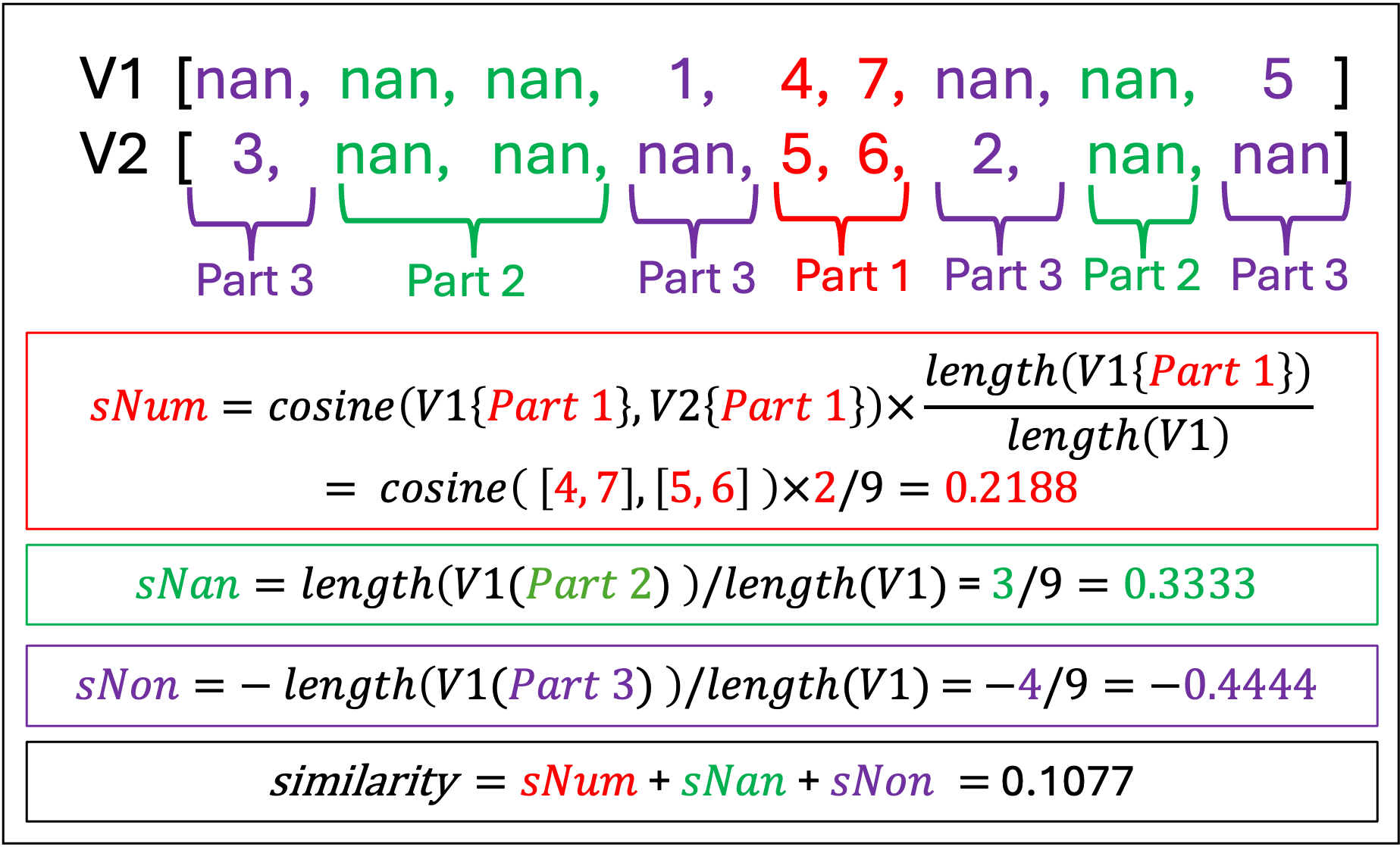}
    \caption{How to calculate weighted similarity}
    \label{fig:simMethod}
\end{figure}

Figure \ref{fig:simMethod} shows a toy example of calculating the weighted similarity between two sparse vectors \textbf{V1} and \textbf{V2}. First, two vectors of equal length are decomposed into three parts. \textcolor{red}{Part 1} represents sub-vectors where both \textbf{V1} and \textbf{V2} have numerical values. This corresponds to \textbf{Vn1} and \textbf{Vn2} in equation \ref{eq:simNum}. \textbf{sNum} is the similarity of \textcolor{red}{Part 1}. \textcolor{green}{Part 2} represents sub-vectors where both \textbf{V1} and \textbf{V2} have missing values. \textbf{sNan} is the similarity of \textcolor{green}{Part 2}. \textcolor{violet}{Part 3} captures positions where either \textbf{V1} or \textbf{V2} has a numerical value and the other has a missing value. \textbf{sNon} is the similarity of \textcolor{violet}{Part 3}.

As an example, \textbf{V1} and \textbf{V2} can represent vectors of sentiments of different topics of shampoo brands 1 and 2, respectively. These topic sentiments can be output of any Natural Language Processing (NLP) model on consumer reviews. Since these reviews are generally short and only covers few topics out of hundreds candidates, the derived vectors of sentiments for each brand are highly sparse data. Assume both brands 1 and 2 provide shampooing and conditioning benefits; and they have apple and mint fragrances, respectively; neither brands offer anti-dandruff and softness benefits. In this case, \textbf{sNum} captures similarity on sentiments of shampooing and conditioning benefits. \textbf{sNan} captures dissimilarity between sentiments of two categorically different fragrances. \textbf{sNon} conveys the idea that brands 1 and 2 are also similar as they both lack anti-dandruff and softness benefits.

\subsection{Similarity \textbf{sNum}}

Similarity \textbf{sNum} is calculated through equation \ref{eq:simNum} based on properties shared by both \textbf{V1} and \textbf{V2}. Cosine similarity is used as an example in the equation, other similarity metrics can also be used. It uses non-missing and aligned numerical values of two vectors, scaled by the proportion of these values. This approach mitigates the issue of data sparsity by focusing on the available values and provides a measure that captures the similarity in terms of direction and proportion of available data.

Similarity \textbf{sNum} is given by:

\begin{equation} \label{eq:simNum}
    \text{sNum} = 
    \begin{cases} 
    0 & \text{if } |\mathbf{Vn1}| = 0 \\
    \frac{\mathbf{sign}(\mathbf{Vn1}\mathbf{Vn2}) \min(\|\mathbf{Vn1}\|, \|\mathbf{Vn2}\|)}{|\mathbf{V1}|\max(\|\mathbf{Vn1}\|, \|\mathbf{Vn2}\|)} & \text{if } |\mathbf{Vn1}| = 1 \\
    \frac{|\mathbf{Vn1}|}{|\mathbf{V1}|}  \cdot \frac{\mathbf{Vn1} \cdot \mathbf{Vn2}}{\|\mathbf{Vn1}\| \|\mathbf{Vn2}\|} & \text{if } |\mathbf{Vn1}| > 1
    \end{cases}
    % \label{eq:simNum}
\end{equation}
where $|\mathbf{V1}|$ is the length of $\mathbf{V1}$; and $\|\mathbf{Vn1}\|$ is the Euclidean norm of  $\mathbf{Vn1}$. 

Equation \ref{eq:simNum} also handles two corner cases. When there is no shared position having numerical values in two vectors,  the \textbf{sNum} is set as 0. When there is only a single shared position having values in two vectors, the \textbf{sNum} depends on both magnitudes and signs of the two scalars.

\subsection{Similarity \textbf{sNan}}

Similarity \textbf{sNan} is calulated using equation \ref{eq:simNan} based on properties lacked in both \textbf{V1} and \textbf{V2}. It uses the proportion of missing values in both vectors. This term captures the pattern of missing data. It can be informative in many real-world datasets where the lack of properties itself carries significant information.

Similarity \textbf{sNan} is given by:

\begin{equation}
\text{sNan} = \frac{cNan}{|\mathbf{V1}|}
\label{eq:simNan}
\end{equation}
where $cNan$ is the count of position having missing value in both vectors.

\subsection{Similarity \textbf{sNon}}

Similarity \textbf{sNon} is calculated using equation \ref{eq:simNon}. This similarity component measures the proportion of non-matching missing values in two vectors. It penalizes similarity based on non-matching missing values. This component is vital for datasets with significant amounts of missing data, as it penalizes positions that differ in their missing data patterns.

Similarity \textbf{sNon} is given by:

\begin{equation}
\text{sNon} = -\frac{cNon}{|\mathbf{V1}|}
\label{eq:simNon}
\end{equation}
where $cNon$ is the count of positions having non-matching missing entries in both vectors.

\subsection{Weighted Similarity}

The weighted similarity combines similarities \textbf{sNum}, \textbf{sNan} and \textbf{sNon} into a single measure in equation \ref{eq:total_sim}. This composite metric provides a comprehensive view of similarity, incorporating both the available numerical data and the missing patterns. It ensures that vectors are compared holistically, considering all aspects of their data.

\begin{equation}
\text{similarity} = \text{sNum} + \text{sNan} + \text{sNon}
\label{eq:total_sim}
\end{equation}

If the similarity metric used for the numerical \textcolor{red}{Part 1} is bounded by -1 and 1, the weighted similarity in equation \ref{eq:total_sim} also has a range of [-1, 1]. In the appendix, we use cosine similarity metric as an example to prove that the range of the weighted similarity metric \(S\), is \(S \in [-1, 1]\). This suggests that the formulation in equation \ref{eq:total_sim} is indeed a proper metric to measure similarity in sparse data without imputation.     

\section{Case Study} \label{sec:case}

A case study was conducted to compare the performance of the weighted similarity against traditional similarity measures and imputation methods in the context of community detection using the Girvan-Newman algorithm \citep{Girvan2002}. We used an anonymized datasets with 78 rows and 44 columns. Each row is a shampoo brand and each column is a topic in the shampoo product category. The value is the sentiment on the corresponding topic and brand. The topics and sentiments were output from BERTopic model based on consumer reviews. Since most reviews are short and only cover few topics, around $70\%$ of topic sentiments are missing. This data is also provided in this paper's GitHub repository.    

First, pairwise brand similarities were calculated using the weighted similarity without imputation and imputed similarities. We investigated traditional Cosine, Euclidean, Canberra and Spearman similarities with mean, KNN (K=4), and MICE imputations. For each similarity method, a graph was built to represent similarities among the brands. Each node in the graph is a brand. Each edge is the similarity between the two connected brands. For each graph, we retained top 100, 600 and 1200 strong edges (i.e., large similarities) to represent different graph complexities. The Girvan-Newman algorithm was then used to detect 5 communities (i.e., clusters) from a graph to group different shampoo brands together.

The performance of each similarity method is evaluated in two folds: general community structures and average community qualities. When evaluating detected communities, it is crucial to consider various metrics that can capture different aspects of community structures. Metrics of general community structures help to assess the overall quality and coherence of the detected communities within a network. These metrics focus on the internal consistency and external separation of communities, providing insights into how well the algorithm can identify and preserve meaningful groups of nodes. Modularity (Q) \citep{Newman2006}, Coverage \citep{Fortunato2010}, Dunn Index \citep{Dunn1974}, Average Clustering Coefficient \citep{Watts1998}, Transitivity \citep{Newman2003}, Modularity Density \citep{Li2008} and Triangle Participation Ratio (TPR) \citep{Jaewon2012} were used to assess general community structures.

Average community quality metrics offer a more detailed assessment of the quality and cohesiveness of each individual community. These metrics focus on specific properties of the communities, such as their internal connectivity, isolation from the rest of the network, and overall density. Conductance \citep{Kannan2004}, Expansion \citep{Leskovec2008}, Normalized Cut \citep{Shi2000}, Density \citep{Radicchi2004}, Internal Density \citep{Fortunato2010} and Local Modularity \citep{Clauset2004} were utilized to assess average community qualities. The appendices have detailed descriptions on these metrics and Girvan-Newman algorithm.

\section{Results} \label{sec:results}

\subsection{Community Network Graph} \label{subsec:cngraph}

\begin{figure*}[!ht] % Options [h!] for placement
    \centering
    \subfigure[Weighted similarity]{
        \includegraphics[width=0.35\linewidth]{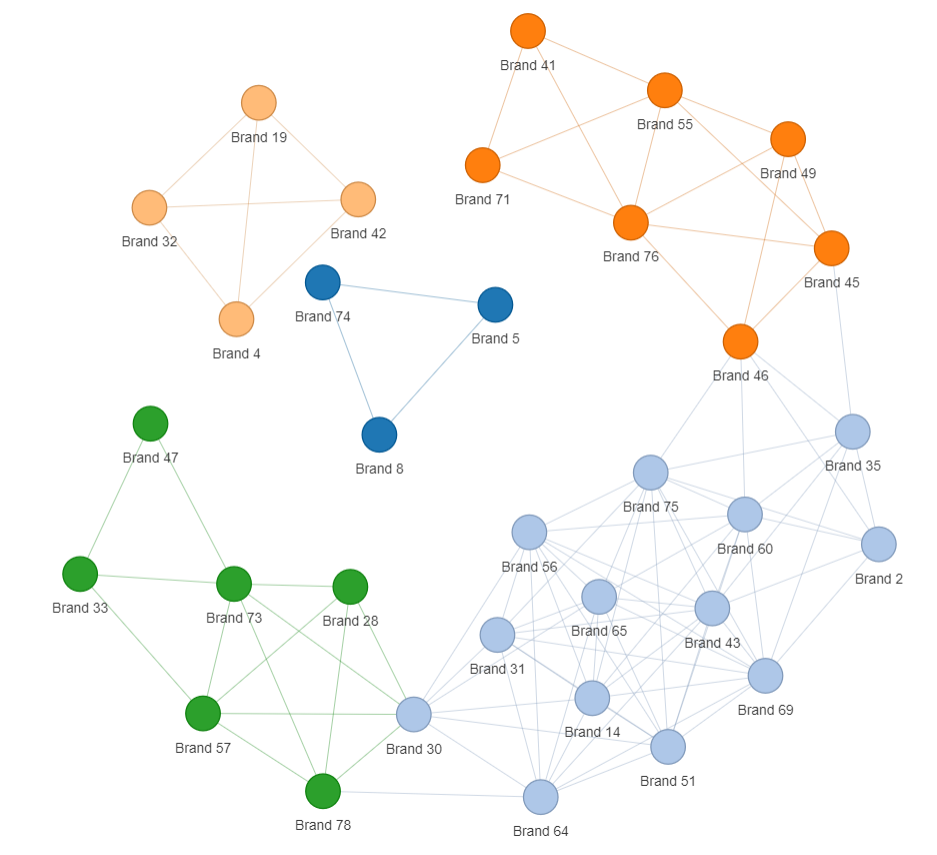}
        \label{fig:weightedSim}}
    \subfigure[Cosine similarity based on KNN imputation]{
        \includegraphics[width=0.35\linewidth]{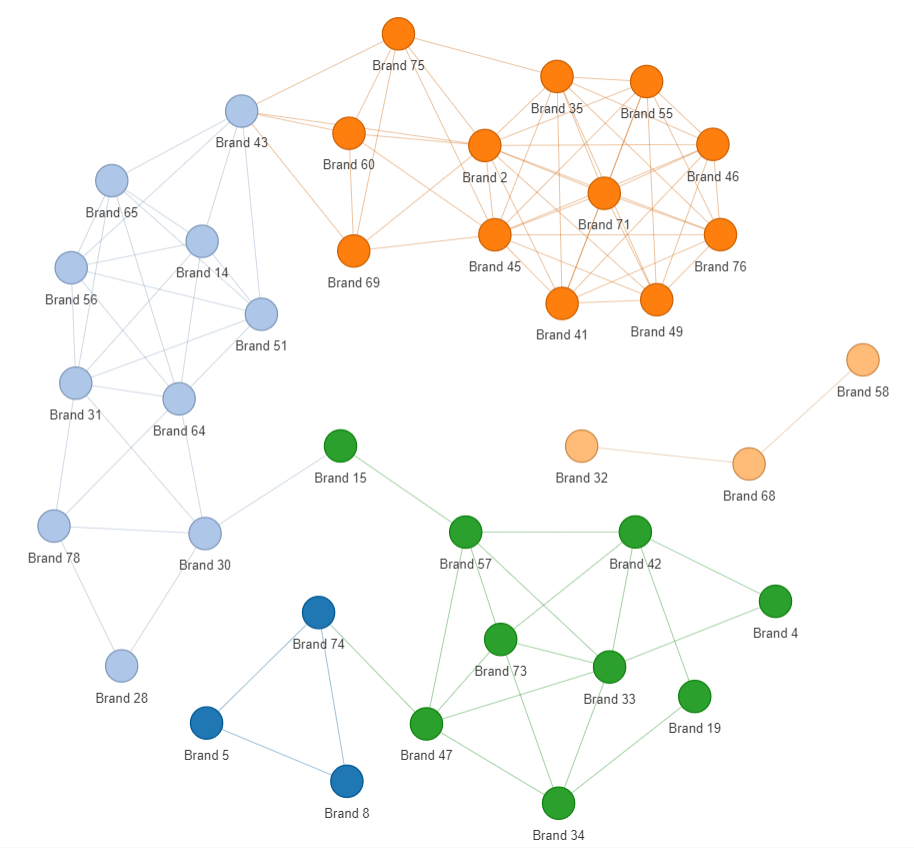}
        \label{fig:KNNImp}}
    \caption{Community structures with 5 clusters and 100 connected edges based on the weighted similarity and imputed cosine similarities using KNN.}
    \label{fig:graph100E5C_Cosine}
\end{figure*}

As an example, figure \ref{fig:graph100E5C_Cosine} shows the network graph and community structures based on the weighted similarity and imputed cosine similarity using KNN. For the network graph with 100 edges and 5 communities, the brand clusters/communities from the weighted similarity and imputed cosine similarity are similar regardless the imputation methods (Figures \ref{fig:graph100E5C_Cosine} and A1). This similar structure is likely due to the fact that cosine similarity is one component in the weighed similarity (equation \ref{eq:simNum}).  It could be also due to the inherent properties of cosine similarity, which focuses on the orientation of the vectors, making it less sensitive to the specific imputation techniques.    

For the rest of the imputed similarity metrics, the community structures differ substantially (Figures A2, A3, A4). These differences can be attributed to the varying ways each method handles missing data and measures similarity, leading to different interpretations of the network's structure. As the number of edges increased from 100 to 600 and then to 1200 (figures A5 and A9), the weighted similarity has a relative better clustering with well-separated clusters, compact intra-cluster distances, higher tendency for nodes to cluster together. These brand structures at different connectivity are consistent with the knowledge from domain experts.

\subsection{General Community Structures}

\begin{figure*}[!ht] % Options [h!] for placement
    \centering
    \subfigure[Imputed cosine similarity vs weighted similarity]{
        \includegraphics[width=0.45\linewidth]{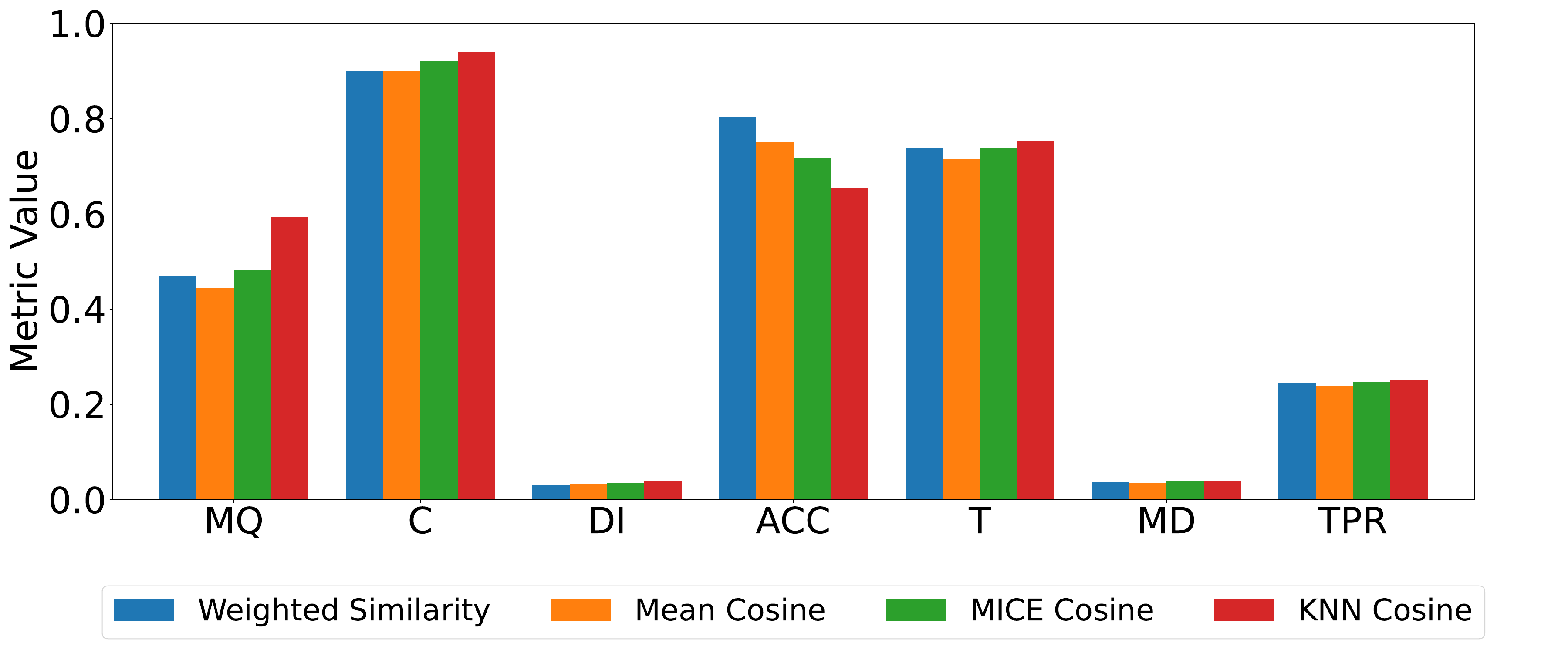}
        \label{fig:genComCosine}}
    \subfigure[Imputed euclidean similarity vs weighted similarity]{
        \includegraphics[width=0.45\linewidth]{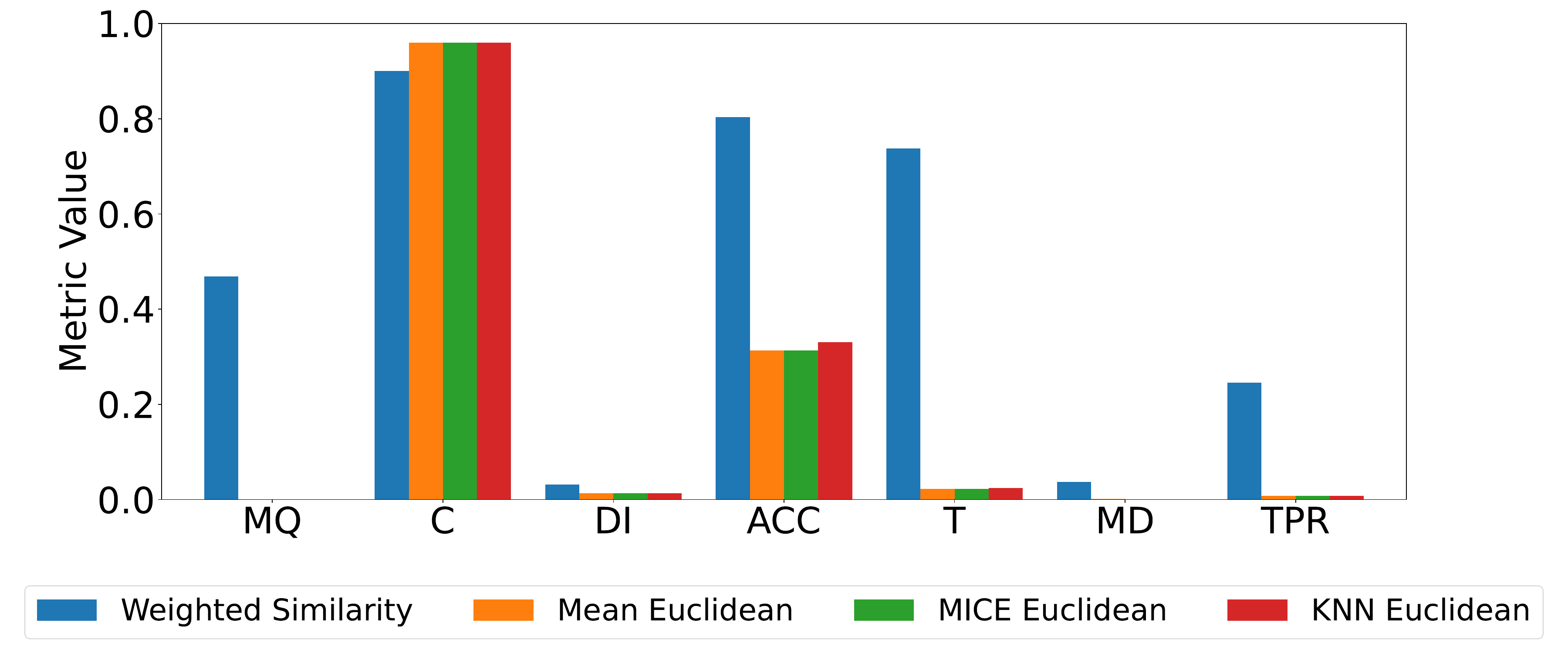}
        \label{fig:genComEuclidean}}
    \subfigure[Imputed canberra similarity vs weighted similarity]{
        \includegraphics[width=0.45\linewidth]{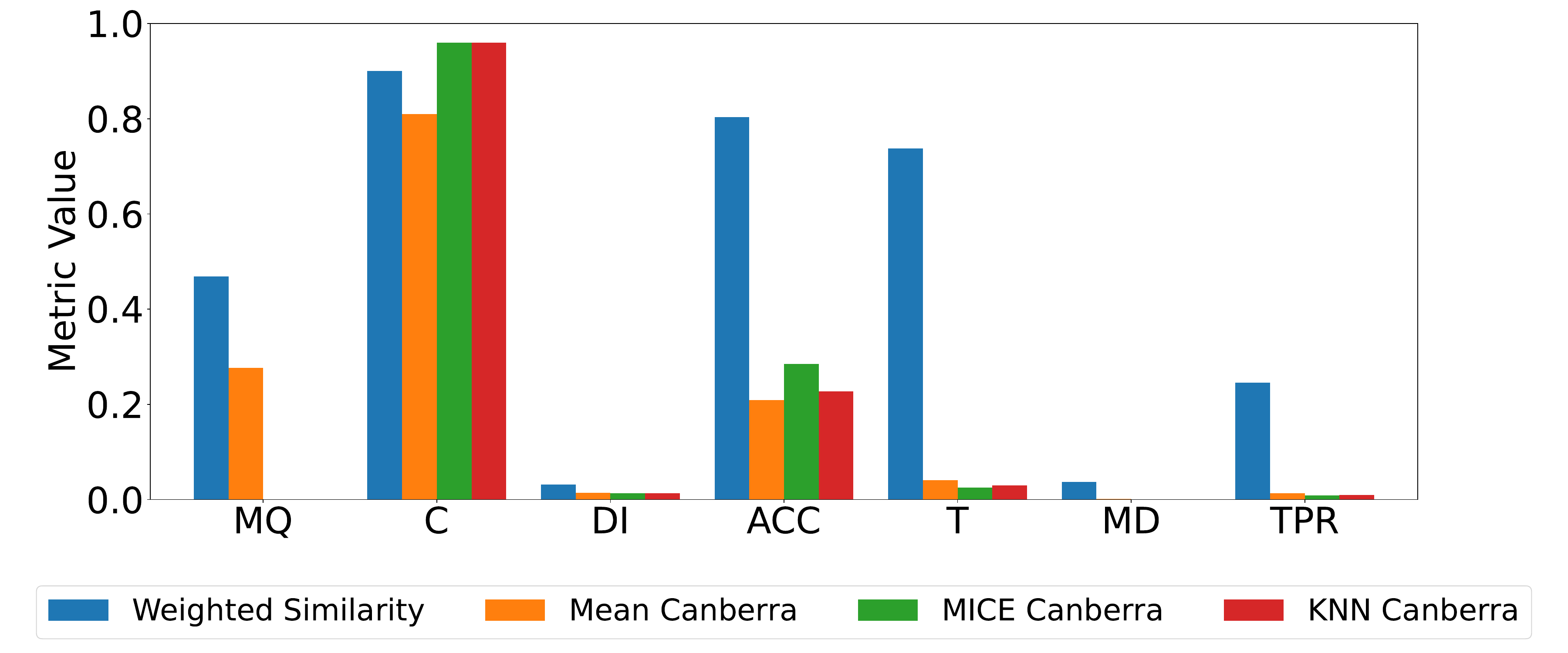}
        \label{fig:genComCanberra}}
    \subfigure[Imputed spearman similarity vs weighted similarity]{
        \includegraphics[width=0.45\linewidth]{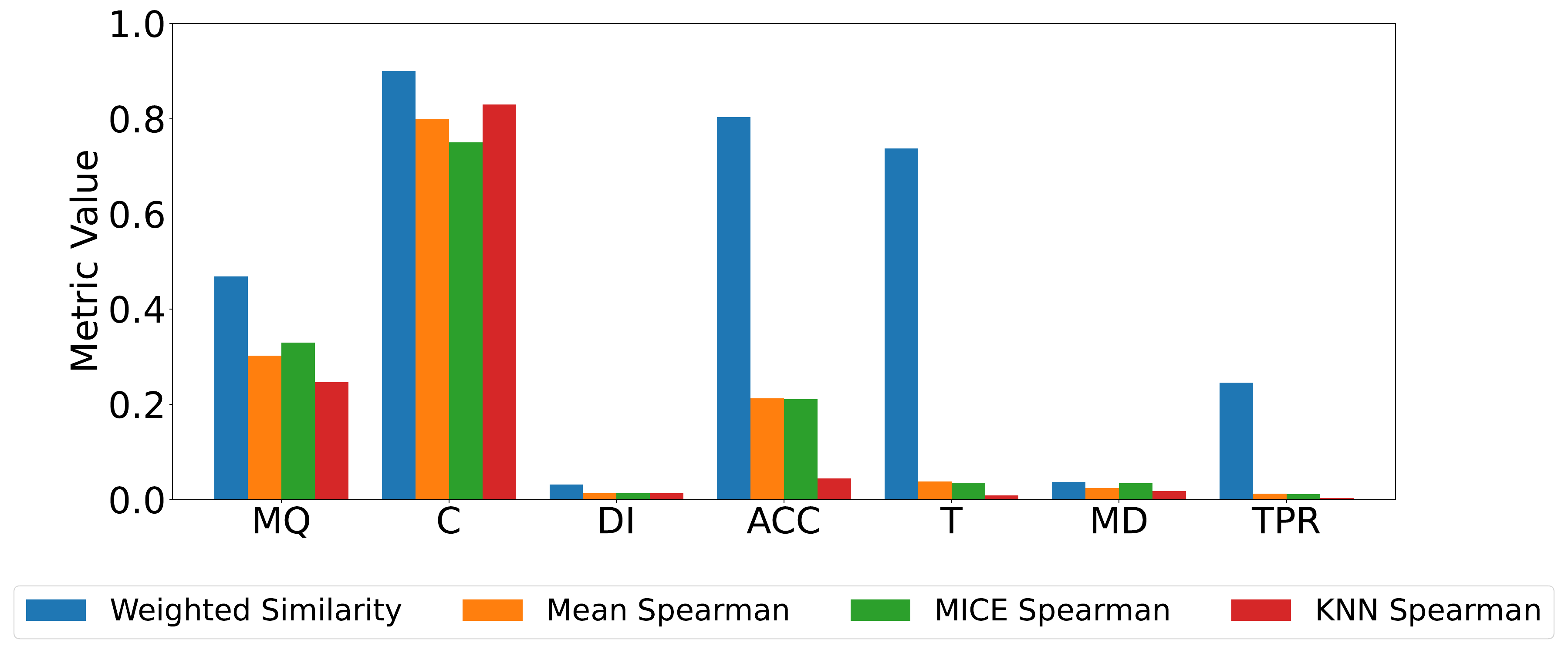}
        \label{fig:genComSpearman}}
    \caption{Performance metrics on general community structures with 100 connected edges and 5 clusters. MQ: Modularity Q, C: Coverage, DI: Dunn Index, ACC: Average Clustering Coefficient, T: Transitivity, MD: Modularity Density, TPR: Triangle Participation Ratio. Higher bar indicates better community structure.}
    \label{fig:performance100E5C_genCom}
\end{figure*}

Figure \ref{fig:performance100E5C_genCom} shows metrics measuring qualities of general community structures derived from the weighted similarity and imputed similarities. The weighted similarity and imputed cosine similarity have comparable qualities on general community structures. They both outperforms imputed similarities based on Euclidean and Canberra distances and Spearman correlation. These are consistent with the structures of community network graphs.    

The weighted similarity has a higher modularity (Q) value. This means the corresponding community structures has a denser connection within communities and sparser connections between communities.  This suggests that the weighted similarity method can identify more cohesive community structures. All similarity measures have relatively high value of coverage. This indicates Girvan-Newman algorithm effectively captures the overall structure of their community networks by ensuring that most edges are within communities. 

While Dunn index is small for all similarity metrics, it is relatively higher for the weighted similarity. The higher Dunn index suggests that the weighted similarity has a relative better clustering. The higher average clustering coefficient indicates that the weighted similarity tends to create tight and neat brand clusters. 

The weighted similarity also has higher values of transitivity, modularity density and triangle participation ratio. These suggests higher density within formed communities and its nodes (i.e., brands) tend to form triangles more frequently. These all indicates a stronger and more cohesive community structure.

\subsection{Average Community Qualities}

\begin{figure*}[!ht] % Options [h!] for placement
    \centering
    \subfigure[Imputed cosine similarity vs weighted similarity]{
        \includegraphics[width=0.45\linewidth]{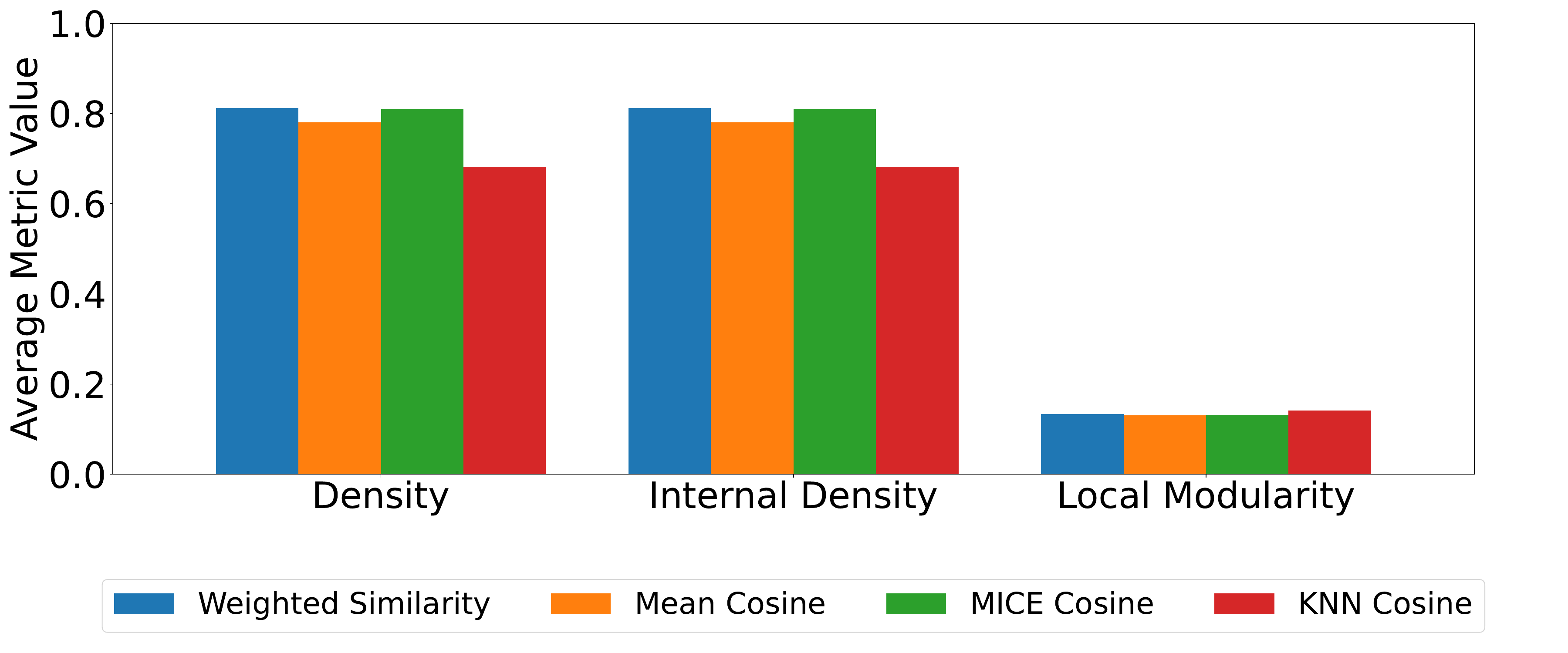}
        \label{fig:aveComHCosine}}
    \subfigure[Imputed euclidean similarity vs weighted similarity]{
        \includegraphics[width=0.45\linewidth]{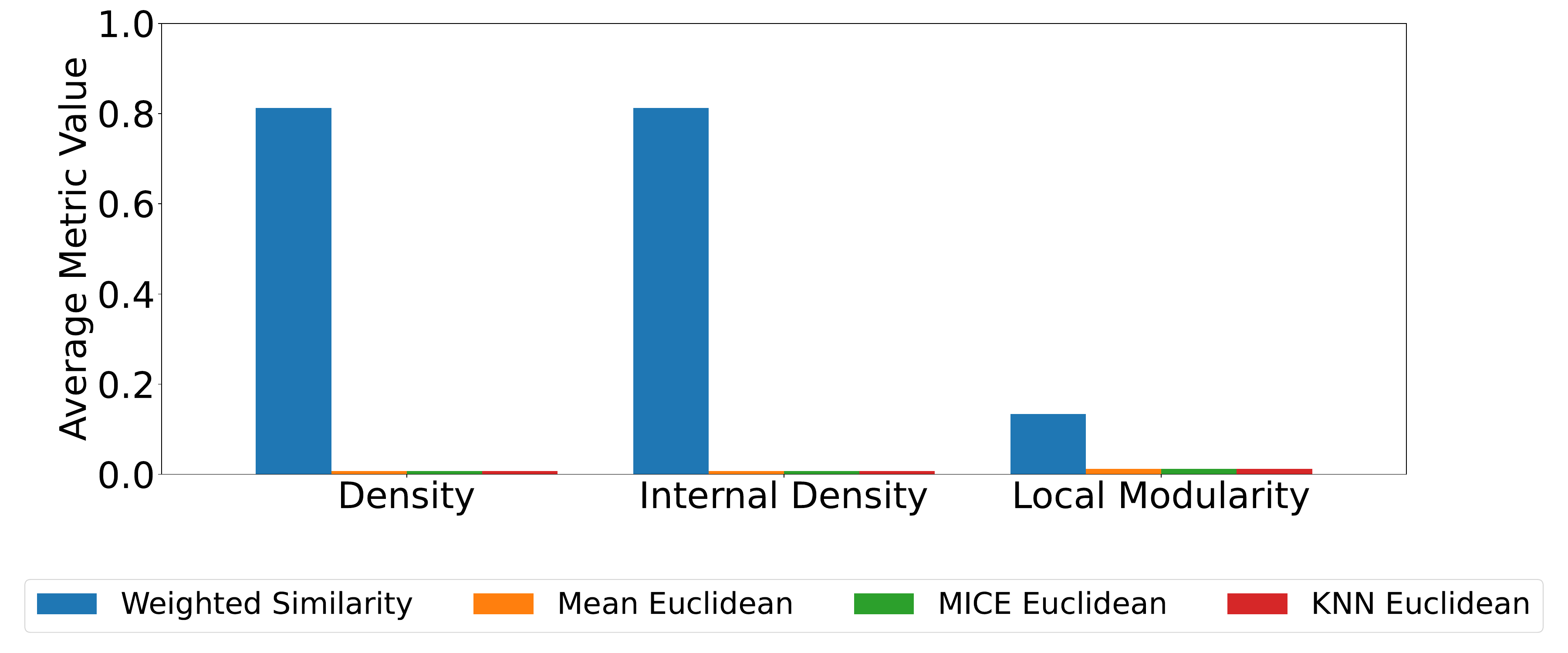}
        \label{fig:aveComHEuclidean}}
    \subfigure[Imputed canberra similarity vs weighted similarity]{
        \includegraphics[width=0.45\linewidth]{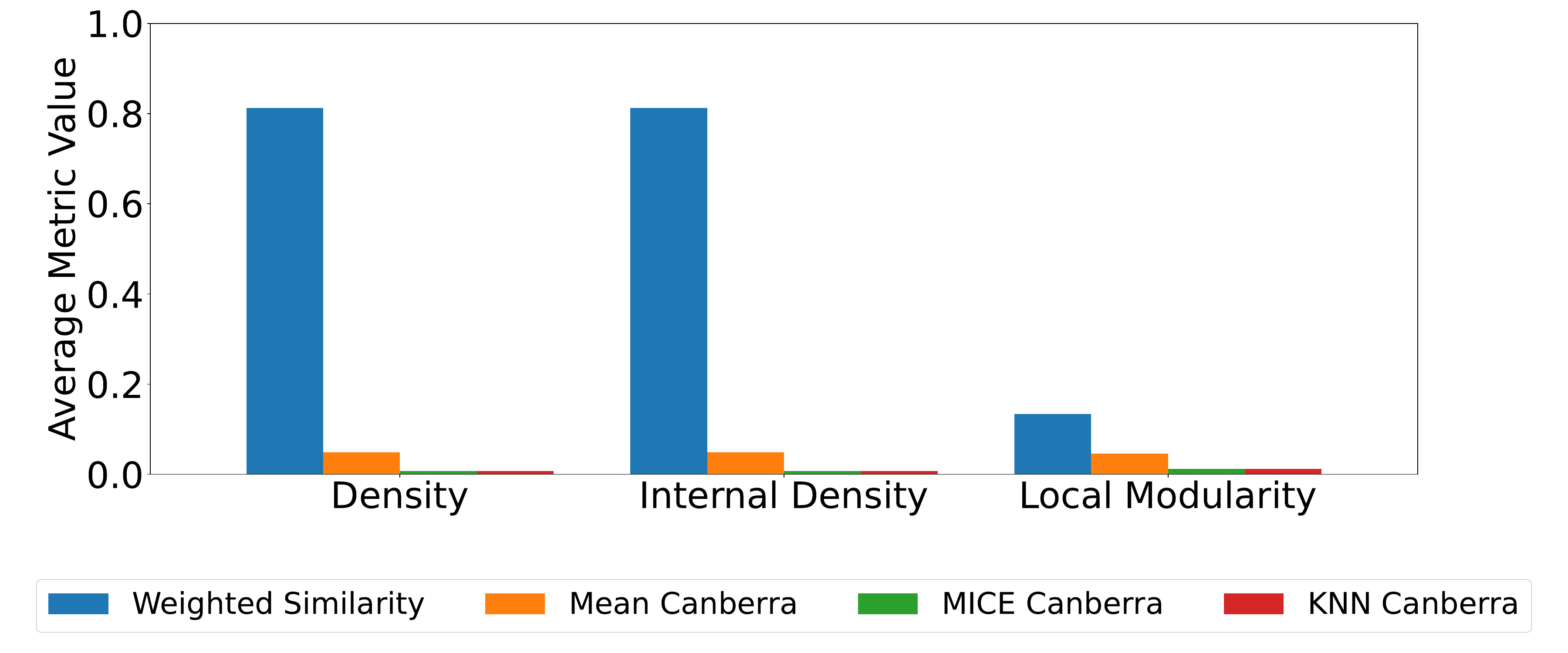}
        \label{fig:aveComHCanberra}}
    \subfigure[Imputed spearman similarity vs weighted similarity]{
        \includegraphics[width=0.45\linewidth]{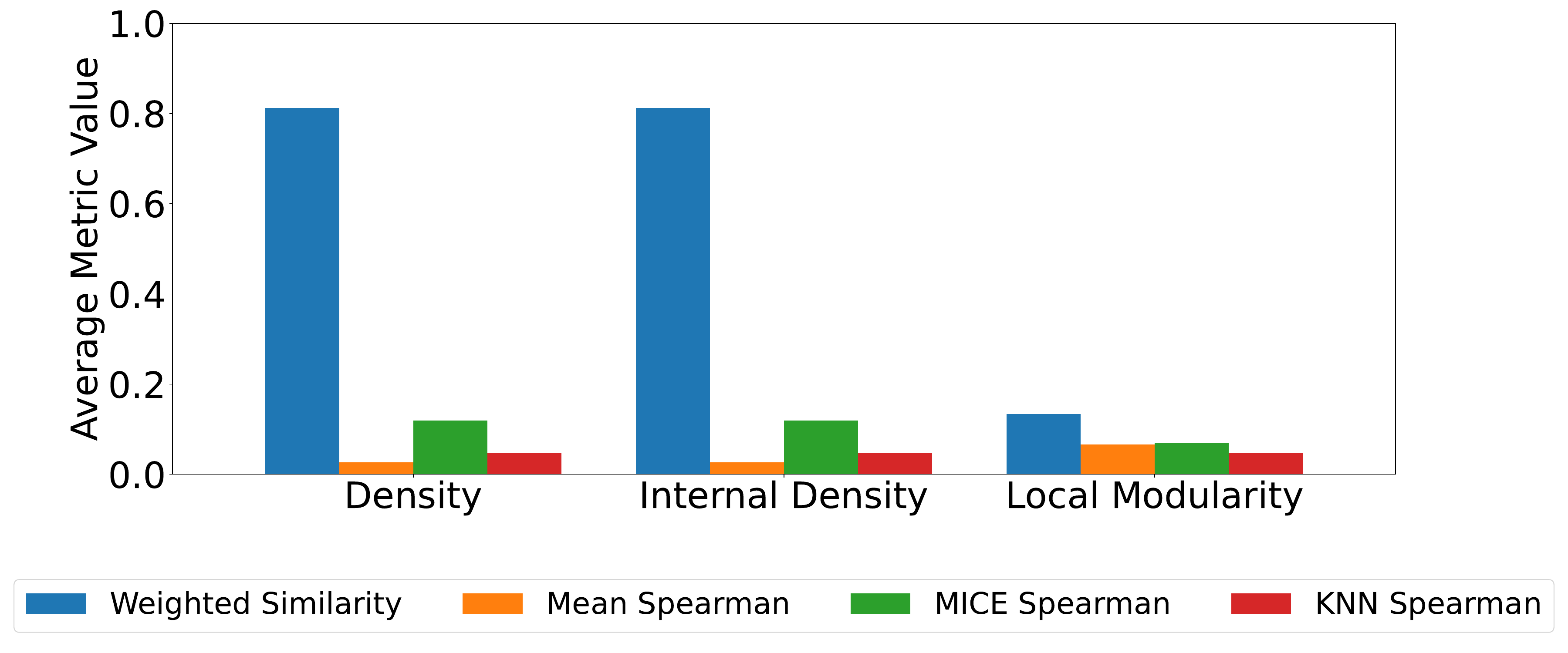}
        \label{fig:aveComHSpearman}}
    \caption{Average community qualities with 100 connected edges and 5 clusters. Higher bar indicates better community.}
    \label{fig:performance100E5C_aveComH}
\end{figure*}

Figure \ref{fig:performance100E5C_aveComH} presents the average of density, internal density and local modularity across individual communities derived from the weighted similarity and imputed similarities. The weighted similarity outperforms all imputed similarities on these three metrics measuring average community qualities. Higher density indicates better internal connectivity within communities. It suggests that the community is more cohesive and well-connected. 

The weighted similarity also has higher values for both internal density and local modularity. It means nodes/brands within each community are more tightly connected. This again suggests a better connectivity within each community in the well-defined and cohesive communities.

\begin{figure*}[!ht] % Options [h!] for placement
    \centering
    \subfigure[Imputed cosine similarity vs weighted similarity]{
        \includegraphics[width=0.45\linewidth]{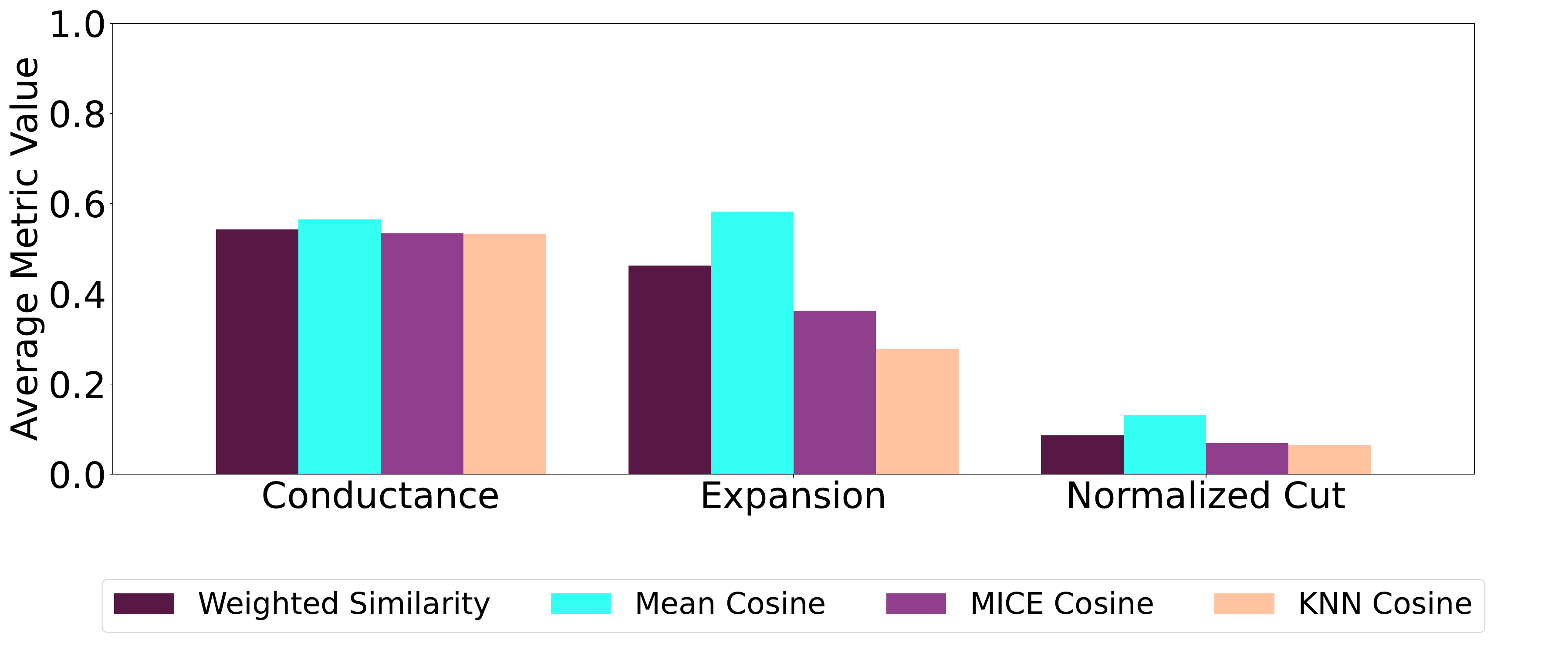}
        \label{fig:aveComLCosine}}
    \subfigure[Imputed euclidean similarity vs weighted similarity]{
        \includegraphics[width=0.45\linewidth]{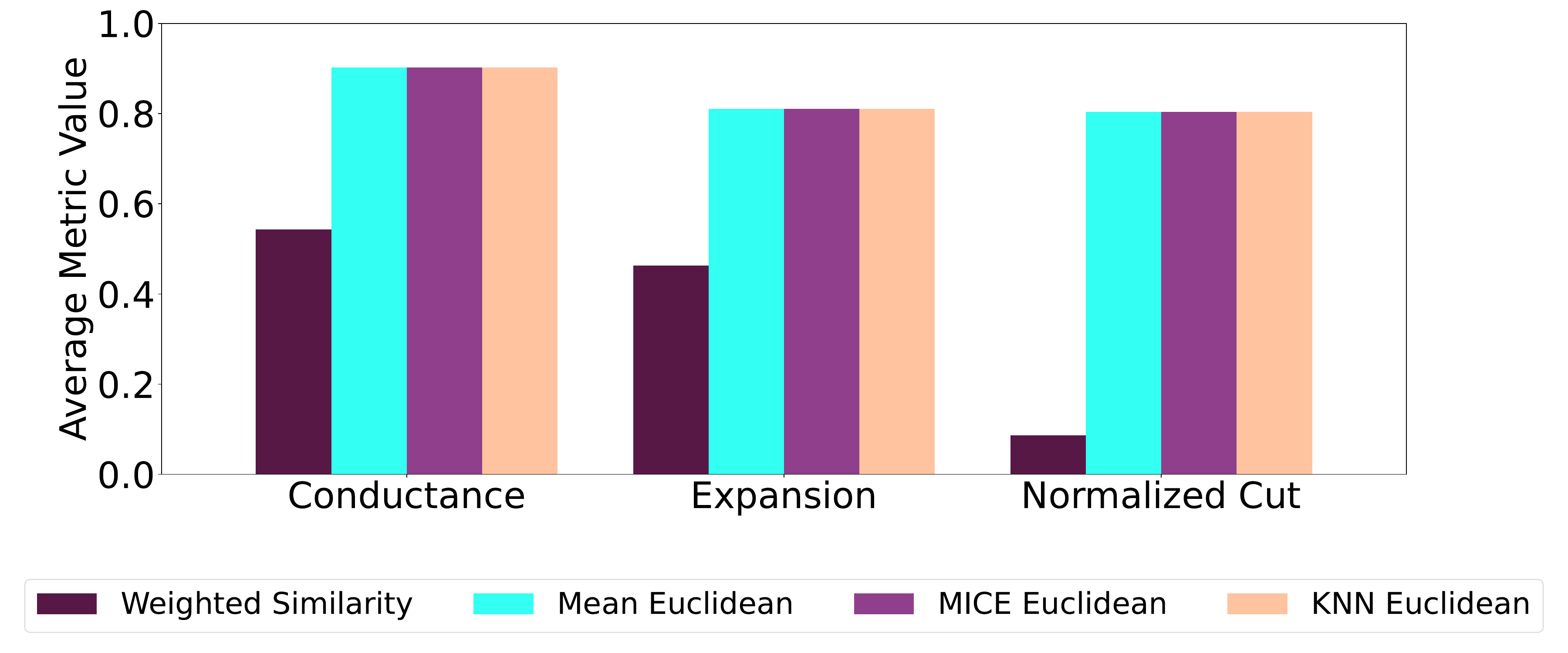}
        \label{fig:aveComLEuclidean}}
    \subfigure[Imputed canberra similarity vs weighted similarity]{
        \includegraphics[width=0.45\linewidth]{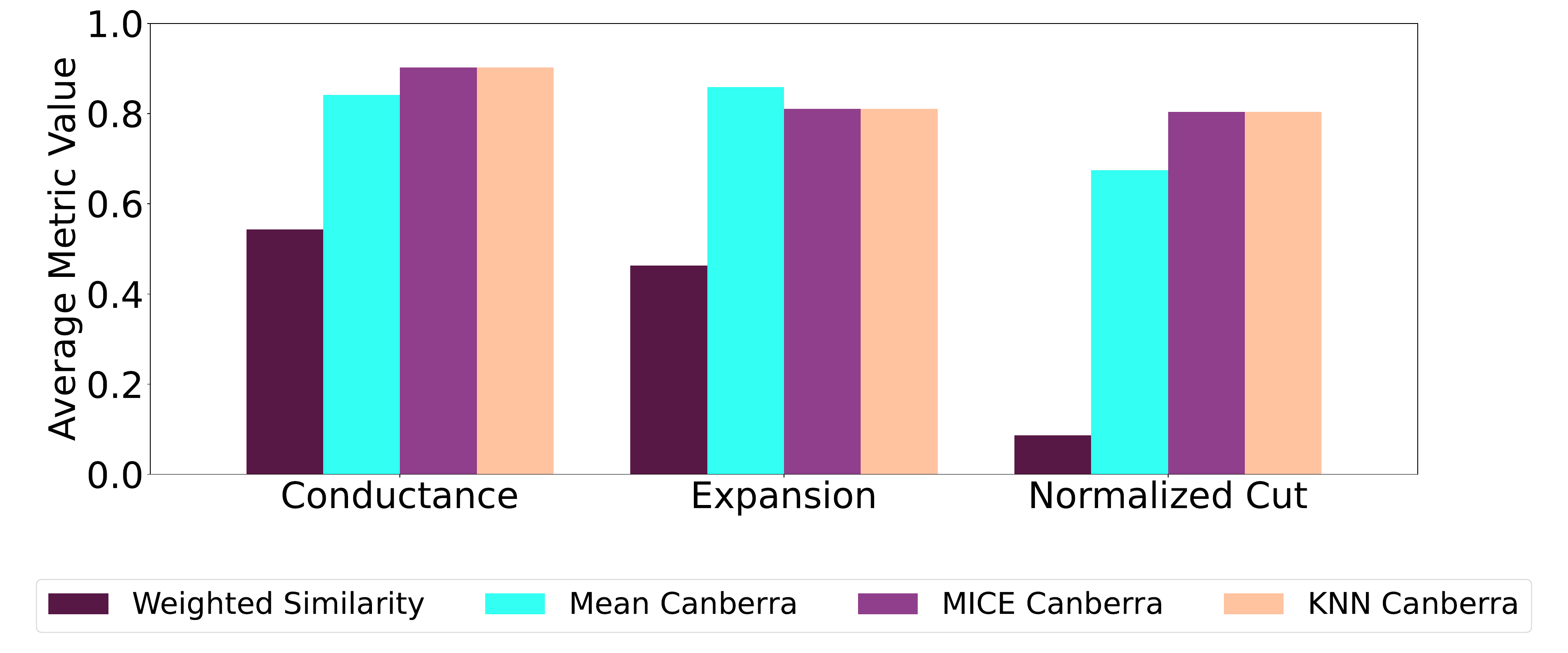}
        \label{fig:aveComLCanberra}}
    \subfigure[Imputed spearman similarity vs weighted similarity]{
        \includegraphics[width=0.45\linewidth]{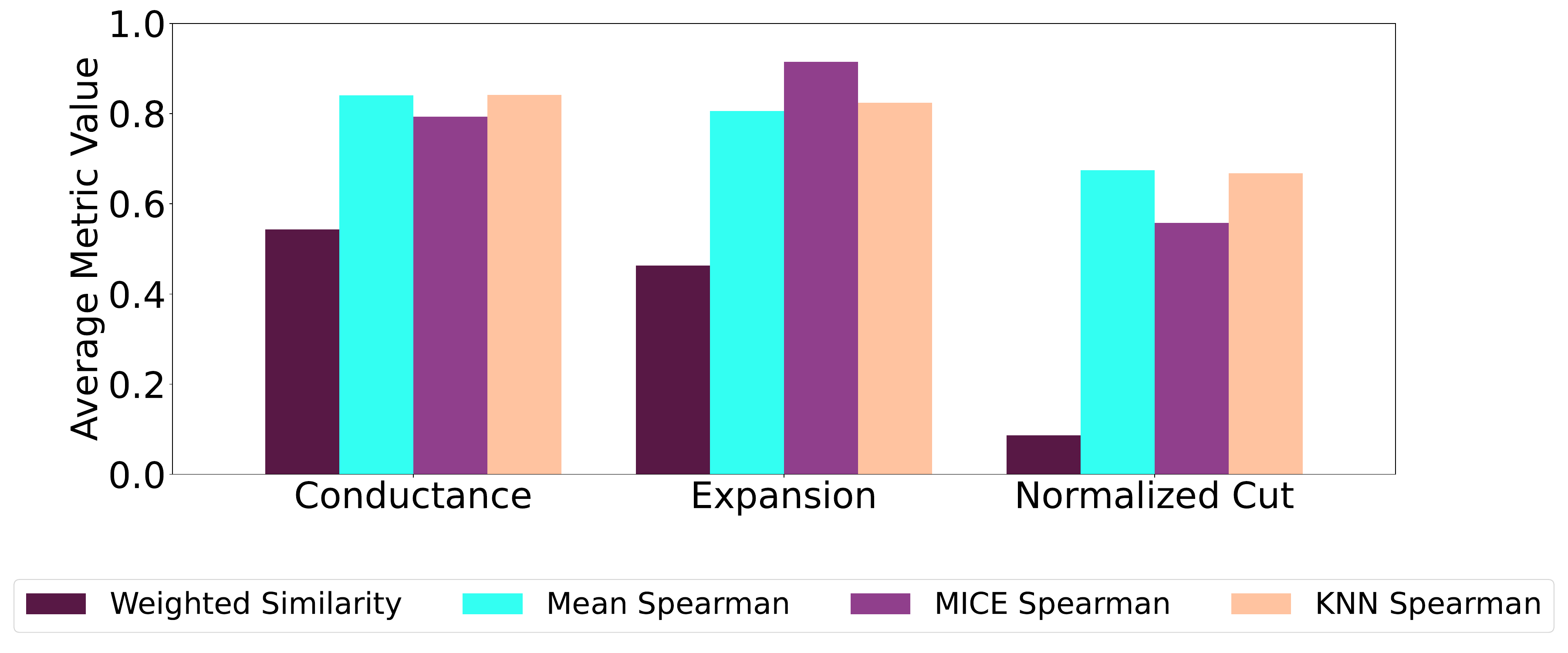}
        \label{fig:aveComLSpearman}}
    \caption{Average community qualities with 100 connected edges and 5 clusters. Shorter bar indicates better community.}
    \label{fig:performance100E5C_aveComL}
\end{figure*}

Figure \ref{fig:performance100E5C_aveComL} displays the average of conductance, expansion and normalized cut across individual communities derived from different similarity measures. A smaller value indicates a better average community quality. Weighted similarity and imputed cosine similarity have lower and comparable values in general. These lower conductance, expansion and normalized cuts indicate that fewer edges leave the community and that the community is more self-contained with fewer connections to the rest of the network. This implies a better community isolation and minimal connections between the communities, and the rest of the network.     

As the number of edges of the community network graph increase from 100 to 600 (figure 5), the weighted similarity starts to outperform all imputed similarities on most metrics for both general community structures and average community qualities (figures A6, A7, A8). This trend becomes even more pronounced as the edges increase to 1200 (figure A9). At this level of complexity of community network graph, the weighted similarity significantly outperforms all imputed similarities across almost all metrics (figures A10, A11, A12). The consistent high performance of the weighted similarity highlights its ability to maintain high quality community detection across varying network complexities.

\section{Discussion} \label{sec:discussion}

\subsection{Performance and Robustness}

Performance of imputed cosine similarity is comparable with that of the weighted similarity at lower network complexity (i.e., 100 edges). The performance of imputed cosine similarity starts to deteriorate at medium network complexity (i.e., 600 edges. figures A6, A7, A8) and underperform dramatically at high network complexity (i.e., 1200 edges. figures A10, A11, A12). This makes senses as high complexity network uses more edges and hence more imputed similarity values. More imputations highly likely causes more deviation from the pattern in real data. The weighted similarity does not rely on imputation and therefore preserves integrity of data.   

The results of the case study demonstrate the robustness and effectiveness of the weighted similarity, particularly in datasets with significant amounts of missing data and dense community structures. The better performance is rooted in the fact that the weighted similarity can directly handle missing data and incorporate it into the similarity calculation. This holistic similarity measurement considers both available data and the patterns of missing data. It helps preserve data integrity and enable downstream application like community detection algorithm to create more cohesive and well-defined communities. Missing data themselves can be informative in many real-world datasets if the missing pattern carries significant information.
 
The better performance observed as the number of edges increased from 100 to 600 and then to 1200 underscores the robustness of the weighted similarity. While other methods showed varying degrees of performance degradation or inconsistency with increasing edge size, the weighted similarity method consistently maintained performance. This robustness can be attributed to its comprehensive approach in handling both available data and patterns of missing data, ensuring accurate and reliable similarity measurements regardless of network complexity. 

\subsection{Impact of Imputation}

Imputation methods such as mean, KNN, and MICE have their weaknesses. Mean imputation can distort the variance and covariance structures of the data, leading to potential biases. KNN imputation, while more sophisticated, is computationally intensive and may not perform well when the nearest neighbors are not representative of the missing values. MICE, though robust, is also computationally intensive and requires careful tuning and validation. These weaknesses can affect the overall performance of community detection when using imputed similarities. 

The weighted similarity method, by directly incorporating the patterns of missing data into the similarity calculation, avoids these pitfalls and provides a more reliable measure of similarity, which is reflected in the superior performance metrics observed in this study. Calculating similarity without requiring imputation reduces the risk of introducing biases and preserves the original data distribution. This capability enhances the robustness of the similarity measure, ensuring that it accurately reflects the true relationships between data points. 

\subsection{Adaption and Application}

The weighted similarity can be easily adapted for different data types and structures. Whether dealing with high-dimensional numerical data, categorical data, or mixed data types, the method can be adapted to the specific characteristics of the dataset, providing a reliable similarity measure. This adaptability makes the method suitable for various applications, from clustering in bioinformatics to community detection in social networks. The method also allows for fine-tuning of the similarity components to suit specific datasets. 

The weighted similarity could be extremely useful for three application scenarios. First, if the missing information is caused by absence of a property or attribute. For example, a shampoo brand does not have the antidandruff benefit; a hotel does not have a swimming pool; and a social media user does not belong to an ethnic group. Lack of properties in this scenario is actually informative. Consumer reviews on the above shampoo brand or hotel naturally do not have topics and sentiments about antidandruff or swimming pool. The weighted similarity can capture these missing information very well without any misleading imputation.

Second, if the data are not missing at random, any imputation method may be misleading in most cases. Practically, it is very hard to infer whether the missing pattern is random or not. It is particularly challenging to impute the missing data if the data are highly sparse regardless missing at random or not. These highly sparse data are quite common in topics and sentiments derived using NLP models and LLMs based on short documents like consumer reviews and social media posts. In both cases, the weighted similarity can capture the missing pattern well for downstream applications like community detection and clustering.

\section{Conclusion}

A weighted similarity has been formulated to calculate similarity without imputation. It constructs similarity based on both existence and lack of common properties and pattern of missing data. It is particularly useful for advanced analytics on highly sparse data such as topics and sentiments from NLP models on consumer reviews and social media posts.

The weighted similarity outperforms a variety of similarity metrics using different imputation methods in the context of community detection. The superior performance is consistent and robust across metrics on both general community structures and average community quality. The superiority is even more pronounced on complex community network graphs.   

The weighted similarity offers a robust and versatile approach to measure similarity in the presence of missing data. It could have broad applications on many advanced data analytics dealing with sparse data.    

\bibliography{aaai25}

\begin{thebibliography}{22}
\providecommand{\natexlab}[1]{#1}

\bibitem[{Brzozowski, Siudem, and Gagolewski(2023)}]{Brzozowski2023CommunityDI}
Brzozowski, L.; Siudem, G.; and Gagolewski, M. 2023.
\newblock Community detection in complex networks via node similarity, graph representation learning, and hierarchical clustering.
\newblock \emph{ArXiv}, abs/2303.12212.

\bibitem[{Clauset, Newman, and Moore(2004)}]{Clauset2004}
Clauset, A.; Newman, M. E.~J.; and Moore, C. 2004.
\newblock Finding community structure in very large networks.
\newblock \emph{Physical Review E}, 70(6): 066111.

\bibitem[{Dunn(1974)}]{Dunn1974}
Dunn, J.~C. 1974.
\newblock Well-separated clusters and optimal fuzzy partitions.
\newblock \emph{Journal of Cybernetics}, 4(1): 95--104.

\bibitem[{Fortunato(2010{\natexlab{a}})}]{FORTUNATO201075}
Fortunato, S. 2010{\natexlab{a}}.
\newblock Community detection in graphs.
\newblock \emph{Physics Reports}, 486(3): 75--174.

\bibitem[{Fortunato(2010{\natexlab{b}})}]{Fortunato2010}
Fortunato, S. 2010{\natexlab{b}}.
\newblock Community detection in graphs.
\newblock \emph{Physics Reports}, 486(3-5): 75--174.

\bibitem[{Girvan and Newman(2002)}]{Girvan2002}
Girvan, M.; and Newman, M. E.~J. 2002.
\newblock Community structure in social and biological networks.
\newblock \emph{Proceedings of the National Academy of Sciences}, 99(12): 7821--7826.

\bibitem[{Jaewon and Jure(2012)}]{Jaewon2012}
Jaewon, Y.; and Jure, L. 2012.
\newblock Defining and Evaluating Network Communities based on Ground-truth.
\newblock In \emph{Proceedings of 2012 IEEE Inter- national Conference on Data Mining (ICDM)}.

\bibitem[{Kannan, Vempala, and Vetta(2004)}]{Kannan2004}
Kannan, R.; Vempala, S.; and Vetta, A. 2004.
\newblock On clusterings: Good, bad and spectral.
\newblock \emph{Journal of the ACM}, 51(3): 497--515.

\bibitem[{Lance and Williams(1967)}]{Lance1967}
Lance, G.~N.; and Williams, W.~T. 1967.
\newblock A general theory of classificatory sorting strategies: 1. Hierarchical systems.
\newblock \emph{The Computer Journal}, 9(4): 373--380.

\bibitem[{Leskovec et~al.(2008)Leskovec, Lang, Dasgupta, and Mahoney}]{Leskovec2008}
Leskovec, J.; Lang, K.~J.; Dasgupta, A.; and Mahoney, M.~W. 2008.
\newblock Statistical properties of community structure in large social and information networks.
\newblock In \emph{Proceedings of the 17th International Conference on World Wide Web}, 695--704.

\bibitem[{Li et~al.(2008)Li, Zhang, Wang, Zhang, and Chen}]{Li2008}
Li, Z.; Zhang, S.; Wang, R.-S.; Zhang, X.-S.; and Chen, L. 2008.
\newblock Quantitative function for community detection.
\newblock \emph{Physical Review E}, 77(3): 036109.

\bibitem[{Little and Rubin(2002)}]{Little2002}
Little, R. J.~A.; and Rubin, D.~B. 2002.
\newblock \emph{Statistical Analysis with Missing Data}.
\newblock Wiley.

\bibitem[{Newman(2003)}]{Newman2003}
Newman, M. E.~J. 2003.
\newblock The structure and function of complex networks.
\newblock \emph{SIAM Review}, 45(2): 167--256.

\bibitem[{Newman(2006)}]{Newman2006}
Newman, M. E.~J. 2006.
\newblock Modularity and community structure in networks.
\newblock \emph{Proceedings of the National Academy of Sciences}, 103(23): 8577--8582.

\bibitem[{Radicchi et~al.(2004)Radicchi, Castellano, Cecconi, Loreto, and Parisi}]{Radicchi2004}
Radicchi, F.; Castellano, C.; Cecconi, F.; Loreto, V.; and Parisi, D. 2004.
\newblock Defining and identifying communities in networks.
\newblock \emph{Proceedings of the National Academy of Sciences}, 101(9): 2658--2663.

\bibitem[{Shi and Malik(2000)}]{Shi2000}
Shi, J.; and Malik, J. 2000.
\newblock Normalized cuts and image segmentation.
\newblock \emph{IEEE Transactions on Pattern Analysis and Machine Intelligence}, 22(8): 888--905.

\bibitem[{Sneath and Sokal(1973)}]{Sneath1973}
Sneath, P. H.~A.; and Sokal, R.~R. 1973.
\newblock \emph{Numerical Taxonomy}.
\newblock Freeman.

\bibitem[{Spearman(1904)}]{Spearman1904}
Spearman, C. 1904.
\newblock The proof and measurement of association between two things.
\newblock \emph{The American Journal of Psychology}, 15(1): 72--101.

\bibitem[{Tan, Steinbach, and Kumar(2005)}]{Tan2005}
Tan, P.-N.; Steinbach, M.; and Kumar, V. 2005.
\newblock \emph{Introduction to Data Mining}.
\newblock Pearson.

\bibitem[{Troyanskaya et~al.(2001)Troyanskaya, Cantor, Sherlock, Brown, Hastie, Tibshirani, Botstein, and Altman}]{Troyanskaya2001}
Troyanskaya, O.; Cantor, M.; Sherlock, G.; Brown, P.; Hastie, T.; Tibshirani, R.; Botstein, D.; and Altman, R.~B. 2001.
\newblock Missing value estimation methods for DNA microarrays.
\newblock \emph{Bioinformatics}, 17(6): 520--525.

\bibitem[{van Buuren and Groothuis-Oudshoorn(2011)}]{Buuren2011}
van Buuren, S.; and Groothuis-Oudshoorn, K. 2011.
\newblock MICE: Multivariate Imputation by Chained Equations in R.
\newblock \emph{Journal of Statistical Software}, 45(3).

\bibitem[{Watts and Strogatz(1998)}]{Watts1998}
Watts, D.~J.; and Strogatz, S.~H. 1998.
\newblock Collective dynamics of 'small-world' networks.
\newblock \emph{Nature}, 393(6684): 440--442.

\end{thebibliography}
\clearpage
%%%
\appendix
\counterwithin*{equation}{section}
\counterwithin*{figure}{section}
\renewcommand\theequation{\thesection\arabic{equation}}
\renewcommand\thefigure{\thesection\arabic{figure}}
\addcontentsline{toc}{section}{Appendices}
% \section*{Appendices}
\section{Appendix} \label{appendix}

\subsection{Range of Weighted Similarity}
\label{apdx:Weighted_Proof}

Assume we use cosine similarity to quantify the similarity for \textcolor{red}{Part 1} where both verctors having numercial value. Here we prove that the range of the weighted similarity metric \(S\), is \(S \in [-1, 1]\).

Define \(|V_{n1}| = \text{length of the numerical \textcolor{red}{Part 1}}\), \(|V_1| = \text{length of vector } V_1 \text{ or } V_2\), \(C_{Nan} = \text{length of missing \textcolor{green}{Part 2} }\) and \(C_{Non}= \text{length of the non-matching \textcolor{violet}{Part 3} }\), consider

\begin{equation}
\label{eq: eqA1}
    -\frac{|V_{n1}|}{|V_1|} \le S_{Num} \le \frac{|V_{n1}|}{|V_1|}
\end{equation}

\begin{equation}
\label{eq: eqA2}
    0 \le S_{Nan} \le \frac{C_{Nan}}{|V_1|}
\end{equation}

\begin{equation}
\label{eq: eqA3}
    -\frac{C_{Non}}{|V_1|} \le S_{Non} \le 0
\end{equation}
and 
\begin{equation}
\label{eq: eqA4}
 C_{Non} + C_{Nan} + |V_{n1}| = |V_1|
\end{equation}

\( (\ref{eq: eqA1}) +  (\ref{eq: eqA2}) + (\ref{eq: eqA3}) \) implies 

\begin{equation}
\label{eq: eqA5}
-\bigg( \frac{|V_{n1}| + C_{Non}}{|V_1|} \bigg) \le S_{Num} + S_{Nan} + S_{Non} \le \frac{|V_{n1}| + C_{Nan}}{|V_1|} 
\end{equation}

Define

\begin{equation}
\label{eq: eqA6}
    S = S_{Num} + S_{Nan} + S_{Non}
\end{equation}

From \((\ref{eq: eqA4})\), it can be deduced that \(|V_{n1}| = |V_1|- C_{Non} - C_{Nan}\). Add \(-\frac{C_{Nan}}{|V_1|}\) to both sides of \( (\ref{eq: eqA5})\) and combine \( (\ref{eq: eqA6})\) gives

\begin{equation}
\label{eq: eqA7}
    -1 \le S \le \frac{|V_{n1}|}{|V_1|}
\end{equation}

Likewise, add \(\frac{C_{Non}}{|V_1|}\) to both sides of \( (\ref{eq: eqA5})\) and combine \( (\ref{eq: eqA6})\) gives

\begin{equation}
\label{eq: eqA8}
    - \frac{|V_{n1}|}{|V_1|} \le S \le 1
\end{equation}

\( \text{ Add } (\ref{eq: eqA7}) \text{ and } (\ref{eq: eqA8}) \text{ gives } \)

\begin{equation}
\label{eq: eqA9}
    - \bigg(1 + \frac{|V_{n1}|}{|V_1|} \bigg) \le 2S \le 1 + \frac{|V_{n1}|}{|V_1|}.
\end{equation}

Then 

\begin{equation}
\label{eq: eqA10}
    - \bigg(\frac{ |V_1| + |V_{n1}|}{2|V_1|} \bigg) \le S \le \frac{ |V_1| + |V_{n1}|}{2|V_1|}.
\end{equation}

Since \(\frac{ |V_1| + |V_{n1}|}{2|V_1|}  > 0 \) then \( (\ref{eq: eqA10}) \) implies 

\begin{equation}
\label{eq: eqA11}
    -1 \le S \le 1.
\end{equation}

Therefore, we have proved that \(S \in [-1, 1].\)

\subsection{Traditional Imputation Methods} \label{apdx:Imputation}
The formula for each of the imputation considered are defined below:
\begin{enumerate}
    \item Mean imputation: This imputation metric is defined by:
    \begin{equation}
    x_i = 
\begin{cases} 
\overline{x} & \text{if } x_i \text{ is missing} \\
x_i & \text{otherwise}
\end{cases}
\label{eq:mean_impute}
\end{equation} where \( \overline{x} \) is the mean of the non-missing values of feature \( x \).

\item  K-Nearest Neighbors (KNN) imputation: This imputation metric is defined by:
\begin{equation}
x_i = \frac{1}{K} \sum_{j \in \text{N}(i)} x_j
\label{eq:KNN_impute}
\end{equation}

where \( \text{N}(i) \) denotes the set of \( K \) nearest neighbors of instance \( i \).

\item MICE: The metric defined for this imputation is given by:

\begin{equation}
\hat{x}_i^{(m)} = f(X_{\text{obs}}) + \epsilon_i^{(m)}
\label{eq:MICE_impute}
\end{equation}

where \( X_{\text{obs}} \) represents observed values, \( f \) is the imputation model, and \( \epsilon_i^{(m)} \) is a random error term. The final imputed value is an average of the multiple imputations.
\end{enumerate}

\subsection{Traditional Similarity Measures}\label{apdx:Similarity}

The formulas of the $4$ similarities considered in this paper are defined below:

\begin{enumerate}
    \item Cosine Similarity: This similarity formula is defined by: 

    \begin{equation}
\text{Cosine Similarity} = \frac{\mathbf{A} \cdot \mathbf{B}}{\|\mathbf{A}\| \|\mathbf{B}\|}
\label{eq:cosine_sim}
\end{equation}

where \( \mathbf{A} \cdot \mathbf{B} \) is the dot product of vectors \( \mathbf{A} \) and \( \mathbf{B} \), and \( \|\mathbf{A}\| \) and \( \|\mathbf{B}\| \) are their magnitudes.

\item Euclidean Distance: The euclidean distance metric is defined by:
\begin{equation}
\text{Euclidean Distance} = \sqrt{\sum_{i=1}^{n} (a_i - b_i)^2}
\label{eq:euclidean_sim}
\end{equation} where \( a_i \) and \( b_i \) are the components of vectors \( \mathbf{A} \) and \( \mathbf{B} \).

\item Canberra Distance: This distance metric is defined by: 

\begin{equation}
\text{Canberra Distance} = \sum_{i=1}^{n} \frac{|a_i - b_i|}{|a_i| + |b_i|}
\label{eq:canberra_sim}
\end{equation}

\item Spearman Correlation: This correlation metric is defined by:

\begin{equation}
\text{Spearman's } \rho = 1 - \frac{6 \sum d_i^2}{n(n^2 - 1)}
\label{eq:spearman_sim}
\end{equation}

where \( d_i \) is the difference between the ranks of corresponding values.
\end{enumerate}

\subsection{The Girvan-Newman Algorithm} \label{apdx:GirvanNewman}

The Girvan-Newman algorithm is a popular method for community detection in complex networks. It identifies communities by iteratively removing edges with the highest betweenness centrality, which measures the number of shortest paths that pass through an edge. The removal of these edges disrupts the network, eventually splitting it into distinct communities \citep{Girvan2002}.

\subsubsection{Betweenness Centrality}

Betweenness centrality is a key concept in the Girvan-Newman algorithm. It quantifies the importance of an edge by counting the number of shortest paths between pairs of nodes that pass through it. Formally, the betweenness centrality \( C_B(e) \) of an edge \( e \) is given by:

\begin{equation}
C_B(e) = \sum_{s \neq t \neq e} \frac{\sigma_{st}(e)}{\sigma_{st}}
\label{eq:bln_cent}
\end{equation}

where \( \sigma_{st} \) is the total number of shortest paths from node \( s \) to node \( t \), and \( \sigma_{st}(e) \) is the number of those paths that pass through edge \( e \).

\subsubsection{Algorithm Steps}

The Girvan-Newman algorithm proceeds as follows:

\begin{enumerate}
    \item \textbf{Compute Betweenness Centrality}: Calculate the betweenness centrality for all edges in the network.
    \item \textbf{Remove Edge}: Remove the edge with the highest betweenness centrality.
    \item \textbf{Recompute Betweenness Centrality}: Recalculate the betweenness centrality for all remaining edges.
    \item \textbf{Repeat}: Repeat steps 2 and 3 until no edges remain or until the desired number of communities is achieved.
\end{enumerate}

This iterative process effectively identifies the most critical edges that connect different communities. By removing these edges, the algorithm reveals the underlying community structure of the network. The Girvan-Newman algorithm is advantageous because it does not require a priori knowledge of the number of communities and can reveal hierarchical community structures. However, it is computationally intensive, especially for large networks, due to the repeated computation of betweenness centrality. Additionally, the algorithm's performance is highly dependent on the quality of the similarity measure used to define the edges. For the purpose of the study, we employ parallel computing in this stage to utilize the algorithm.

\subsection{Performance Metrics of the Community Detection} \label{apdx:performanceMetrics}

As already introduced in the early section, the metrics are grouped into two folds: general community structure performance and average community quality metrics. The formulas are defined in the subsequent sections.

\subsubsection{General Community Structures Performance Metrics}

\begin{enumerate}
    \item \textbf{Modularity (Q)} \citep{Newman2006}: One of the most widely used metrics for assessing community structure is modularity (Q) which measures the strength of the division of a network into communities as defined by \eqref{eq:modularity}. High modularity indicates dense connections within communities and sparse connections between communities. The metric is defined as
    \begin{equation}
        Q = \frac{1}{2m} \sum _{i,j} \left[A_{ij} - \frac{k_i k_j}{2m} \right]\delta (c_i, c_j)
        \label{eq:modularity}
    \end{equation}
    where $A_{ij}$ is the adjacency matrix, $k_i$ is the degree of node $i$, $m$ is the total number of edges in the network, $c_i$ is the community to which node $i$ belongs, and $\delta (c_i, c_j)$ is the Kronecker delta, which is $1$ if $c_i = c_j$ and $0$ otherwise. Modularity values range from -1 to 1. Higher values indicate a better-defined community structure, where nodes within the same community are more densely connected than nodes in different communities (stronger community structure with dense intra-community edges and sparse inter-community edges).

    \item \textbf{Coverage} \citep{Fortunato2010}: Coverage is the fraction of edges that fall within the detected communities compared to the total number of edges in the network and is defined by:
    \begin{equation}
        \text{Coverage} = \frac{\text{Number of intra-community edges}}{\text{Total number of edges}}
        \label{eq:coverage}
    \end{equation}
    Coverage values range from 0 to 1. High coverage values suggest that the algorithm effectively captures the overall structure of the network by ensuring that most edges are within communities. This metric is useful for understanding how well the algorithm can preserve the internal connectivity of communities.

    \item \textbf{Dunn Index} \citep{Dunn1974}: As defined by \eqref{eq:DI}, Dunn index is a ratio of the minimum inter-cluster distance to the maximum intra-cluster distance. Higher values indicate better clustering.
    \begin{equation}
        D = \frac{\min\{ d(i,j) | 1\le i \le j \le n \}}{\max\{ \delta(k) | 1 \le k \le n \}}
        \label{eq:DI}
    \end{equation}
    where $d(i,j)$ is the distance between clusters $i$ and $j$, and $\delta(k)$ is the diameter of cluster $k$. Higher Dunn index values indicate better clustering with well-separated clusters and compact intra-cluster distances. This metric helps assess the distinctiveness and cohesiveness of the detected communities. 

    \item \textbf{Average Clustering Coefficient} \citep{Watts1998}: This coefficient measures the degree to which nodes in a graph tend to cluster together. Values range from 0 to 1, with higher values indicating a higher degree of clustering. It is defined by:

    \begin{equation}
        C = \frac{1}{n} \sum_{i=1}^{n} C_i
        \label{eq:clustering_coefficient}
    \end{equation}
    where $C_i$ is the local clustering coefficient of node $i$, and $n$ is the total number of nodes.

    \item \textbf{Transitivity} \citep{Newman2003}: This metric measures the probability that the adjacent vertices of a vertex are connected as defined by \eqref{eq:transitivity}. Values range from 0 to 1, with higher values indicating that nodes tend to form triangles, suggesting strong community structure.

    \begin{equation}
        T = \frac{3 \times \text{number of triangles}}{\text{number of connected triples of vertices}}
        \label{eq:transitivity}
    \end{equation}

    \item \textbf{Modularity Density} \citep{Li2008}: This metric considers the density of communities. Higher values indicate better community structure with consideration for density and is defined by \eqref{eq:modularity_density}.

    \begin{equation}
        Q_D = \sum_{i=1}^{k} \left( \frac{2m_i}{n_i(n_i-1)} \right) \frac{d_i}{m}
        \label{eq:modularity_density}
    \end{equation}
    where $m_i$ is the number of edges in community $i$, $n_i$ is the number of nodes in community $i$, $d_i$ is the internal density of community $i$, and $m$ is the total number of edges in the network.

    \item \textbf{Triangle Participation Ratio (TPR)} \citep{Jaewon2012}: As defined by \eqref{eq:tpr}, this ratio measures the fraction of nodes participating in at least one triangle. Values range from 0 to 1, with higher values indicating more participation in triangles, suggesting a stronger community structure.
    
     \begin{equation}
        \text{TPR} = \frac{|\{v \in V : v \text{ participates in at least one triangle}\}|}{|V|}
        \label{eq:tpr}
    \end{equation}
    where $V$ is the set of all nodes.
    
\end{enumerate}

\subsubsection{Average Community Quality Metrics}

\begin{enumerate}
    \item \textbf{Conductance} \citep{Kannan2004}: Conductance is a crucial metric for evaluating the quality of individual communities. It measures the fraction of total edge volume that points outside the community, providing an indication of how well-defined and isolated the community is. Conductance values range from 0 to 1. Lower conductance values suggest better community quality, as fewer edges leave the community, indicating that the community is more self-contained. The metric is defined by
    \begin{equation}
        \phi (S) = \frac{c_S}{2m_S + c_S}
        \label{eq:conductance}
    \end{equation}
    where $c_S$ is the number of edges leaving the community $S$ and $m_S$ is the number of edges inside the community $S$. 

    \item \textbf{Expansion} \citep{Leskovec2008}: This metric measures the number of edges per node that point outside the community. Lower values indicate better community isolation with fewer connections to the rest of the network. This metric helps evaluate how well the community is separated from the rest of the network, which is crucial for understanding the distinctiveness of the community and is defined by:

    \begin{equation}
        \text{Expansion}(S) = \frac{|\{(u,v) \in E : u \in S, v \notin S\}|}{|S|}
        \label{eq:expansion}
    \end{equation}
    where $|S|$ is the number of nodes in community $S$.

    \item \textbf{Normalized Cut} \citep{Shi2000}: This metric measures the total edge weight connecting the nodes in the community to the rest of the graph. Lower values indicate better community isolation and minimal connections between the community and the rest of the network. This metric provides insights into the overall quality of the community structure by considering both the internal and external connections of the community and is defined by:

    \begin{equation}
        \text{Ncut}(S) = \frac{c_S}{2m_S + c_S} + \frac{c_T}{2m_T + c_T}
        \label{eq:normalized_cut}
    \end{equation}
    where $c_S$ is the number of edges leaving community $S$, $m_S$ is the number of edges inside community $S$, $c_T$ is the number of edges leaving the complement of community $S$, and $m_T$ is the number of edges inside the complement of community $S$.

    \item \textbf{Density} \citep{Radicchi2004}: This metric measures the fraction of possible edges within the community that are actually present. Values range from 0 to 1, with higher values indicating denser communities, suggesting that the nodes within the community are more tightly connected. This metric helps assess the internal connectivity of the community and its overall cohesiveness and is defined by:

    \begin{equation}
        \text{Density}(S) = \frac{2m_S}{|S|(|S|-1)}
        \label{eq:density}
    \end{equation}
    where $m_S$ is the number of edges within community $S$ and $|S|$ is the number of nodes in community $S$.

    \item \textbf{Internal Density} \citep{Fortunato2010}: This metric measures the density of edges within the community. Values range from 0 to 1, with higher values indicating better internal connectivity within communities, suggesting that the community is more cohesive and well-connected. This metric provides insights into the quality of the internal structure of the community. Internal Density is defined by:

    \begin{equation}
        \text{Internal Density}(S) = \frac{2m_S}{|S|(|S|-1)}
        \label{eq:internal_density}
    \end{equation}
    where $m_S$ is the number of edges within community $S$ and $|S|$ is the number of nodes in community $S$.

    \item \textbf{Local Modularity} \citep{Clauset2004}: This metric measures the quality of the division of a network into communities on a local scale. Higher values indicate better community structure, suggesting that the community is well-defined and cohesive. This metric is particularly useful for understanding the local properties of the community and its overall quality. It is defined by:

    \begin{equation}
        Q_{\text{local}} = \sum_{i=1}^{k} \left( \frac{l_i}{m} - \left( \frac{d_i}{2m} \right)^2 \right)
        \label{eq:local_modularity}
    \end{equation}
    where $l_i$ is the number of edges within community $i$, $d_i$ is the sum of the degrees of the nodes in community $i$, and $m$ is the total number of edges in the network.
\end{enumerate}

\begin{figure*}[!ht] % Options [h!] for placement
    \centering
    \subfigure[Cosine similarity based on mean imputation]{
        \includegraphics[width=0.35\linewidth]{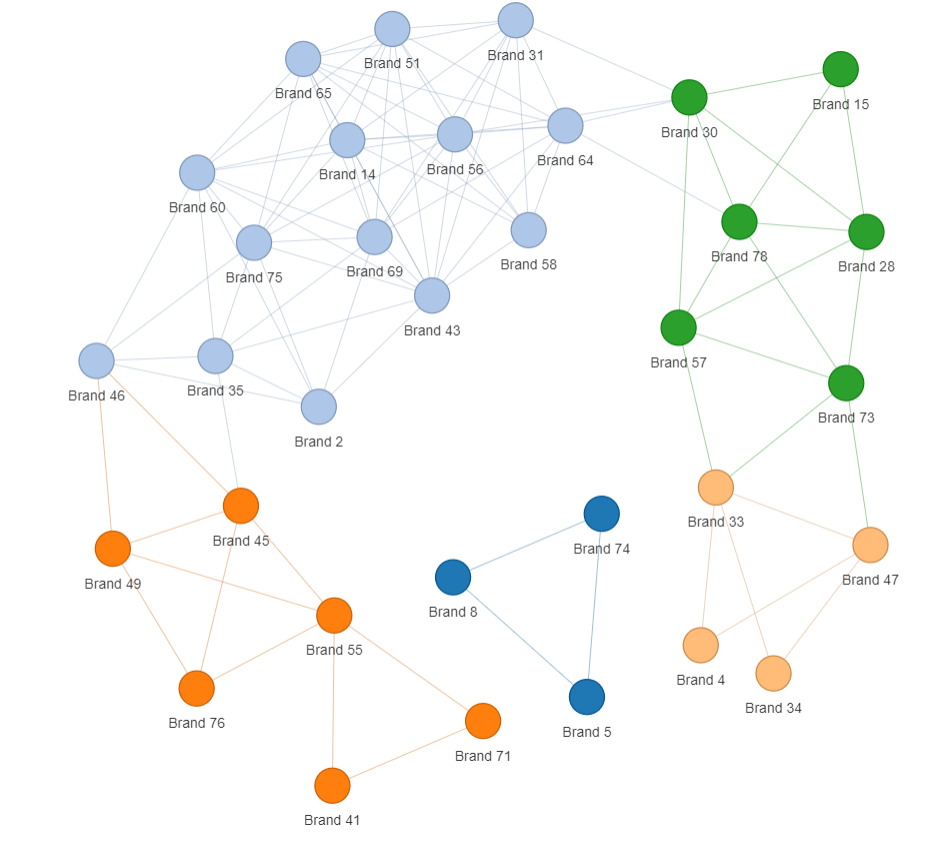}
        \label{fig:meanImp}}
    \subfigure[Cosine similarity based on MICE imputation]{
        \includegraphics[width=0.35\linewidth]{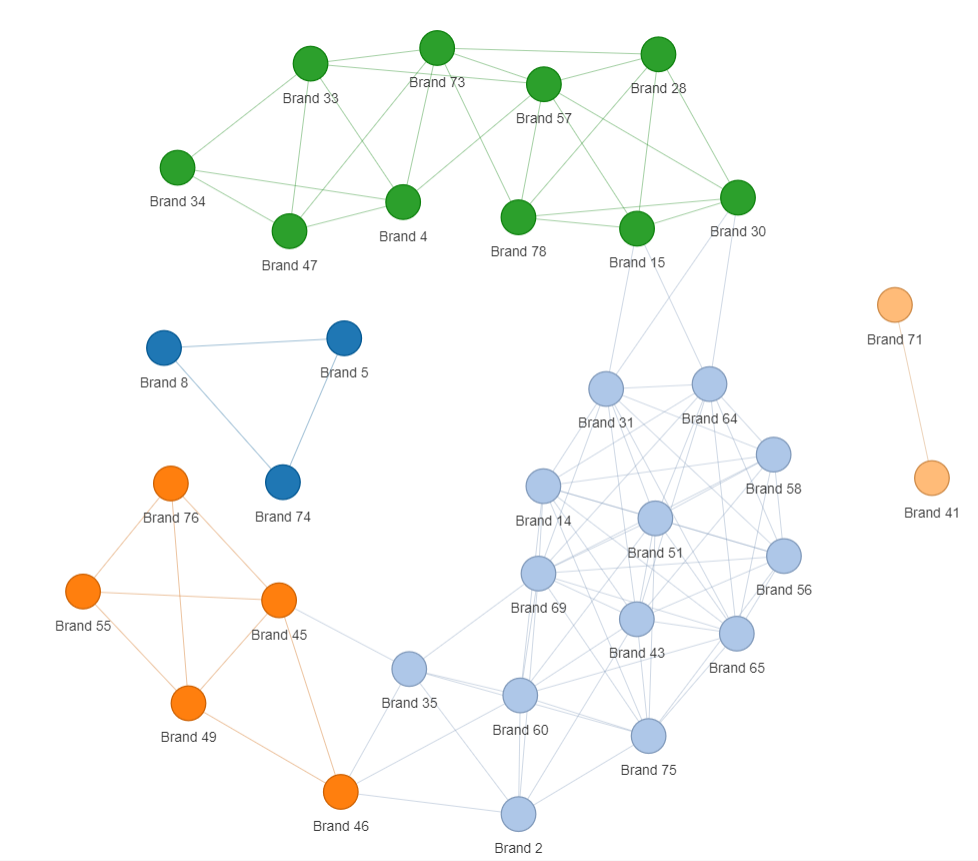}
        \label{fig:MICEImp}}
    \caption{Community structures with 5 clusters and 100 connected edges based on the weighted similarity and imputed cosine similarities using mean and MICE.}
    \label{fig:graph100E5C_CosineAppendix}
\end{figure*}

\begin{figure*}[!ht] % Options [h!] for placement
    \centering
    \subfigure[Weighted similarity]{
        \includegraphics[width=0.485\linewidth, height=0.36\textwidth]{New1.PNG}
        \label{fig:weightedSim}}
    \subfigure[Euclidean similarity based on mean imputation]{
        \includegraphics[width=0.485\linewidth, height=0.36\textwidth]{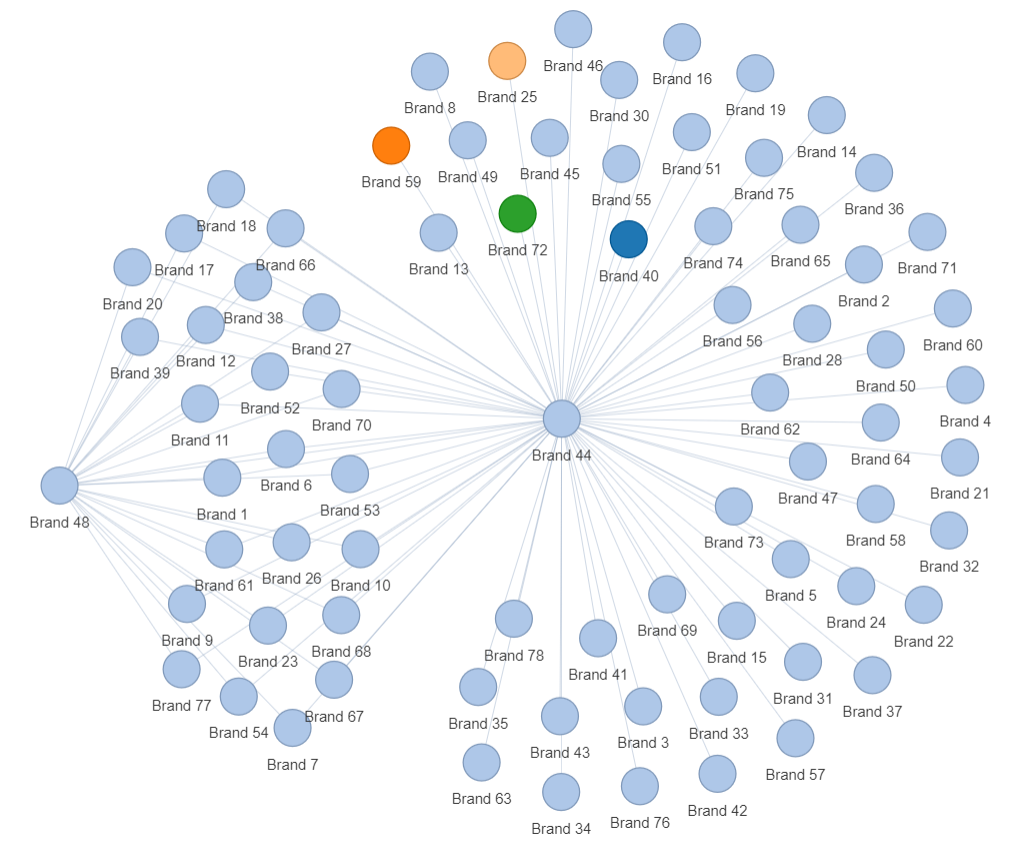}
        \label{fig:meanImp}}
    \subfigure[Euclidean similarity based on MICE imputation]{
        \includegraphics[width=0.485\linewidth, height=0.36\textwidth]{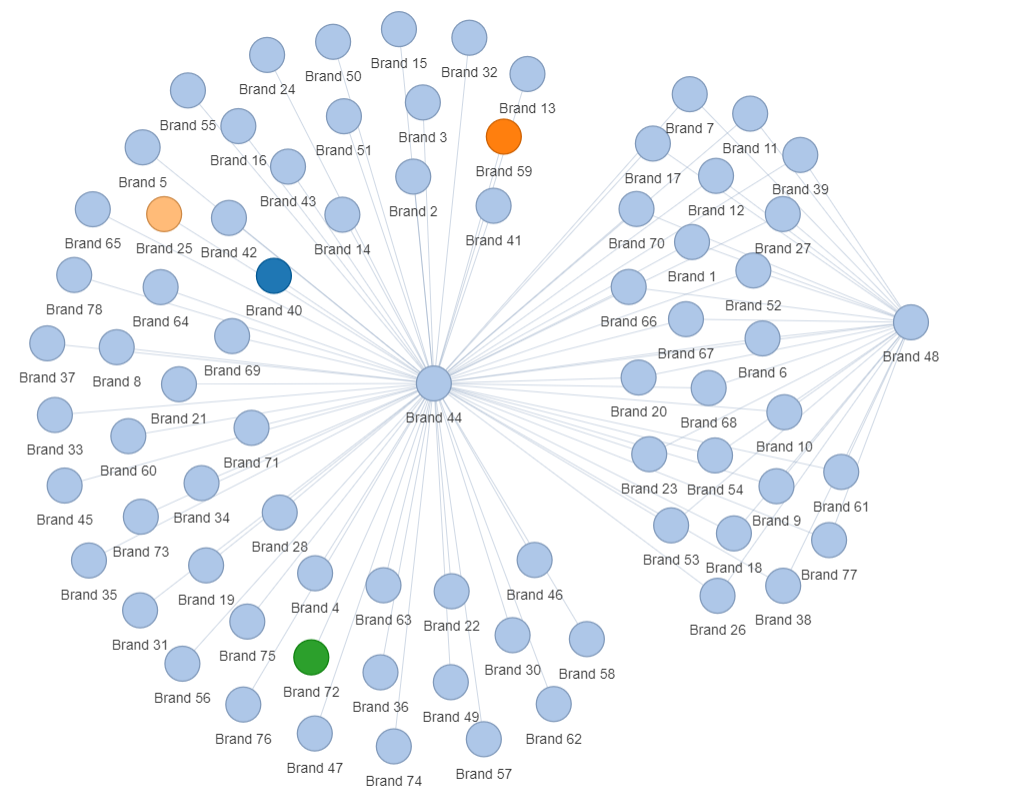}
        \label{fig:MICEImp}}
    \subfigure[Euclidean similarity based on KNN(k=4) imputation]{
        \includegraphics[width=0.485\linewidth, height=0.36\textwidth]{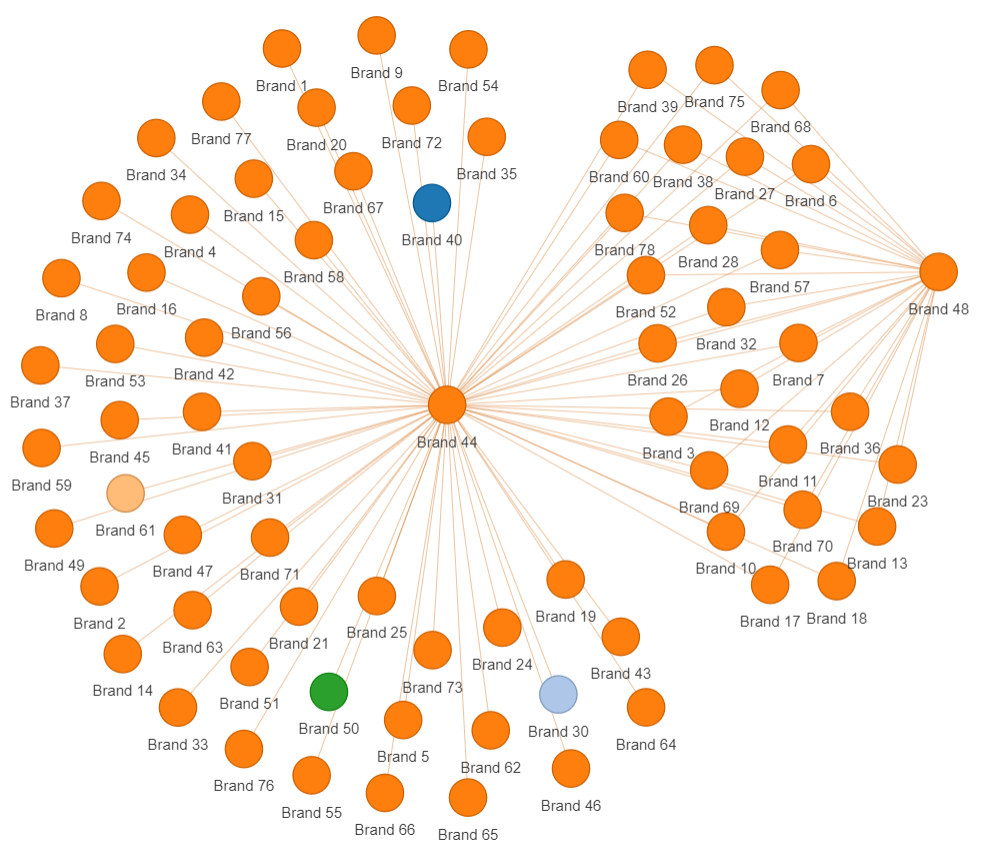}
        \label{fig:KNNImp}}
    \caption{Community structures with 5 clusters and 100 connected edges based on the weighted similarity and imputed euclidean similarities.}
    \label{fig:graph100E5C_Euclidea}
\end{figure*}

\begin{figure*}[!ht] % Options [h!] for placement
    \centering
    \subfigure[Weighted similarity]{
        \includegraphics[width=0.485\linewidth, height=0.36\textwidth]{New1.PNG}
        \label{fig:weightedSim}}
    \subfigure[Canberra similarity based on mean imputation]{
        \includegraphics[width=0.485\linewidth, height=0.36\textwidth]{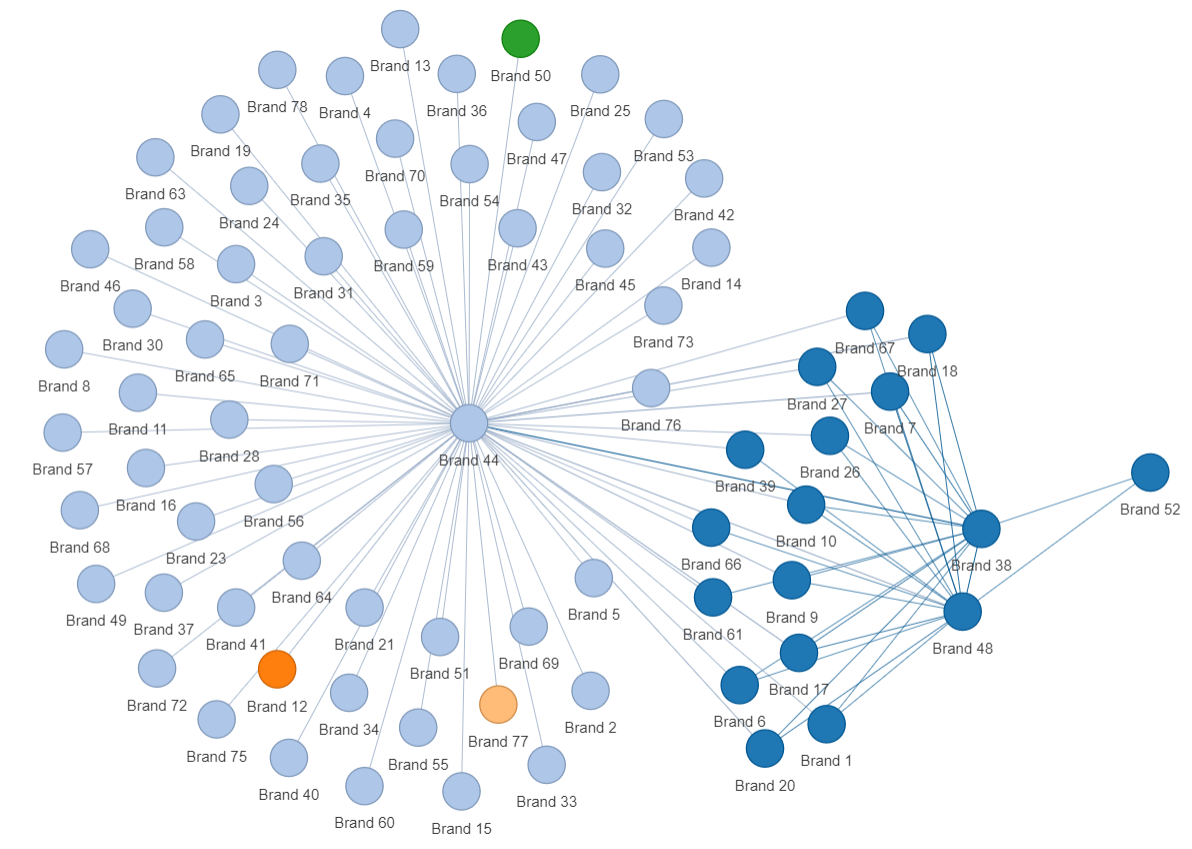}
        \label{fig:meanImp}}
    \subfigure[Canberra similarity based on MICE imputation]{
        \includegraphics[width=0.485\linewidth, height=0.36\textwidth]{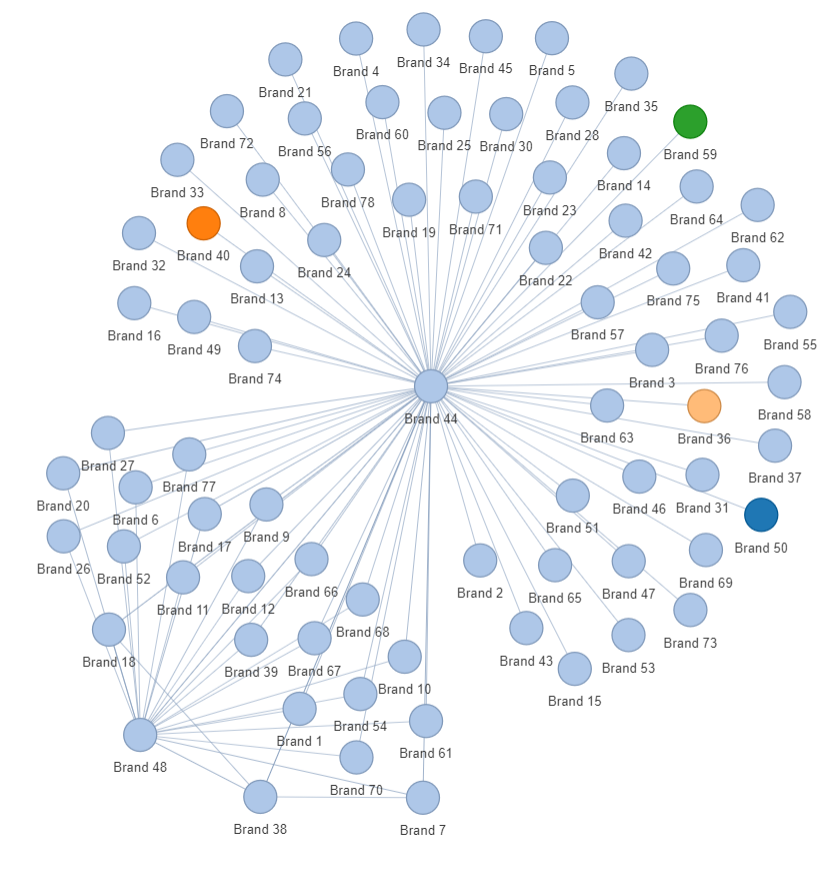}
        \label{fig:MICEImp}}
    \subfigure[Canberra similarity based on KNN(k=4) imputation]{
        \includegraphics[width=0.485\linewidth, height=0.36\textwidth]{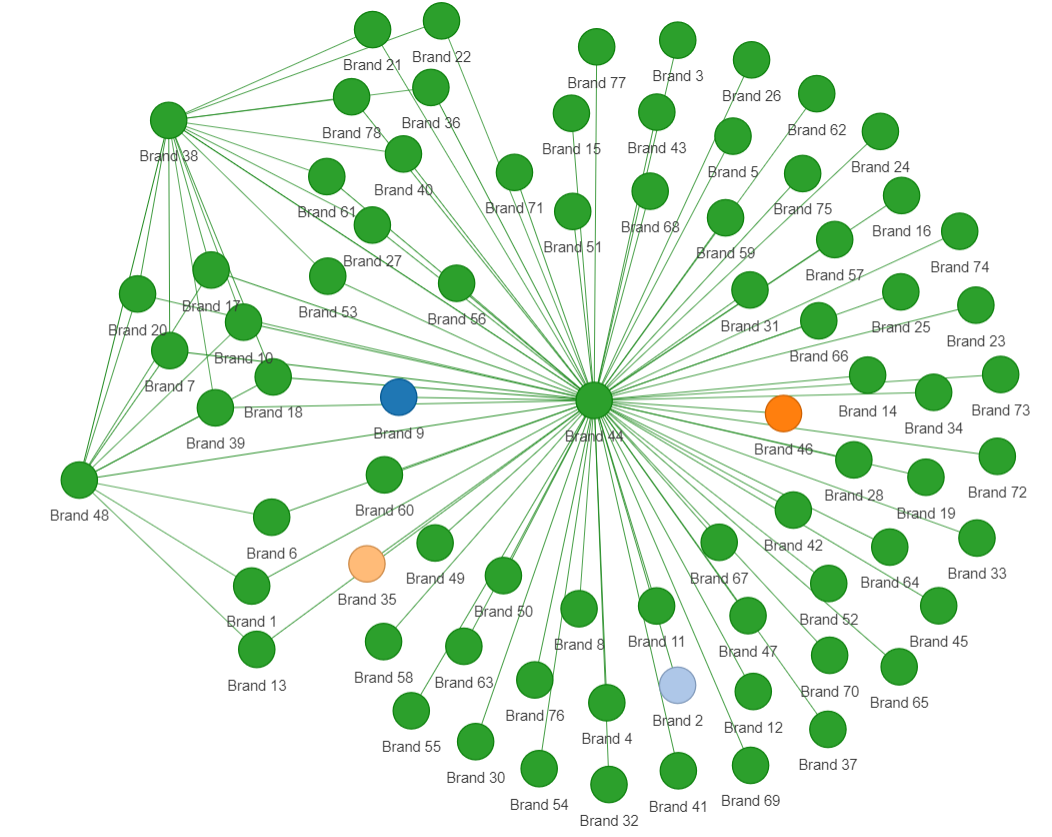}
        \label{fig:KNNImp}}
    \caption{Community structures with 5 clusters and 100 connected edges based on the weighted similarity and imputed canberra similarities.}
    \label{fig:graph100E5C_Canberra}
\end{figure*}

\begin{figure*}[!ht] % Options [h!] for placement
    \centering
    \subfigure[Weighted similarity]{
        \includegraphics[width=0.485\linewidth, height=0.36\textwidth]{New1.PNG}
        \label{fig:weightedSim}}
    \subfigure[Spearman similarity based on mean imputation]{
        \includegraphics[width=0.485\linewidth, height=0.36\textwidth]{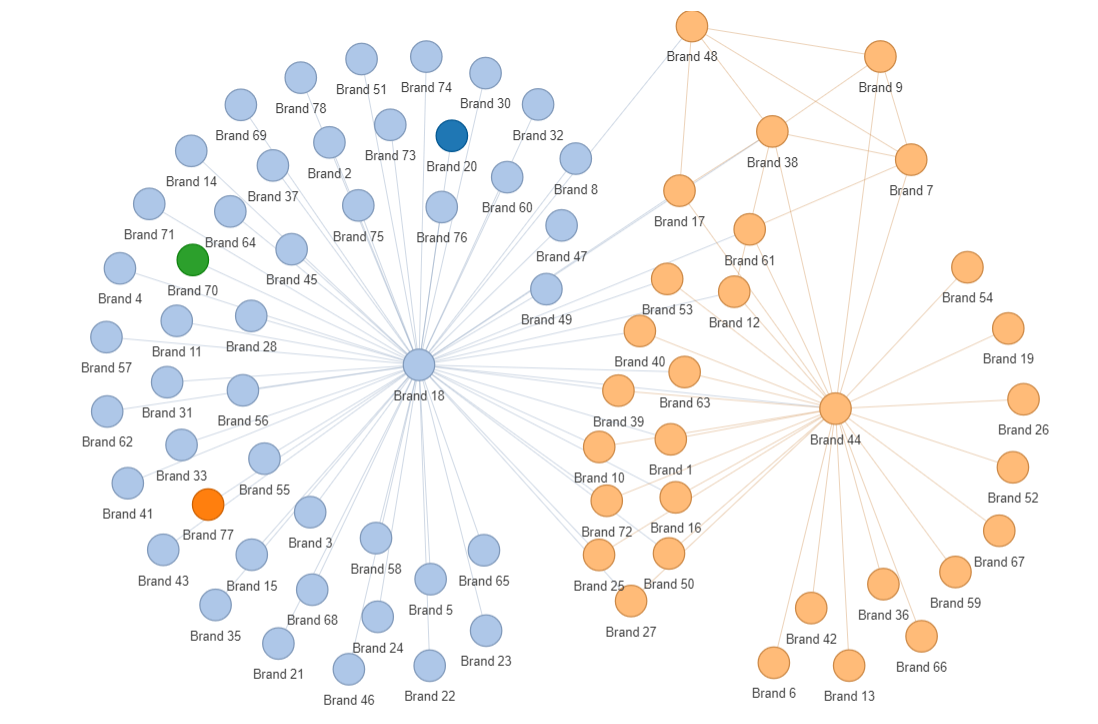}
        \label{fig:meanImp}}
    \subfigure[Spearman similarity based on MICE imputation]{
        \includegraphics[width=0.485\linewidth, height=0.36\textwidth]{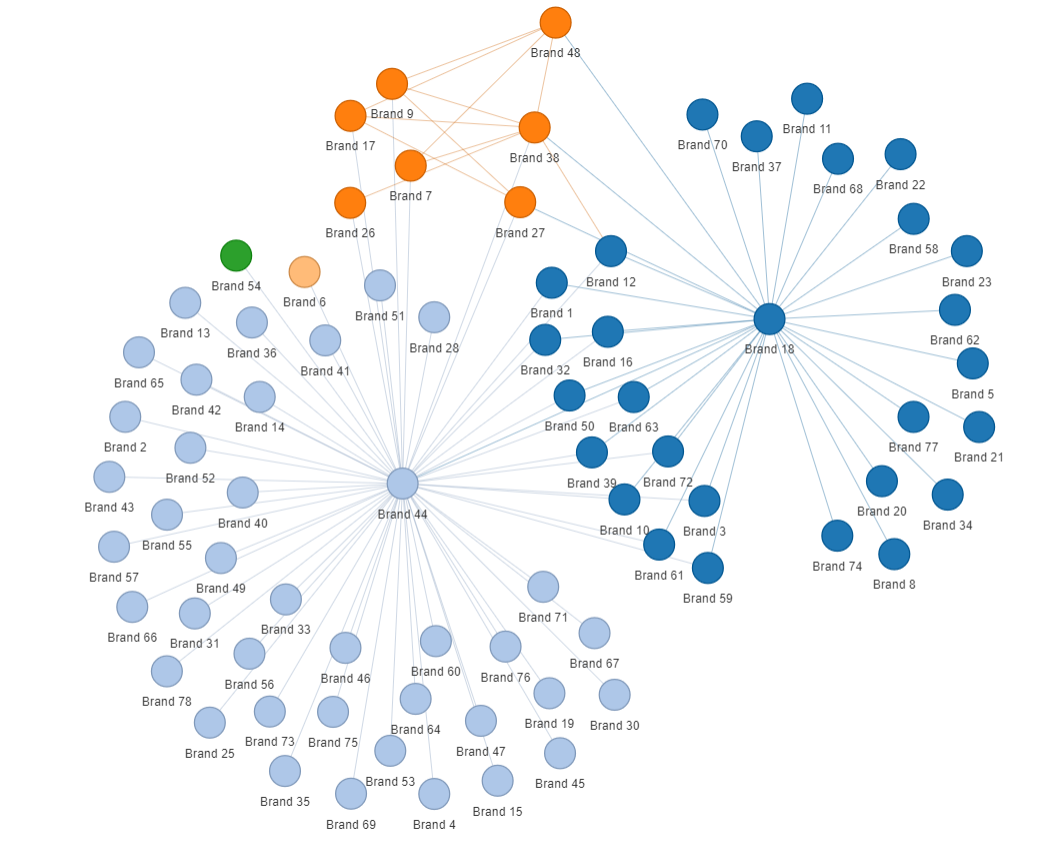}
        \label{fig:MICEImp}}
    \subfigure[Spearman similarity based on KNN(k=4) imputation]{
        \includegraphics[width=0.485\linewidth, height=0.36\textwidth]{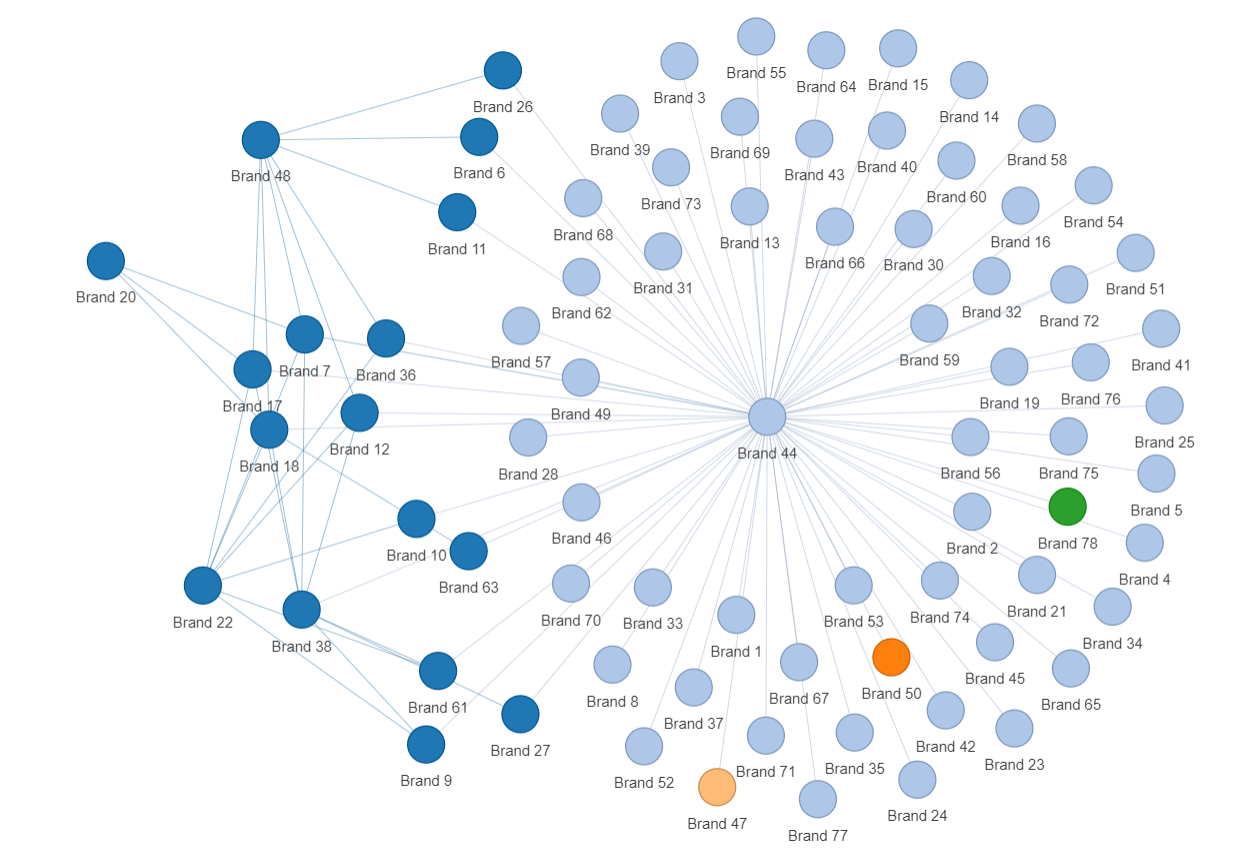}
        \label{fig:KNNImp}}
    \caption{Community structures with 5 clusters and 100 connected edges based on the weighted similarity and imputed spearman similarities.}
    \label{fig:graph100E5C_Spearman}
\end{figure*}

\begin{figure*}[!ht] % Options [h!] for placement
    \centering
    \subfigure[Weighted similarity]{
        \includegraphics[width=0.485\linewidth, height=0.36\textwidth]{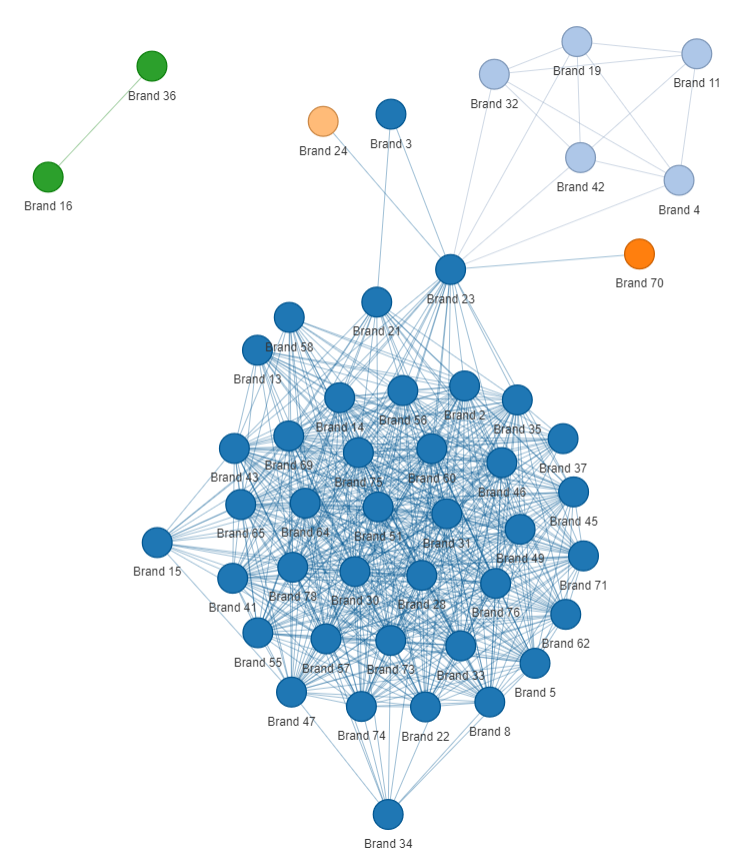}
        \label{fig:left}}
    \subfigure[Cosine similarity based on mean imputation]{
        \includegraphics[width=0.485\linewidth, height=0.36\textwidth]{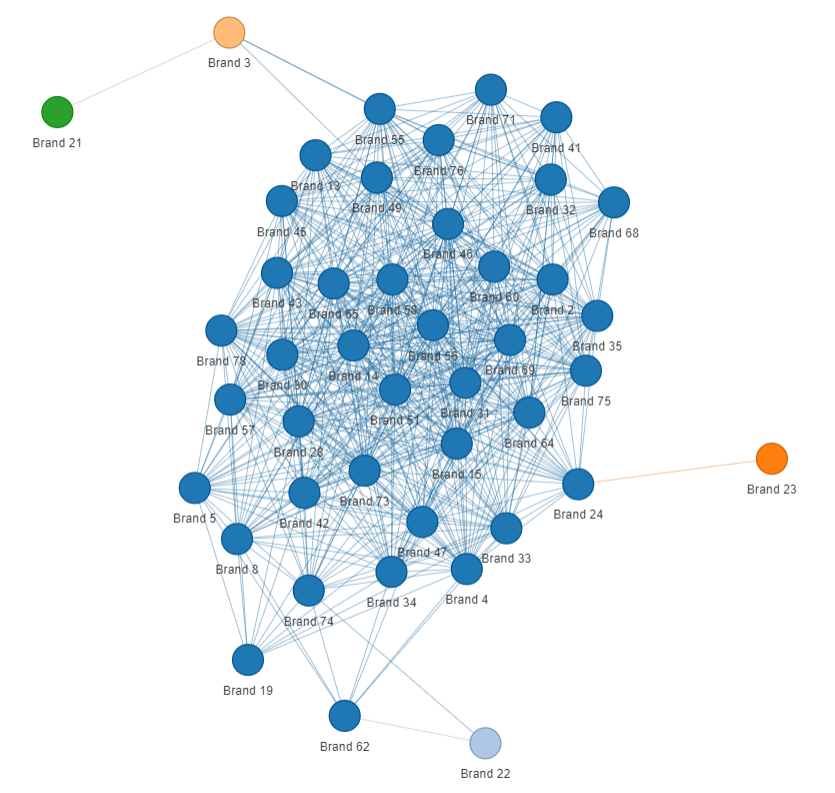}
        \label{fig:right}}
    \subfigure[Cosine similarity based on MICE imputation]{
        \includegraphics[width=0.485\linewidth, height=0.36\textwidth]{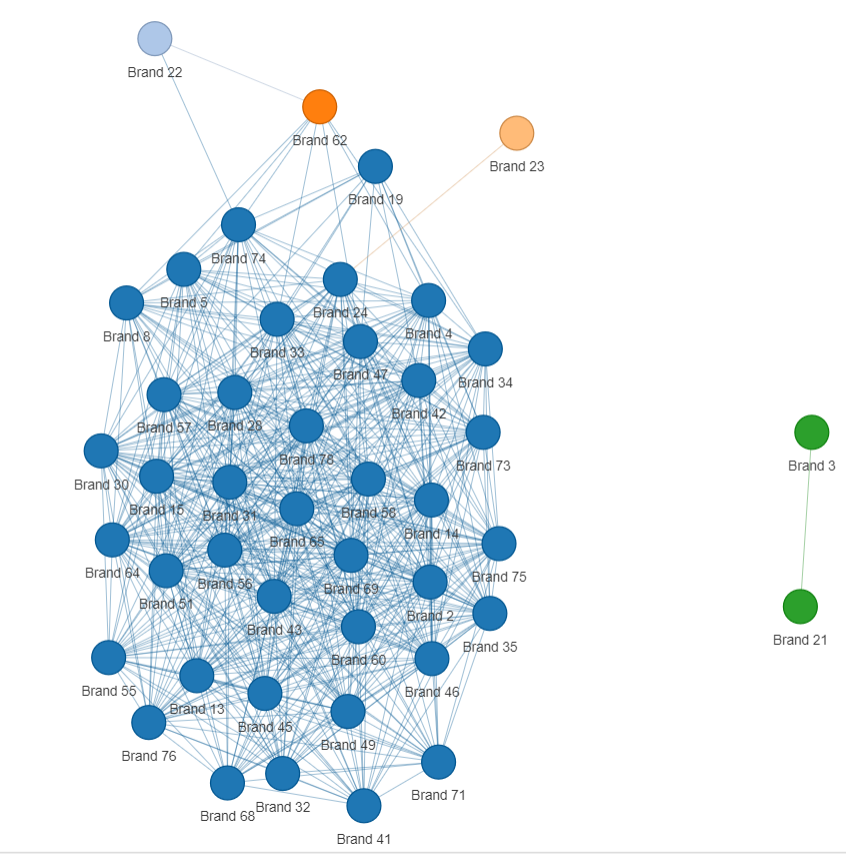}
        \label{fig:left}}
    \subfigure[Cosine similarity based on KNN(k=4) imputation]{
        \includegraphics[width=0.485\linewidth, height=0.36\textwidth]{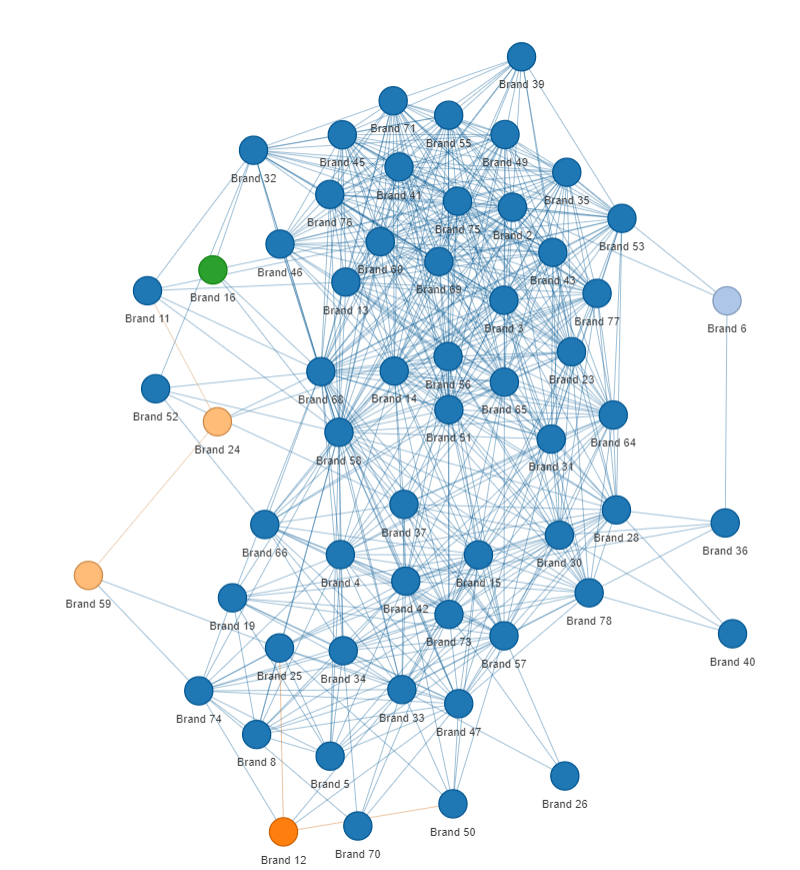}
        \label{fig:right}}
    \caption{Community structures with 5 clusters and 600 connected edges based on the weighted similarity and imputed cosine similarities.}
    \label{fig:graph600E5C_Cosine}
\end{figure*}

\begin{figure*}[!ht] % Options [h!] for placement
    \centering
    \subfigure[Imputed cosine similarity vs. weighted similarity]{
        \includegraphics[width=0.485\linewidth, height=0.36\textwidth]{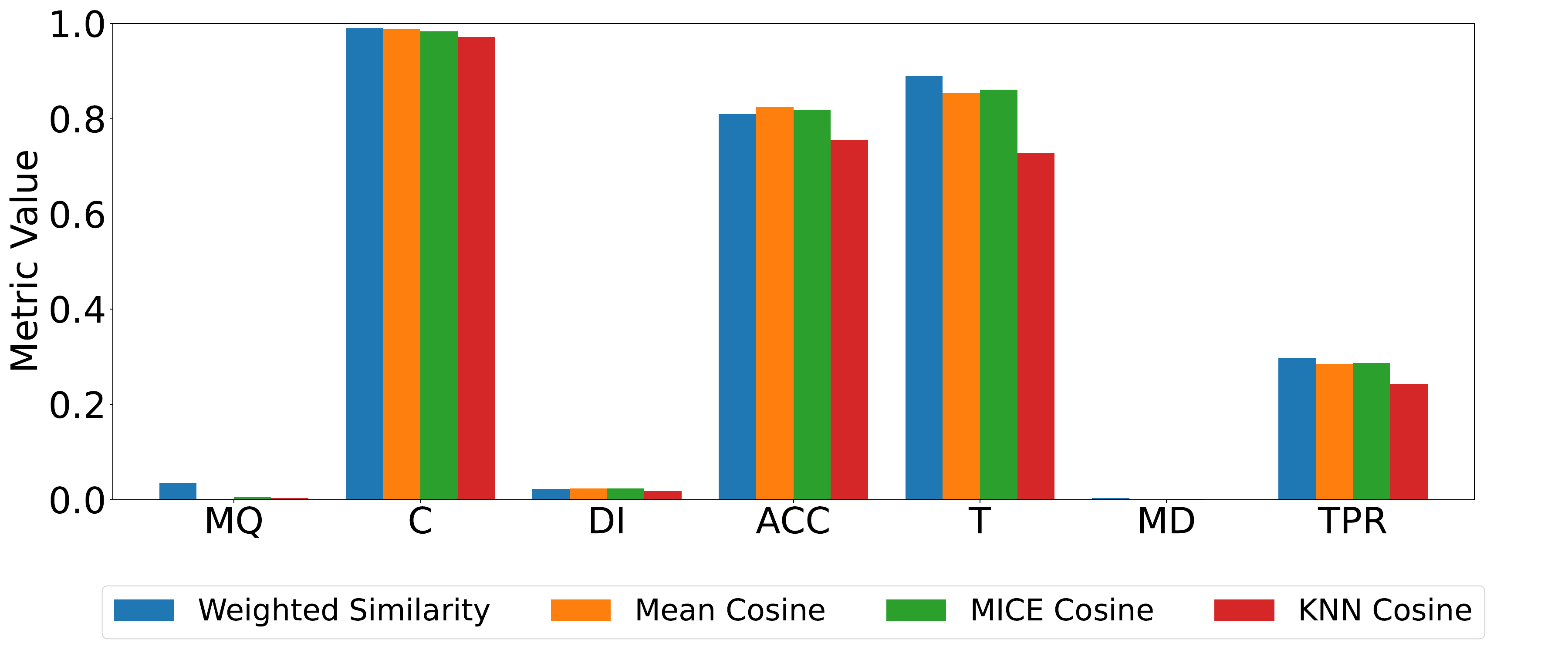}
        \label{fig:left}}
    \subfigure[Imputed euclidean similarity vs. weighted similarity]{
        \includegraphics[width=0.485\linewidth, height=0.36\textwidth]{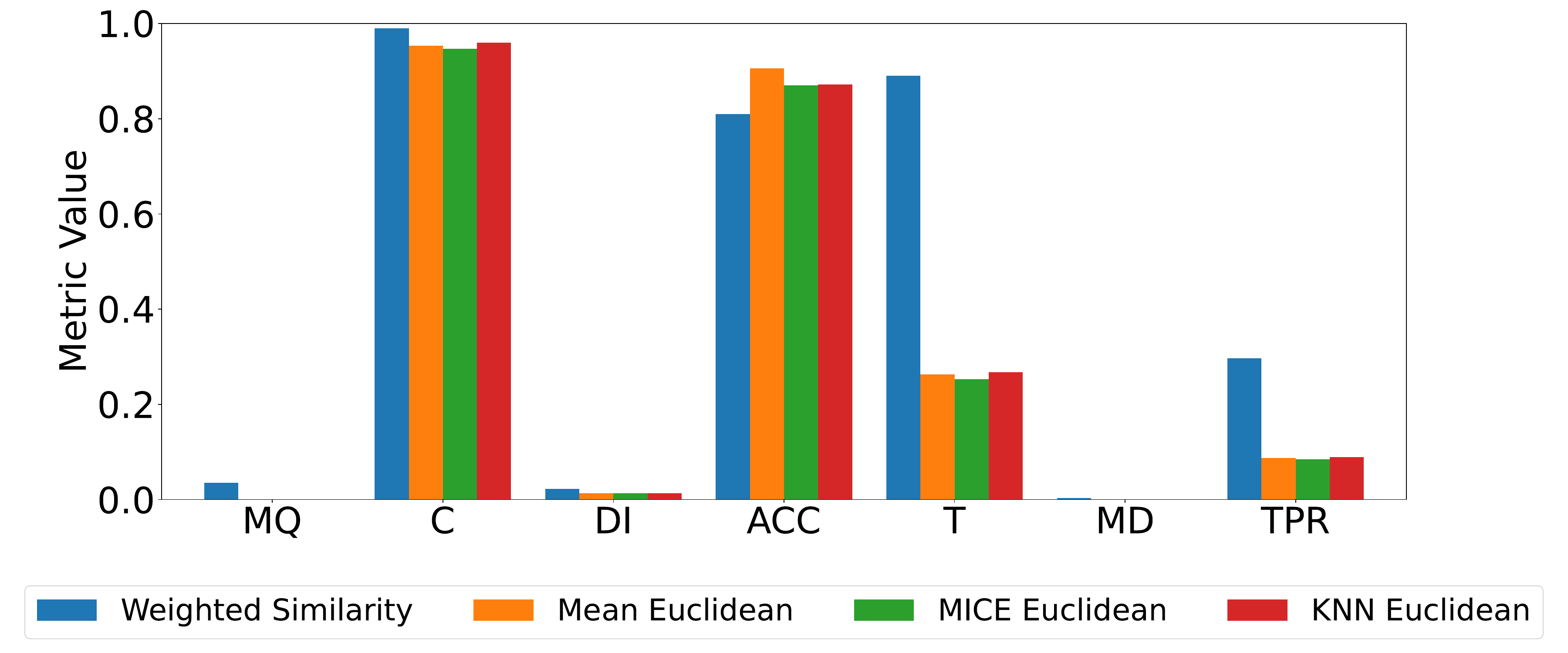}
        \label{fig:right}}
    \subfigure[Imputed canberra similarity vs. weighted similarity]{
        \includegraphics[width=0.485\linewidth, height=0.36\textwidth]{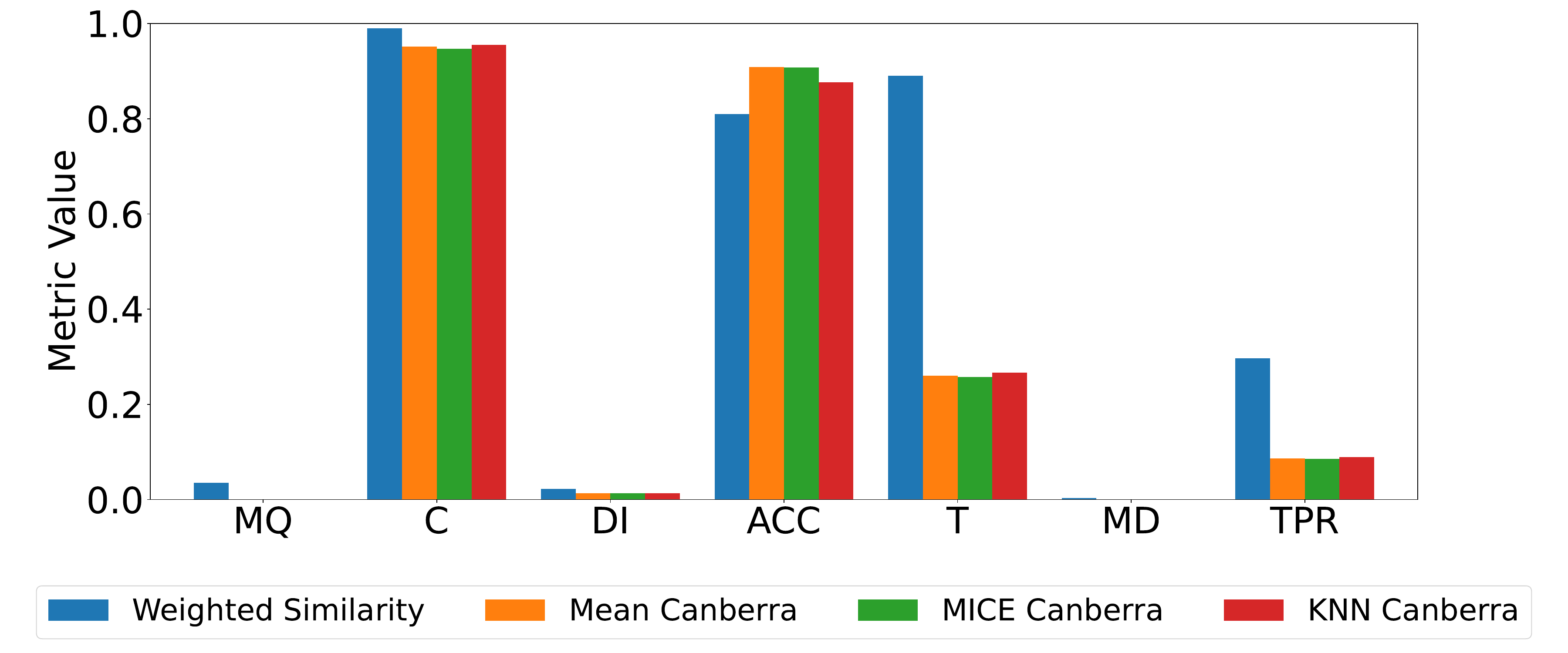}
        \label{fig:left}}
    \subfigure[Imputed spearman vs. weighted similarity]{
        \includegraphics[width=0.485\linewidth, height=0.36\textwidth]{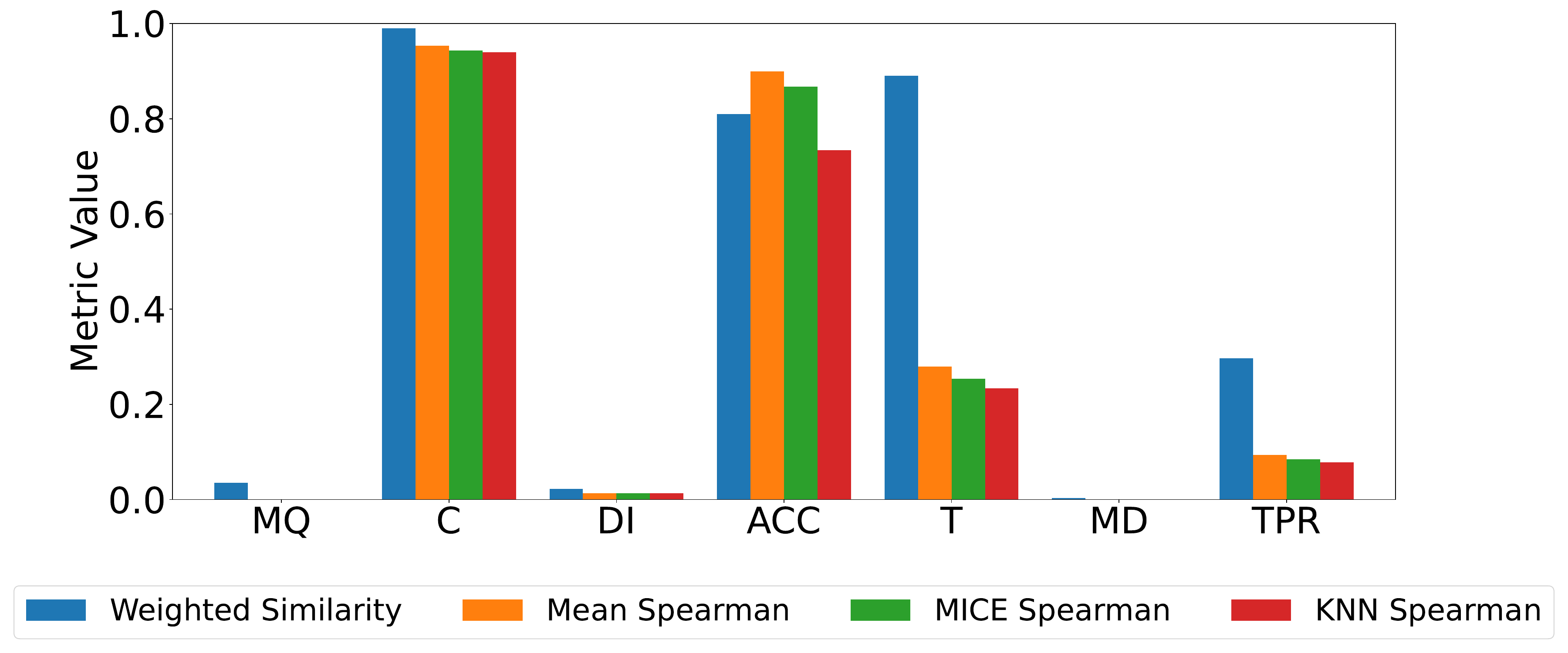}
        \label{fig:right}}
    \caption{Performance metrics on general community structures with 600 connected edges and 5 clusters. MQ: Modularity Q, C: Coverage, DI: Dunn Index, ACC: Average Clustering Coefficient, T: Transitivity, MD: Modularity Density, TPR: Triangle Participation Ratio. Higher bar indicates better community structure.}
    \label{fig:performance600E5C_genCom}
\end{figure*}

\begin{figure*}[!ht] % Options [h!] for placement
    \centering
    \subfigure[Imputed cosine similarity vs. weighted similarity]{
        \includegraphics[width=0.485\linewidth, height=0.36\textwidth]{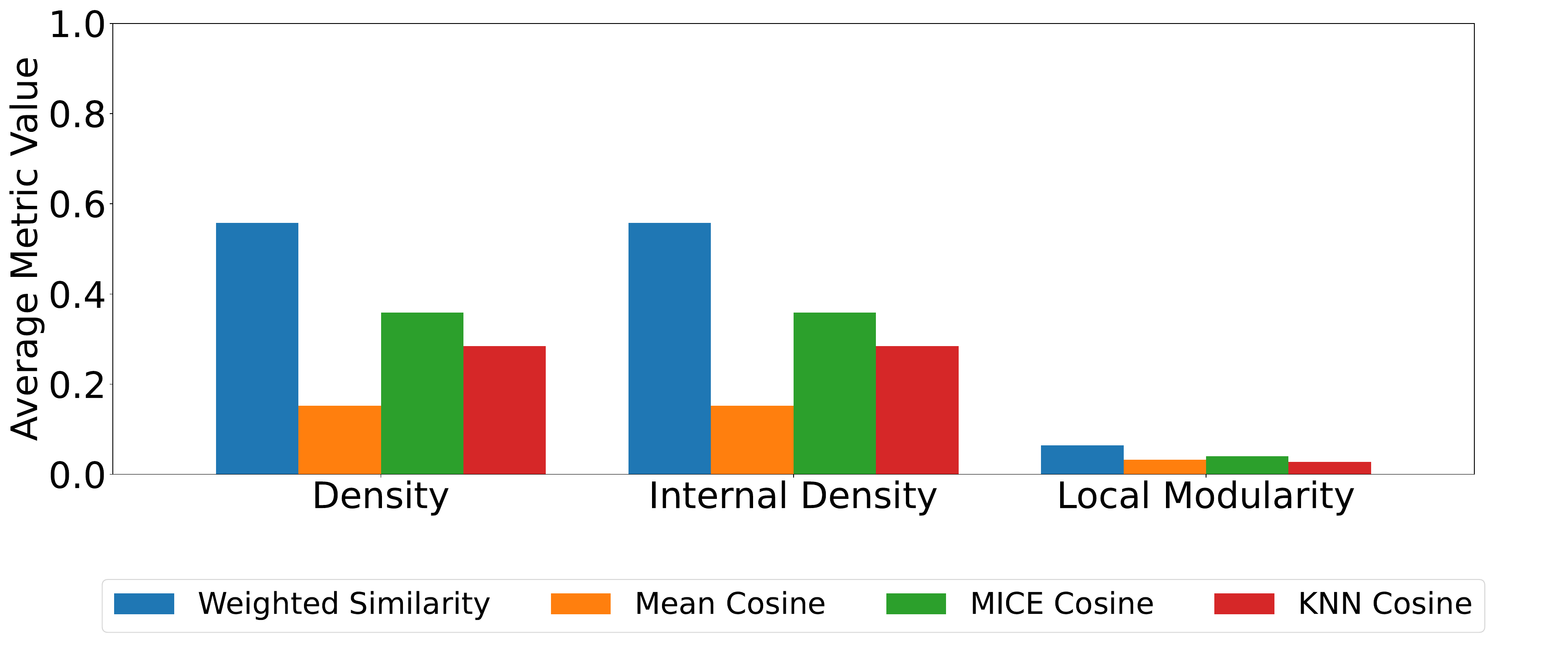}
        \label{fig:left}}
    \subfigure[Imputed euclidean similarity vs. weighted similarity]{
        \includegraphics[width=0.485\linewidth, height=0.36\textwidth]{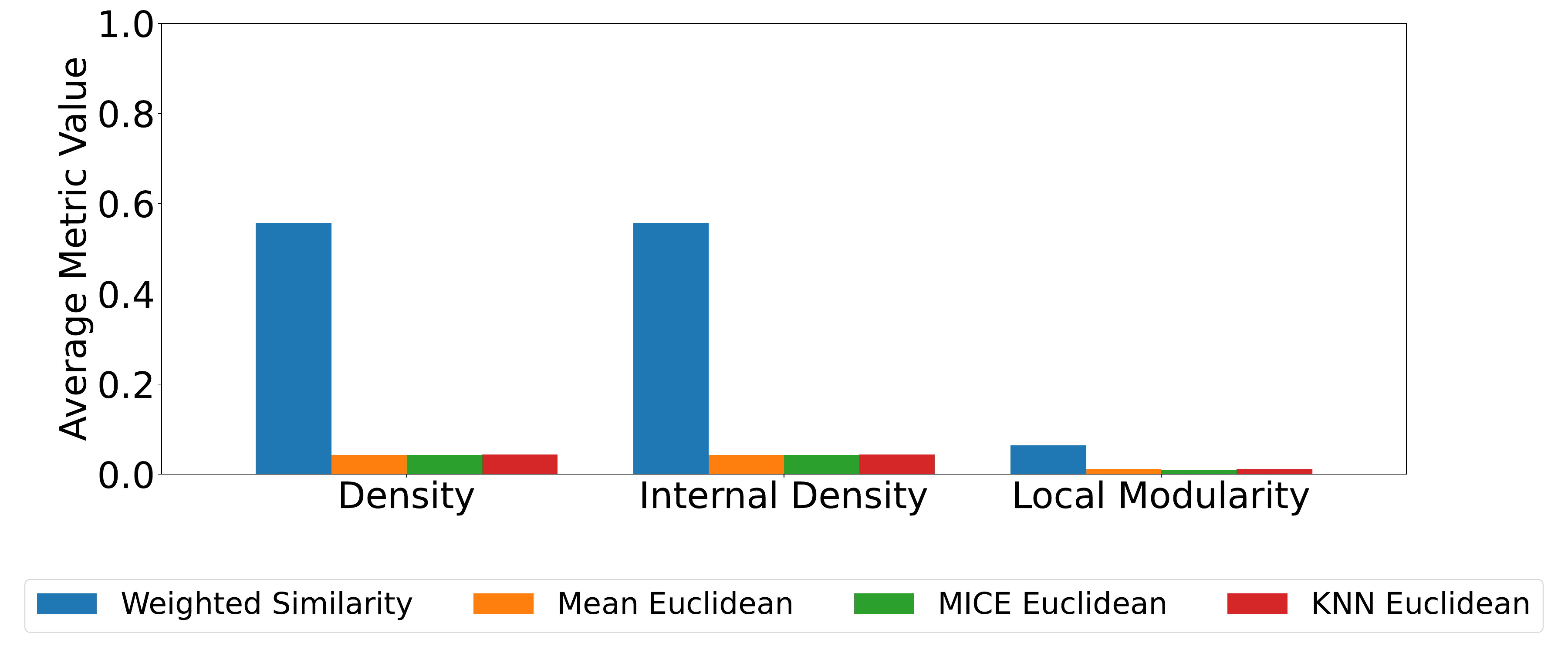}
        \label{fig:right}}
    \subfigure[Imputed canberra similarity vs. weighted similarity]{
        \includegraphics[width=0.485\linewidth, height=0.36\textwidth]{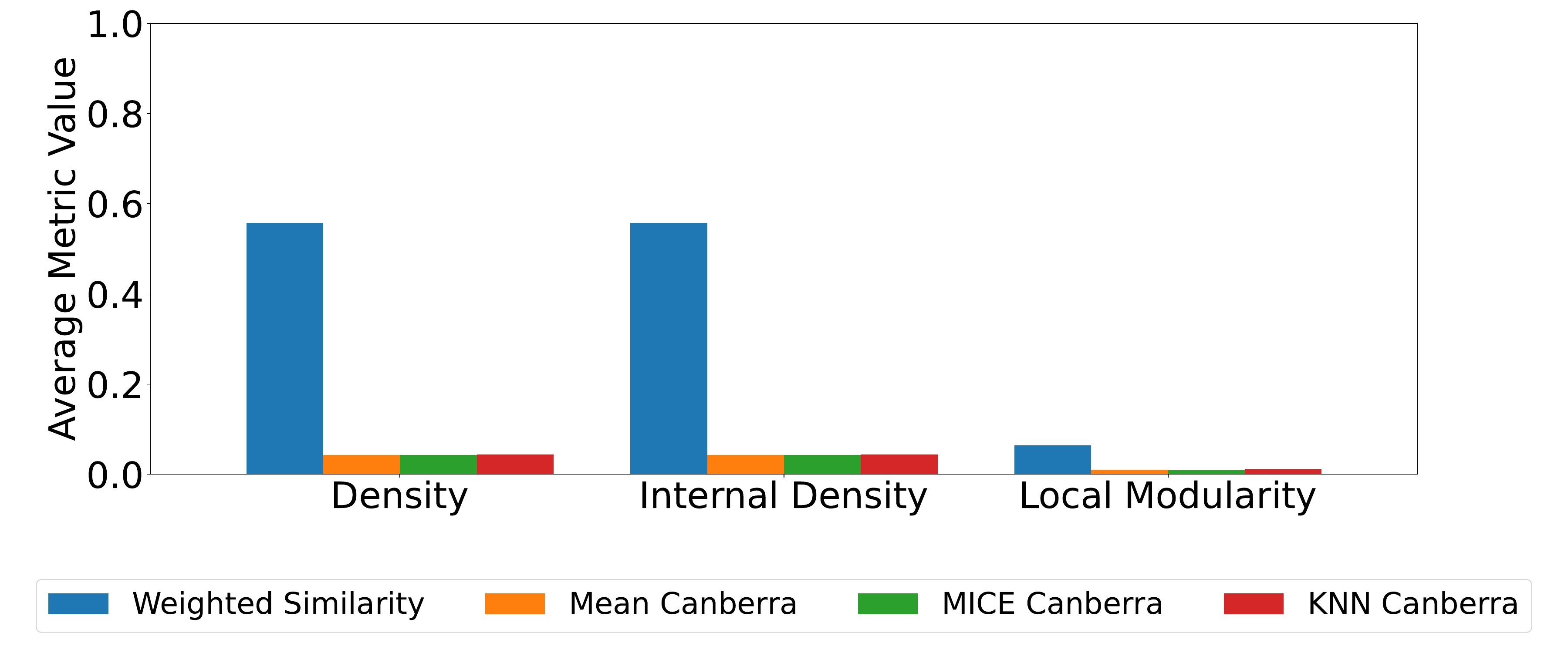}
        \label{fig:left}}
    \subfigure[Imputed spearman similarity vs. weighted similarity]{
        \includegraphics[width=0.485\linewidth, height=0.36\textwidth]{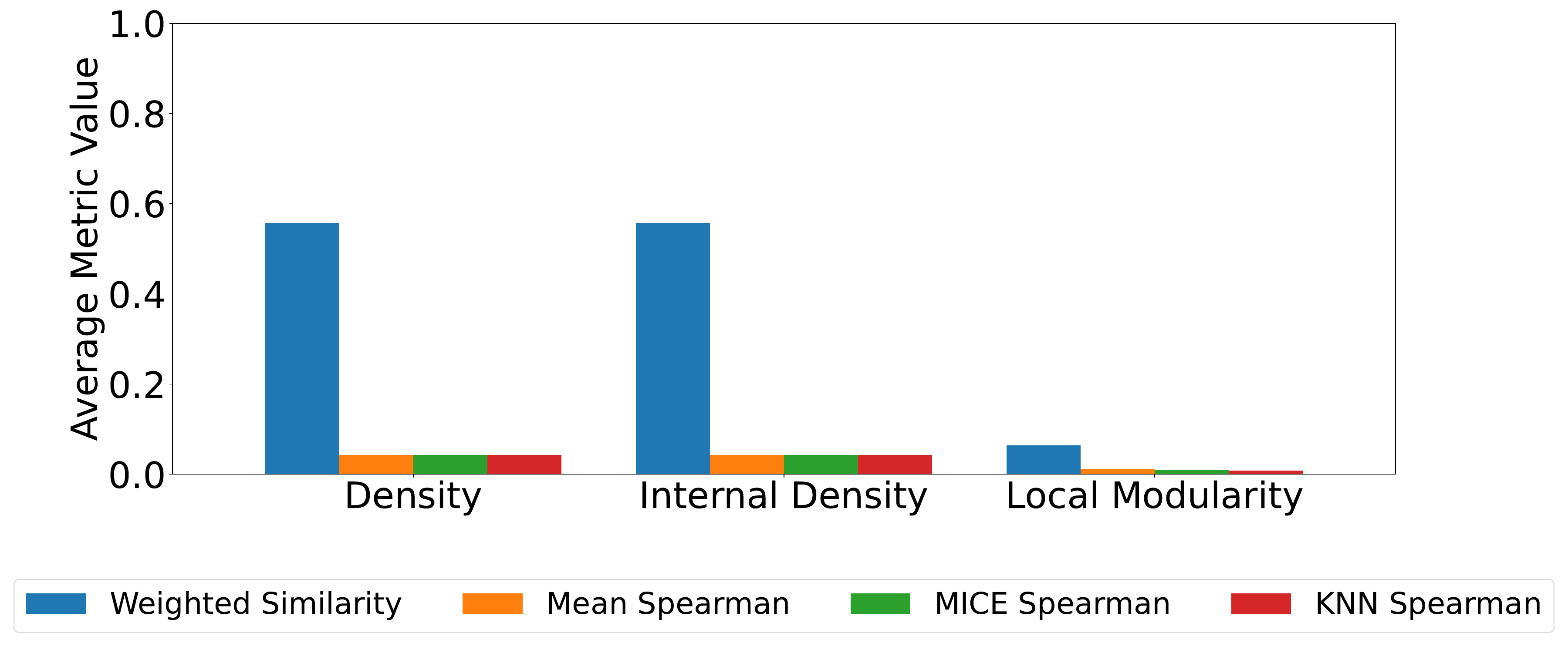}
        \label{fig:right}}
    \caption{Average community qualities with 600 connected edges and 5 clusters. Higher bar indicates better community.}
    \label{fig:performance600E5C_aveComH}
\end{figure*}

\begin{figure*}[!ht] % Options [h!] for placement
    \centering
    \subfigure[Imputed cosine similarity vs. weighted similarity]{
        \includegraphics[width=0.485\linewidth, height=0.36\textwidth]{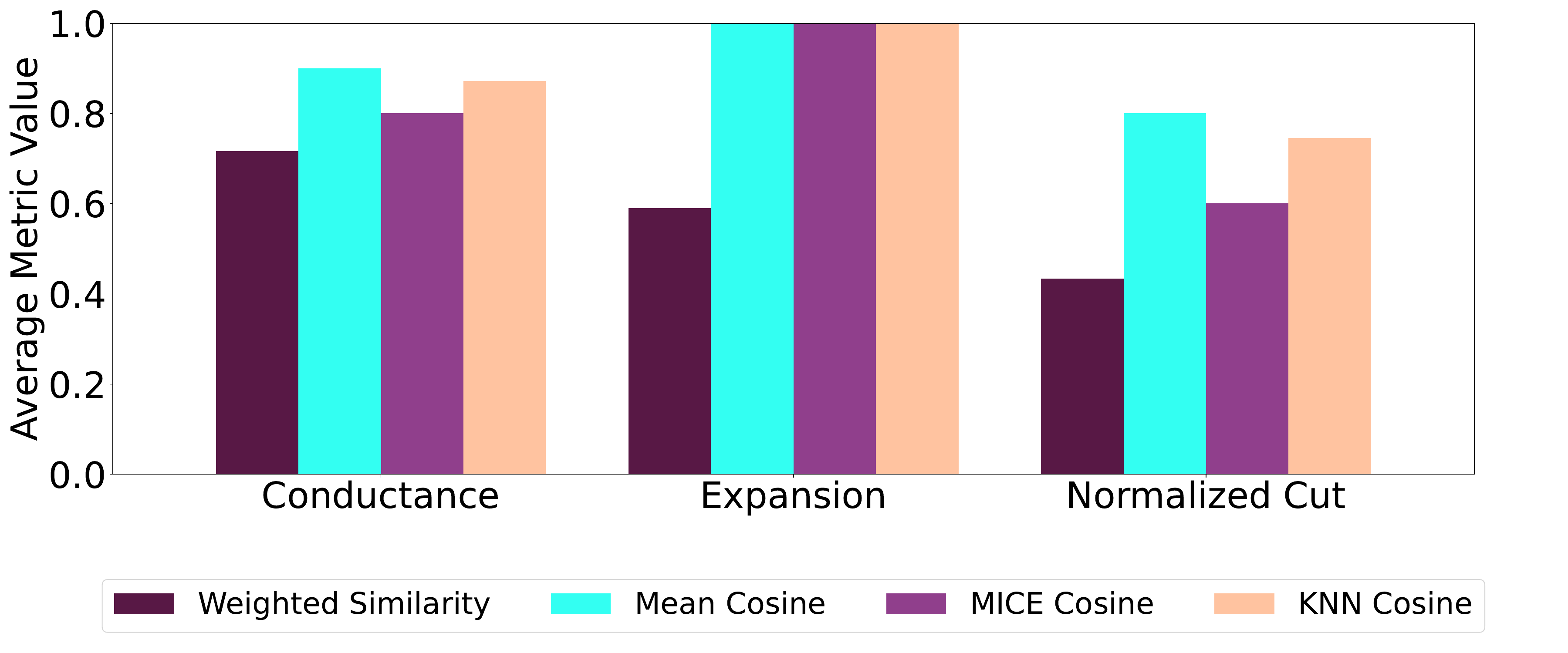}
        \label{fig:left}}
    \subfigure[Imputed euclidean similarity vs. weighted similarity]{
        \includegraphics[width=0.485\linewidth, height=0.36\textwidth]{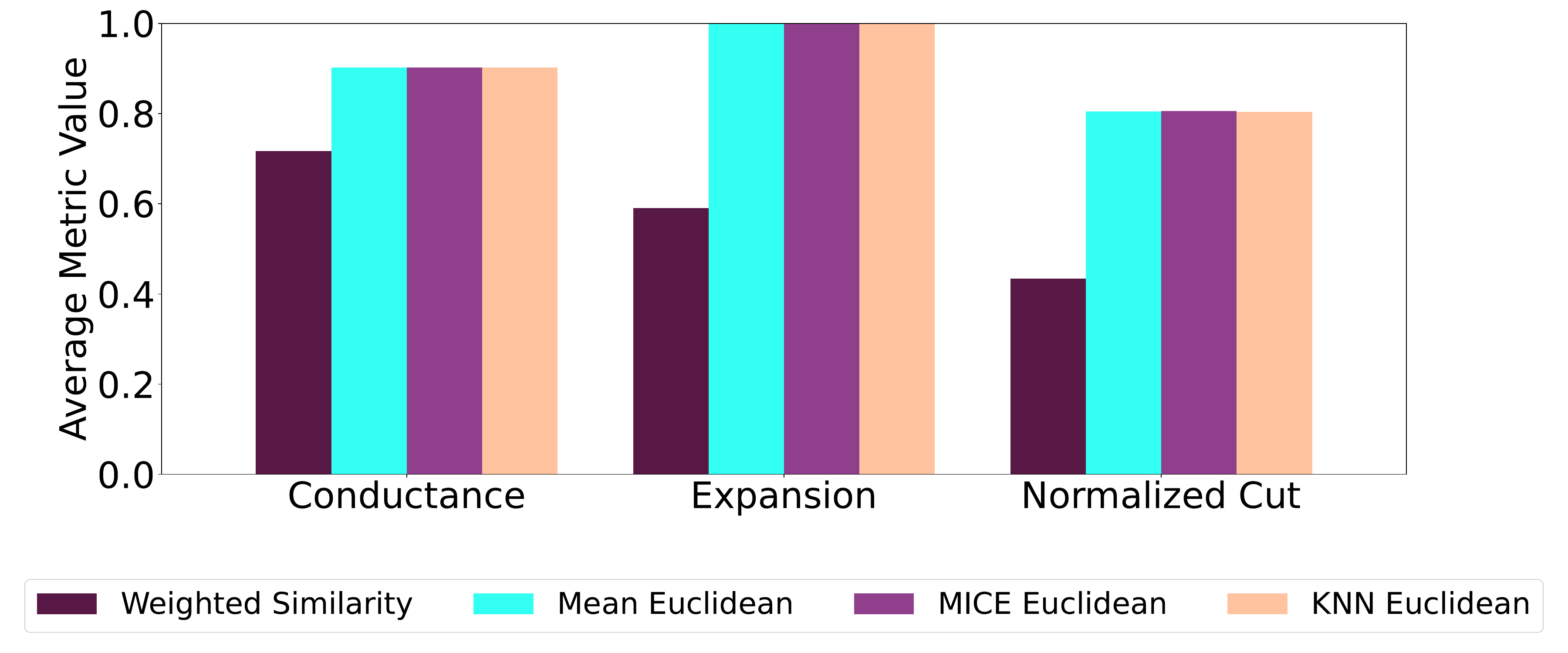}
        \label{fig:right}}
    \subfigure[Imputed canberra similarity vs. weighted similarity]{
        \includegraphics[width=0.485\linewidth, height=0.36\textwidth]{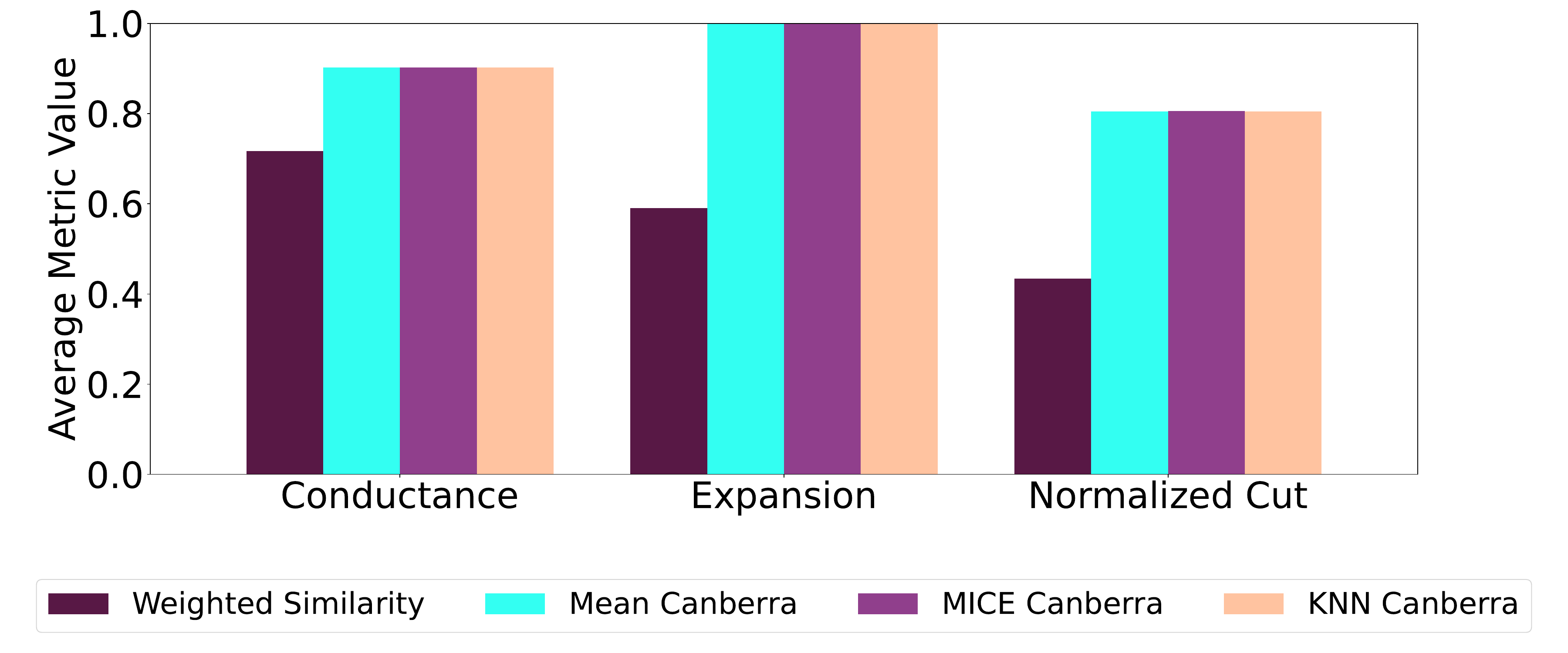}
        \label{fig:left}}
    \subfigure[Imputed spearman similarity vs. weighted similarity]{
        \includegraphics[width=0.485\linewidth, height=0.36\textwidth]{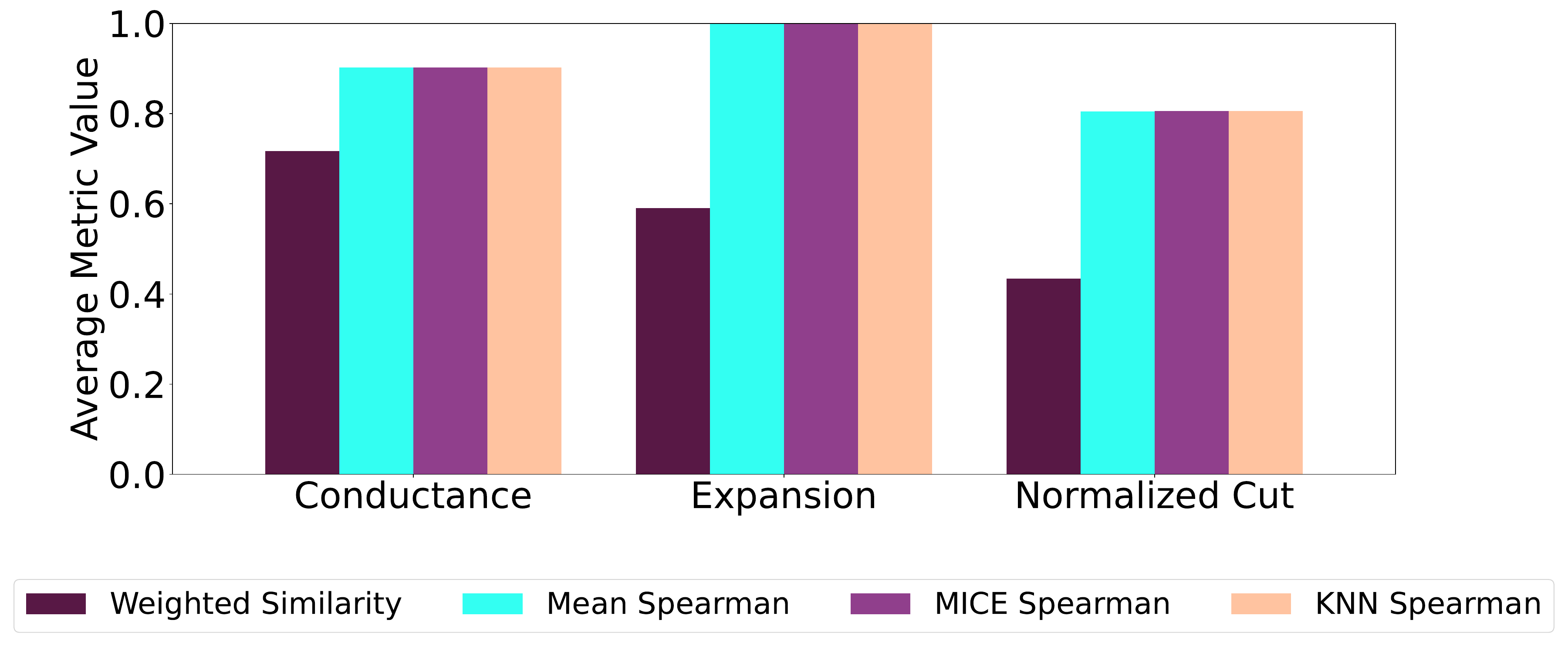}
        \label{fig:right}}
    \caption{Average community qualities with 600 connected edges and 5 clusters. Shorter bar indicates better community.}
    \label{fig:performance600E5C_aveComL}
\end{figure*}

\begin{figure*}[!ht] % Options [h!] for placement
    \centering
    \subfigure[Weighted similarity]{
        \includegraphics[width=0.485\linewidth, height=0.36\textwidth]{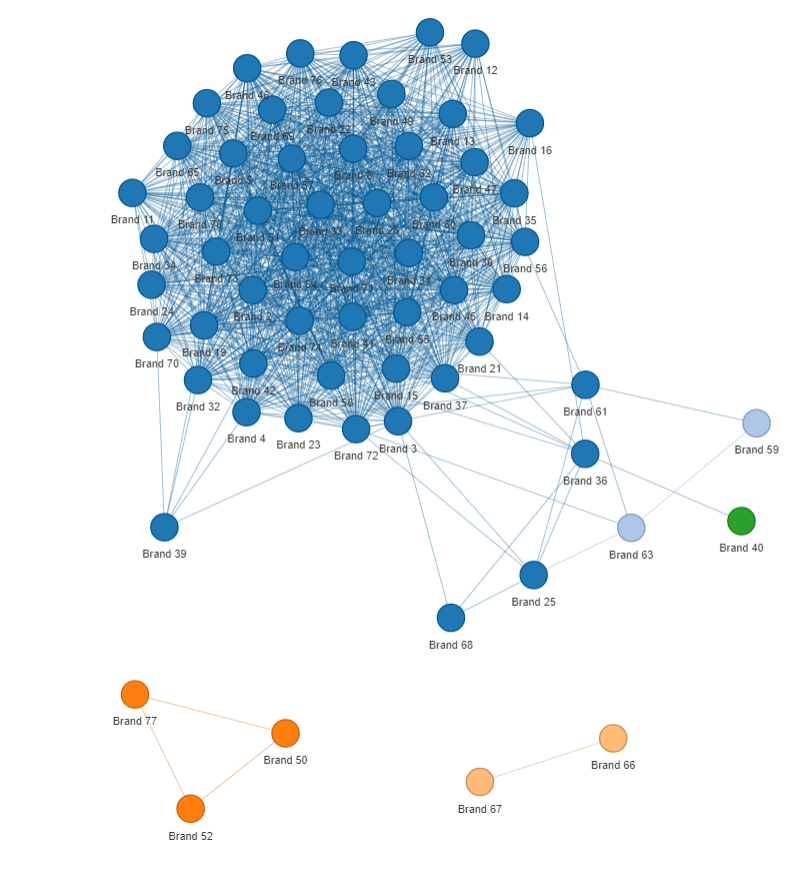}
        \label{fig:left}}
    \subfigure[Cosine similarity based on mean imputation]{
        \includegraphics[width=0.485\linewidth, height=0.36\textwidth]{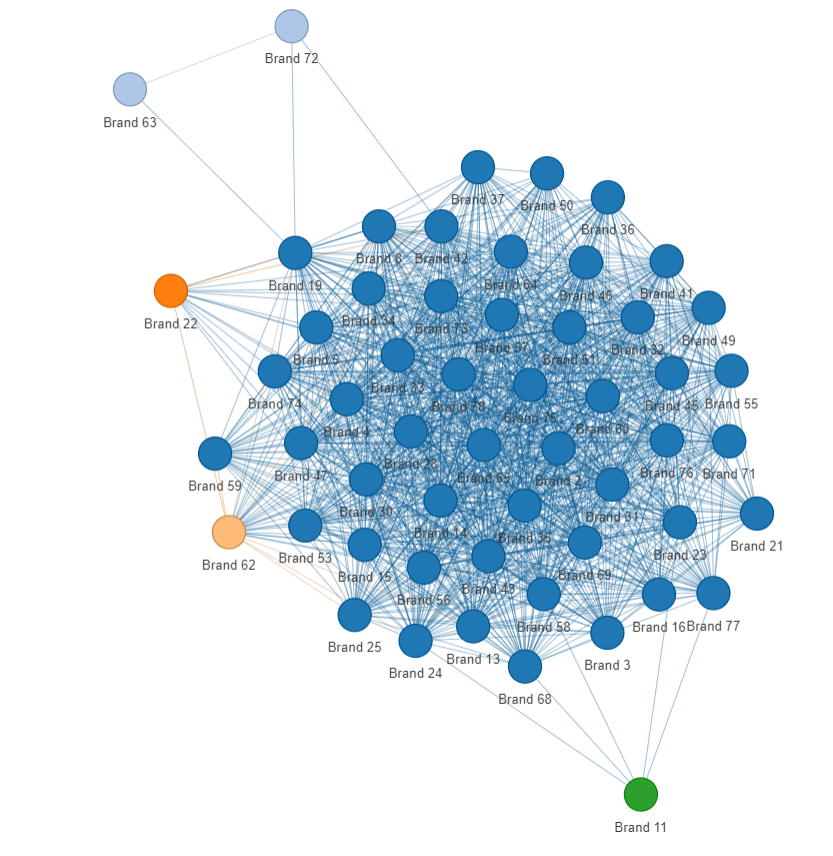}
        \label{fig:right}}
    \subfigure[Cosine similarity based on MICE imputation]{
        \includegraphics[width=0.485\linewidth, height=0.36\textwidth]{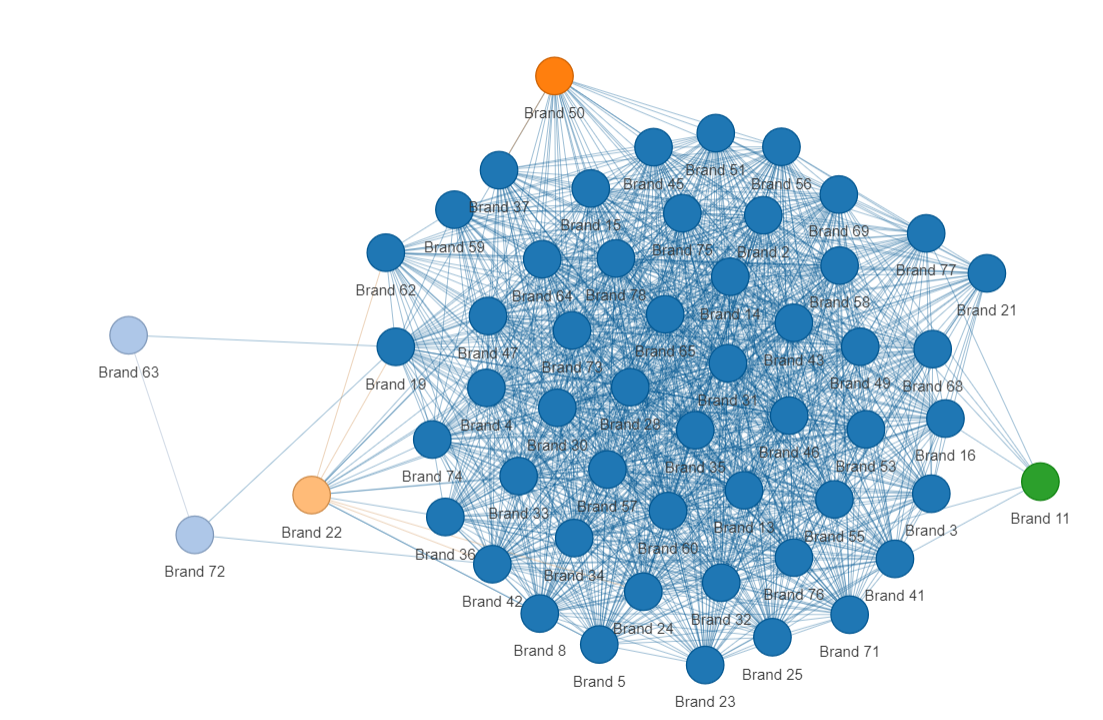}
        \label{fig:left}}
    \subfigure[Cosine similarity based on KNN(k=4) imputation]{
        \includegraphics[width=0.485\linewidth, height=0.36\textwidth]{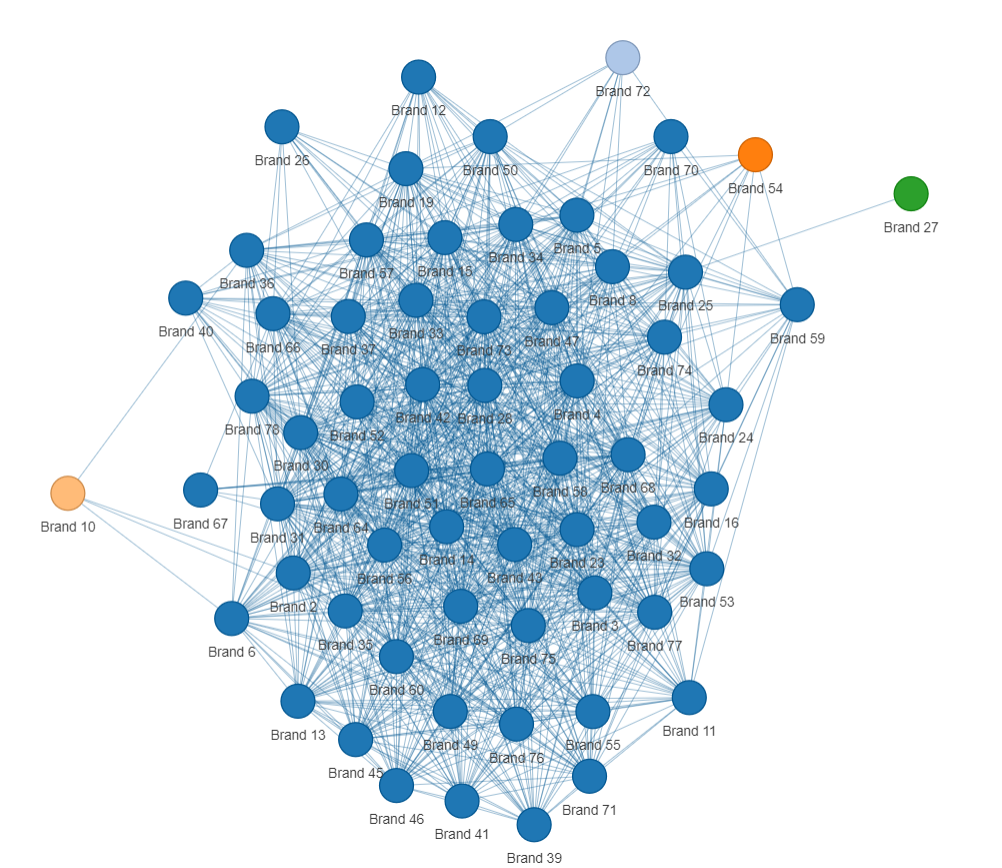}
        \label{fig:right}}
    \caption{Community structures with 5 clusters and 1200 connected edges based on the weighted similarity and imputed cosine similarities.}
    \label{fig:graph1200E5C_Cosine}
\end{figure*}

\begin{figure*}[!ht] % Options [h!] for placement
    \centering
    \subfigure[Imputed cosine similarity vs. weighted similarity]{
        \includegraphics[width=0.485\linewidth, height=0.36\textwidth]{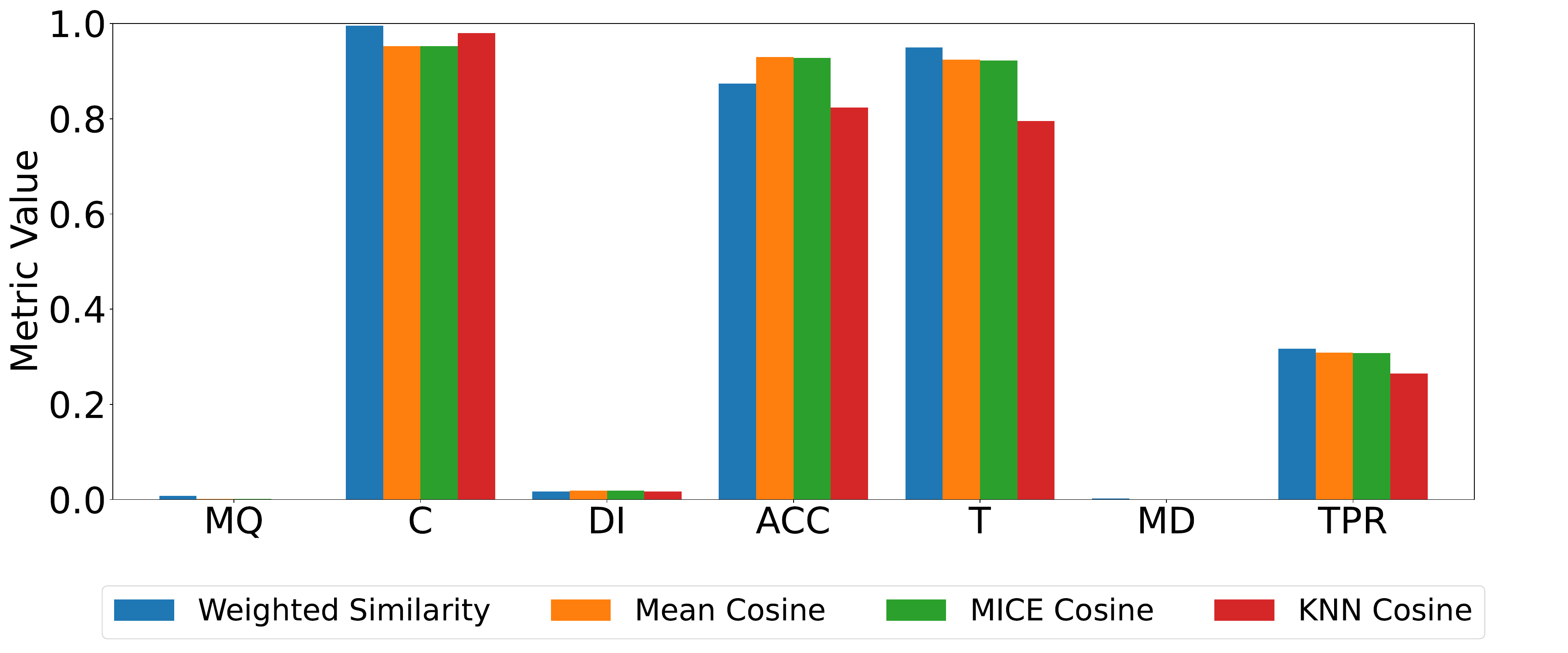}
        \label{fig:left}}
    \subfigure[Imputed euclidean similarity vs.weighted similarity]{
        \includegraphics[width=0.485\linewidth, height=0.36\textwidth]{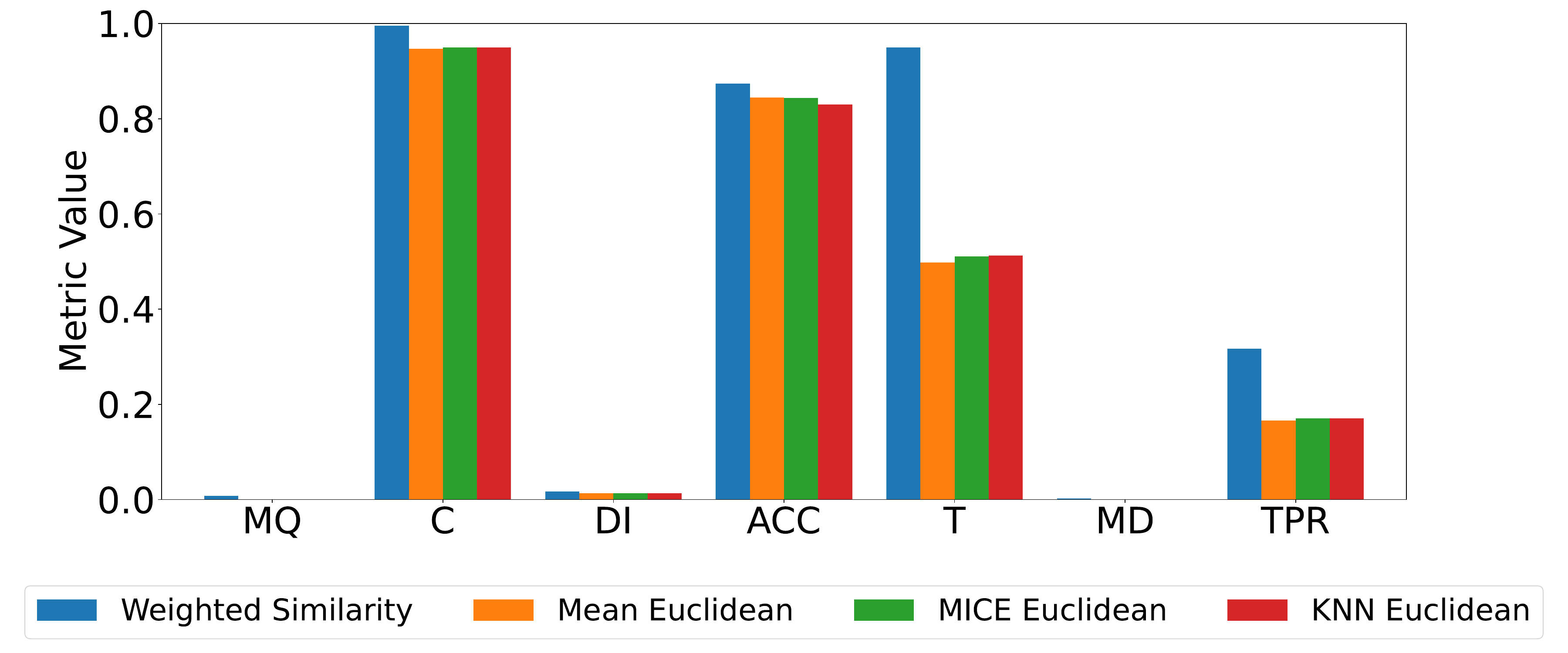}
        \label{fig:right}}
    \subfigure[Imputed canberra similarity vs. weighted similarity]{
        \includegraphics[width=0.485\linewidth, height=0.36\textwidth]{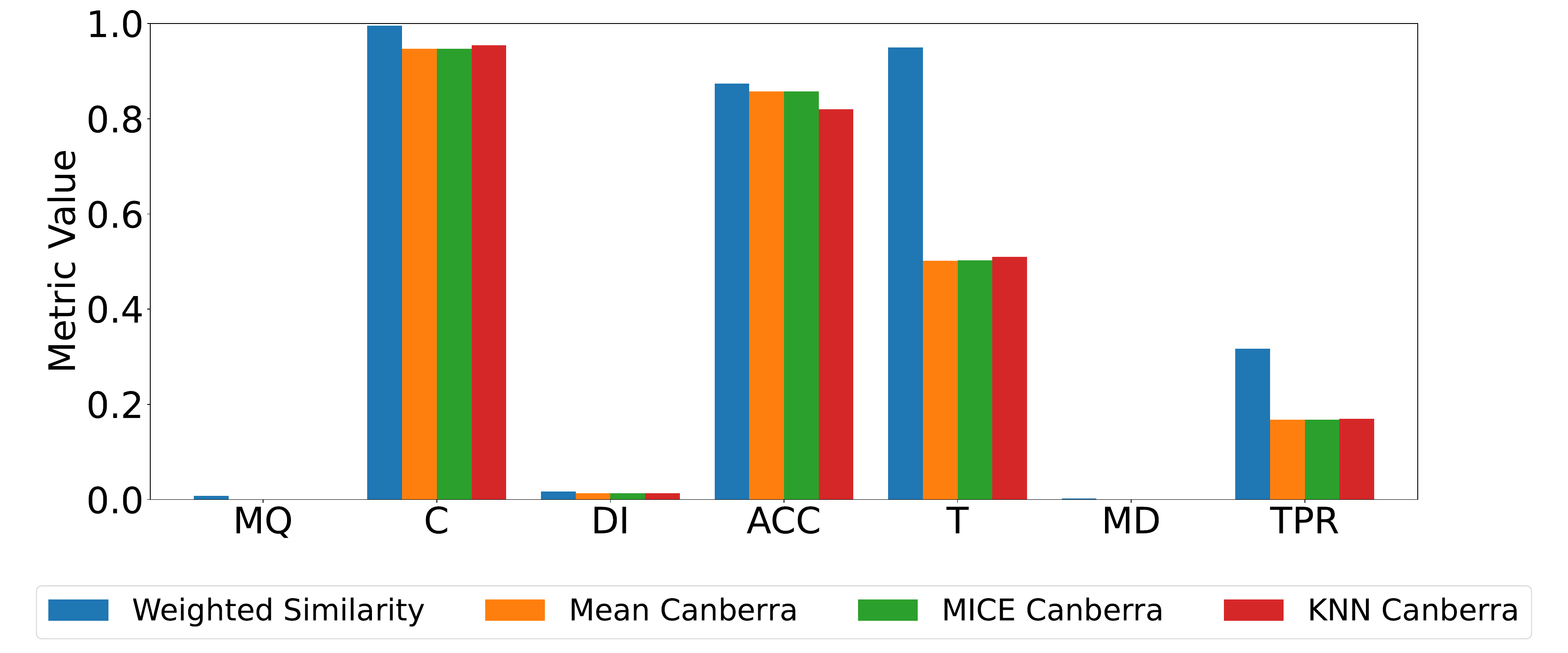}
        \label{fig:left}}
    \subfigure[Imputed spearman similarity vs. weighted similarity]{
        \includegraphics[width=0.485\linewidth, height=0.36\textwidth]{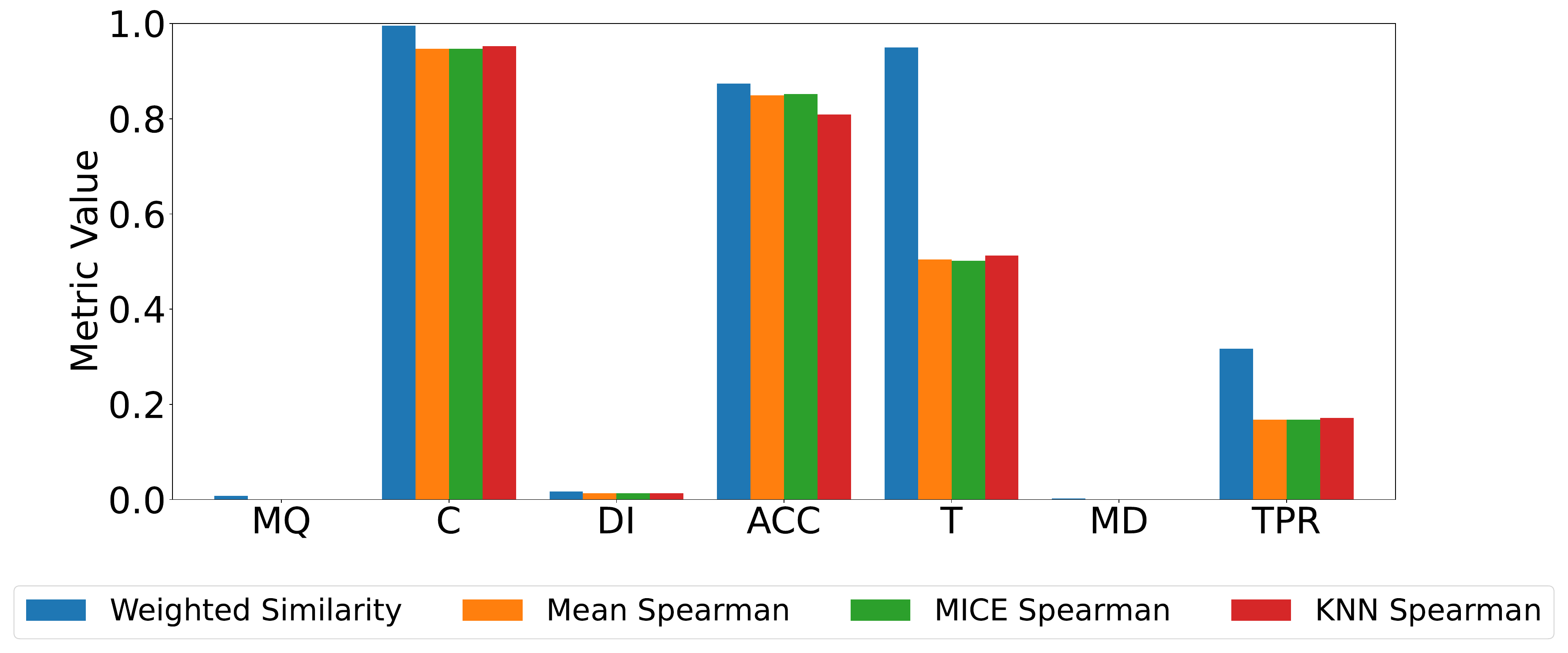}
        \label{fig:right}}
    \caption{Performance metrics on general community structures with 1200 connected edges and 5 clusters. MQ: Modularity Q, C: Coverage, DI: Dunn Index, ACC: Average Clustering Coefficient, T: Transitivity, MD: Modularity Density, TPR: Triangle Participation Ratio. Higher bar indicates better community structure.}
    \label{fig:performance1200E5C_genCom}
\end{figure*}

\begin{figure*}[!ht] % Options [h!] for placement
    \centering
    \subfigure[Imputed cosine similarity vs. weighted similarity]{
        \includegraphics[width=0.485\linewidth, height=0.36\textwidth]{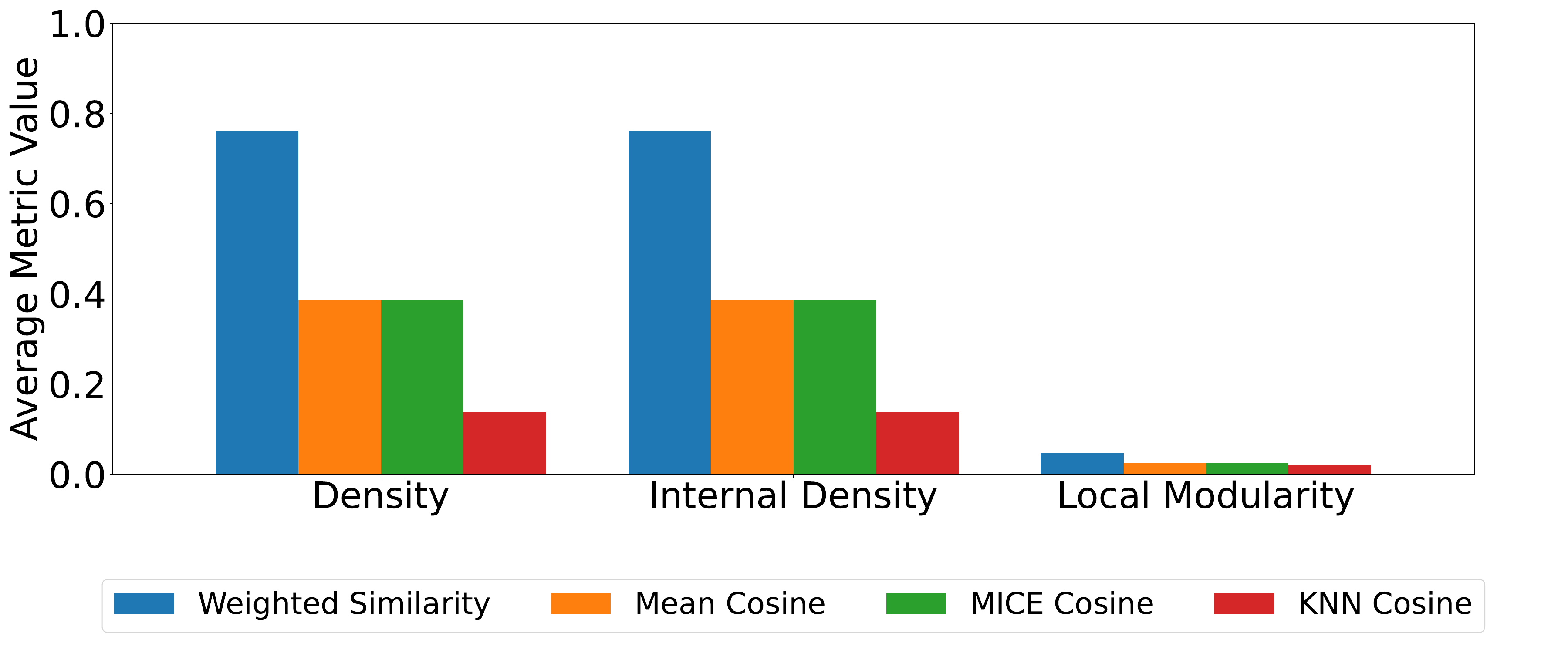}
        \label{fig:left}}
    \subfigure[Imputed euclidean similarity vs. weighted similarity]{
        \includegraphics[width=0.485\linewidth, height=0.36\textwidth]{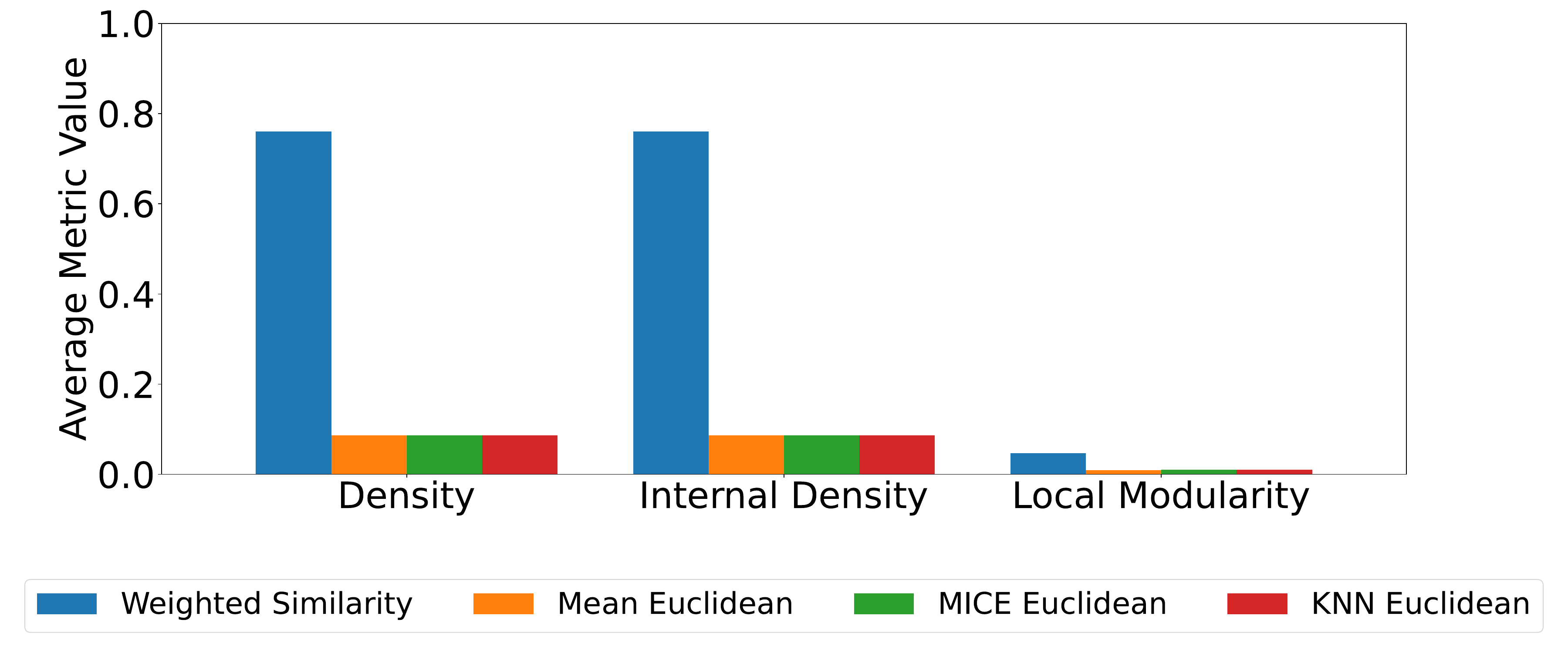}
        \label{fig:right}}
    \subfigure[Imputed canberra similarity vs. weighted similarity]{
        \includegraphics[width=0.485\linewidth, height=0.36\textwidth]{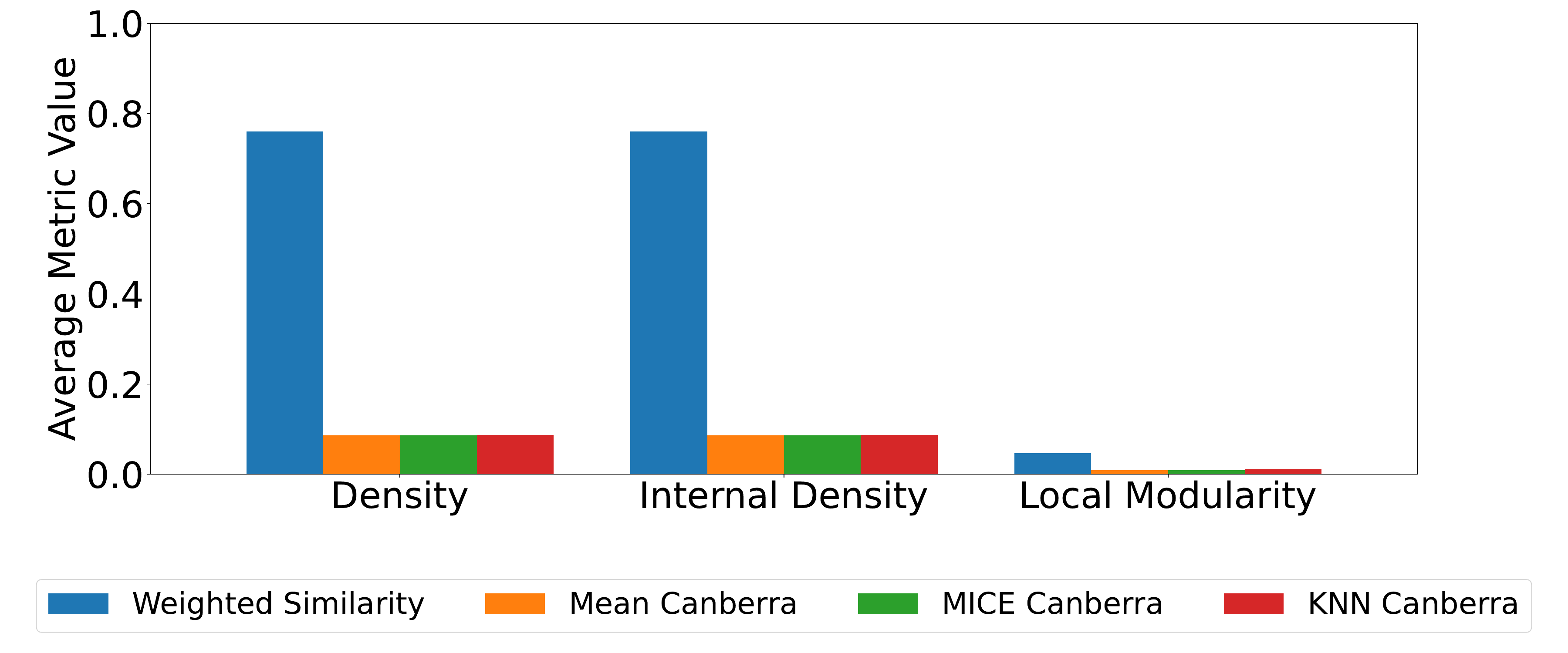}
        \label{fig:left}}
    \subfigure[Imputed spearman similarity vs. weighted similarity]{
        \includegraphics[width=0.485\linewidth, height=0.36\textwidth]{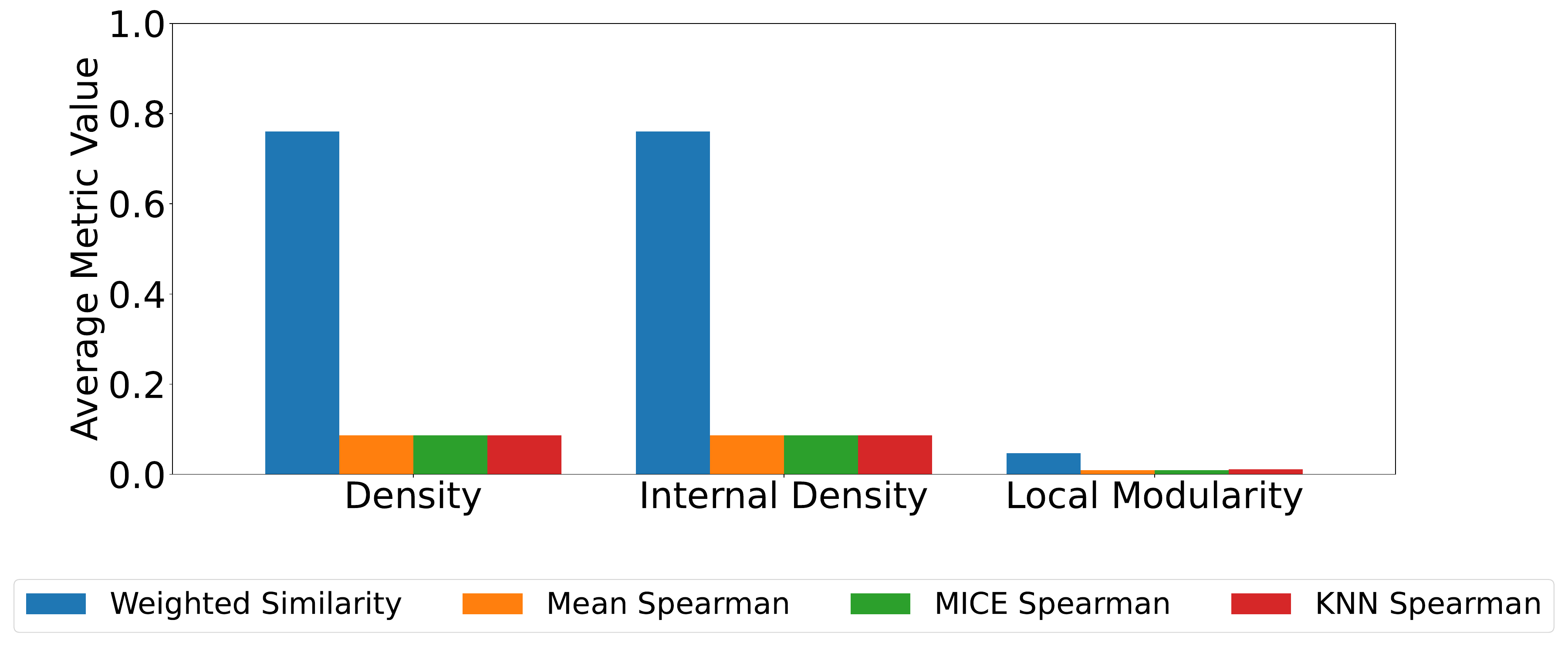}
        \label{fig:right}}
    \caption{Average community qualities with 1200 connected edges and 5 clusters. Higher bar indicates better community.}
    \label{fig:performance1200E5C_aveComH}
\end{figure*}

\begin{figure*}[!ht] % Options [h!] for placement
    \centering
    \subfigure[Imputed cosine similarity vs. weighted similarity]{
        \includegraphics[width=0.485\linewidth, height=0.36\textwidth]{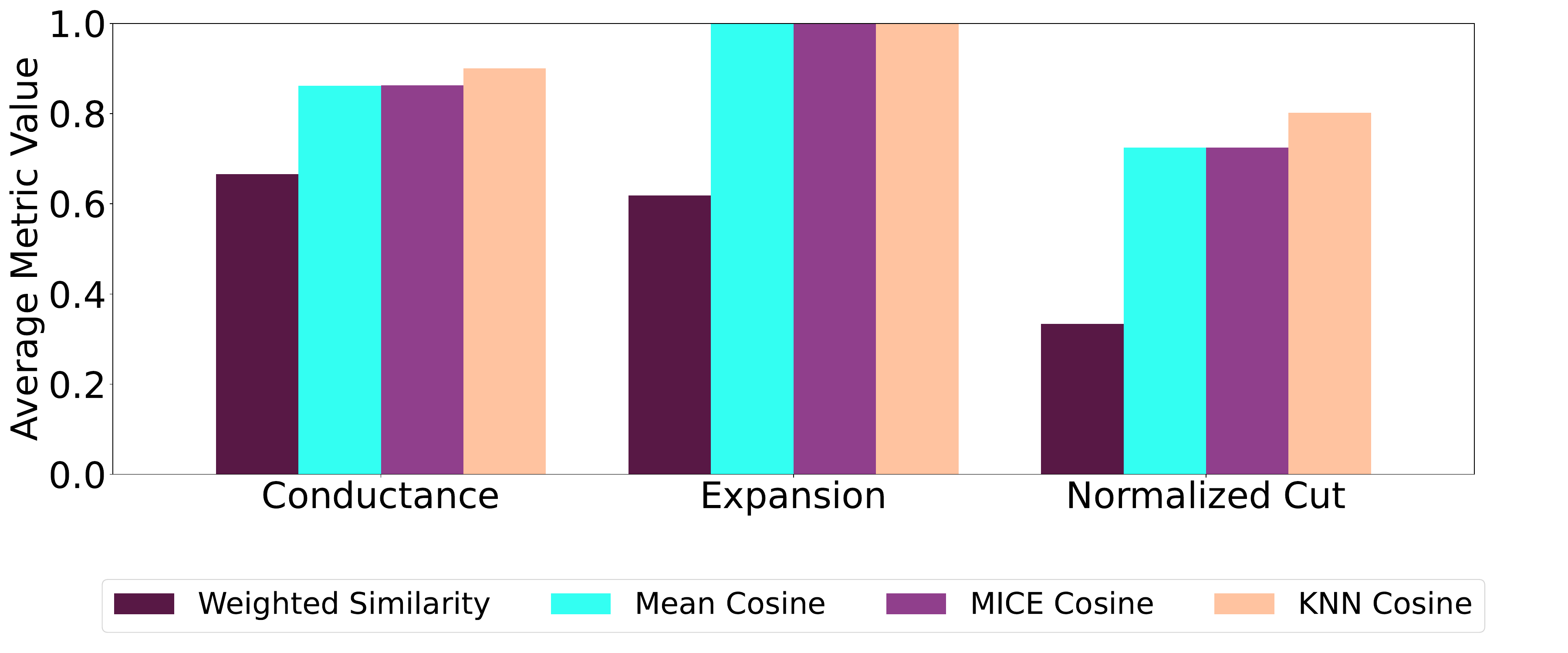}
        \label{fig:left}}
    \subfigure[Imputed euclidean similarity based vs. weighted similarity]{
        \includegraphics[width=0.485\linewidth, height=0.36\textwidth]{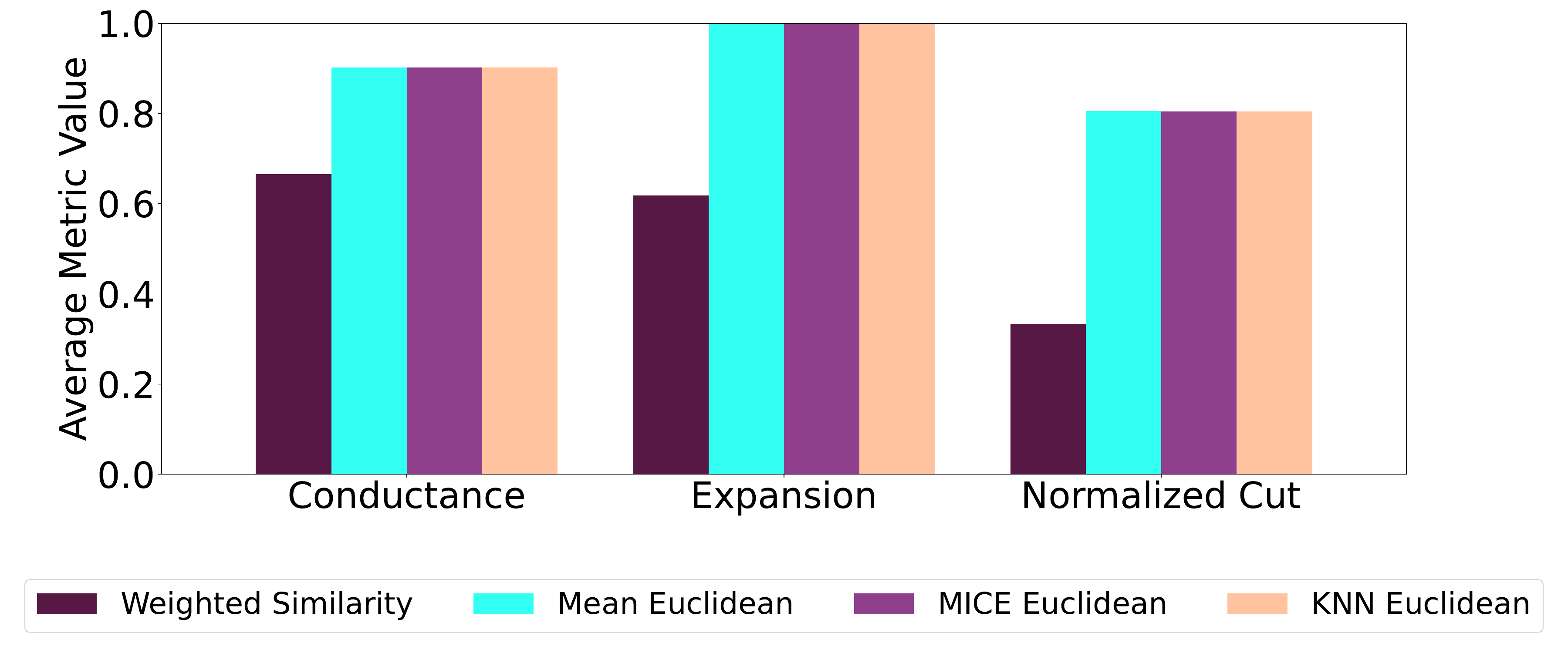}
        \label{fig:right}}
    \subfigure[Imputed canberra similarity vs. weighted similarity]{
        \includegraphics[width=0.485\linewidth, height=0.36\textwidth]{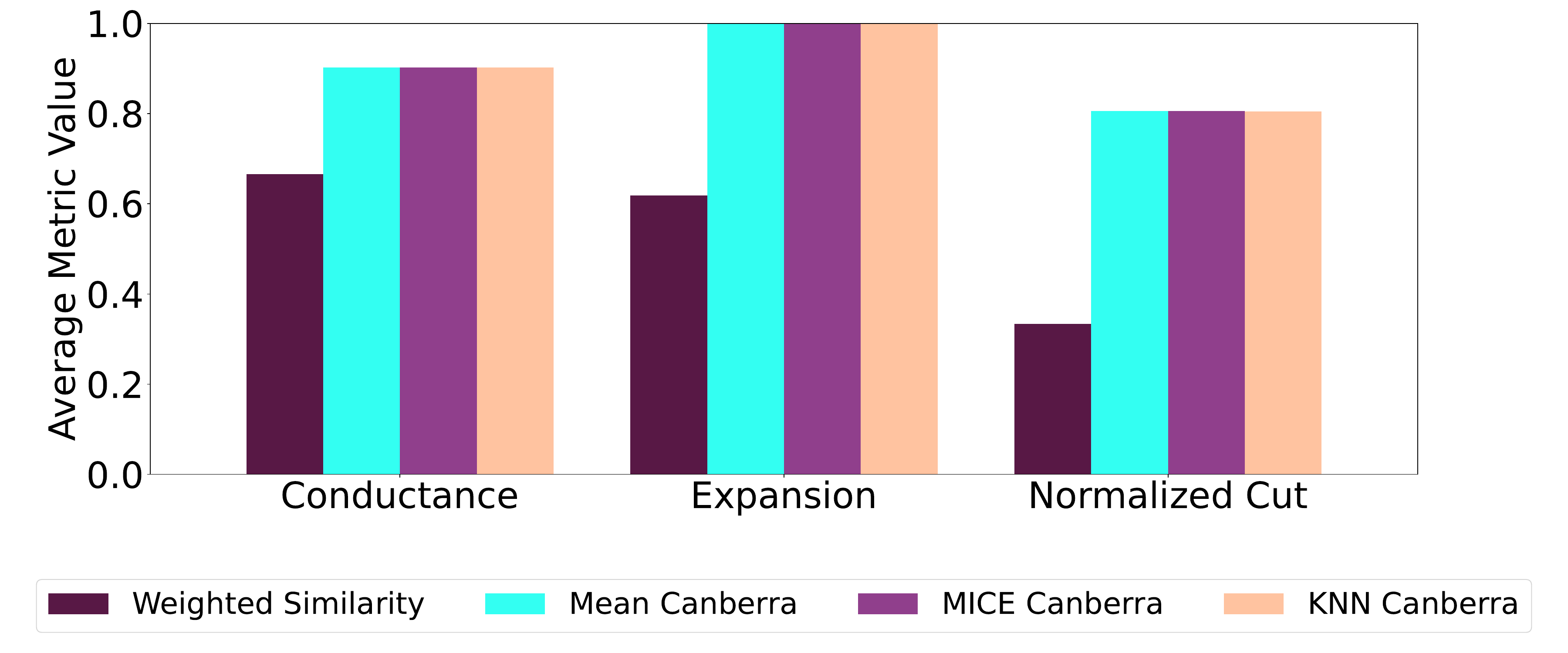}
        \label{fig:left}}
    \subfigure[Imputed spearman similarity vs. weighted similarity]{
        \includegraphics[width=0.485\linewidth, height=0.36\textwidth]{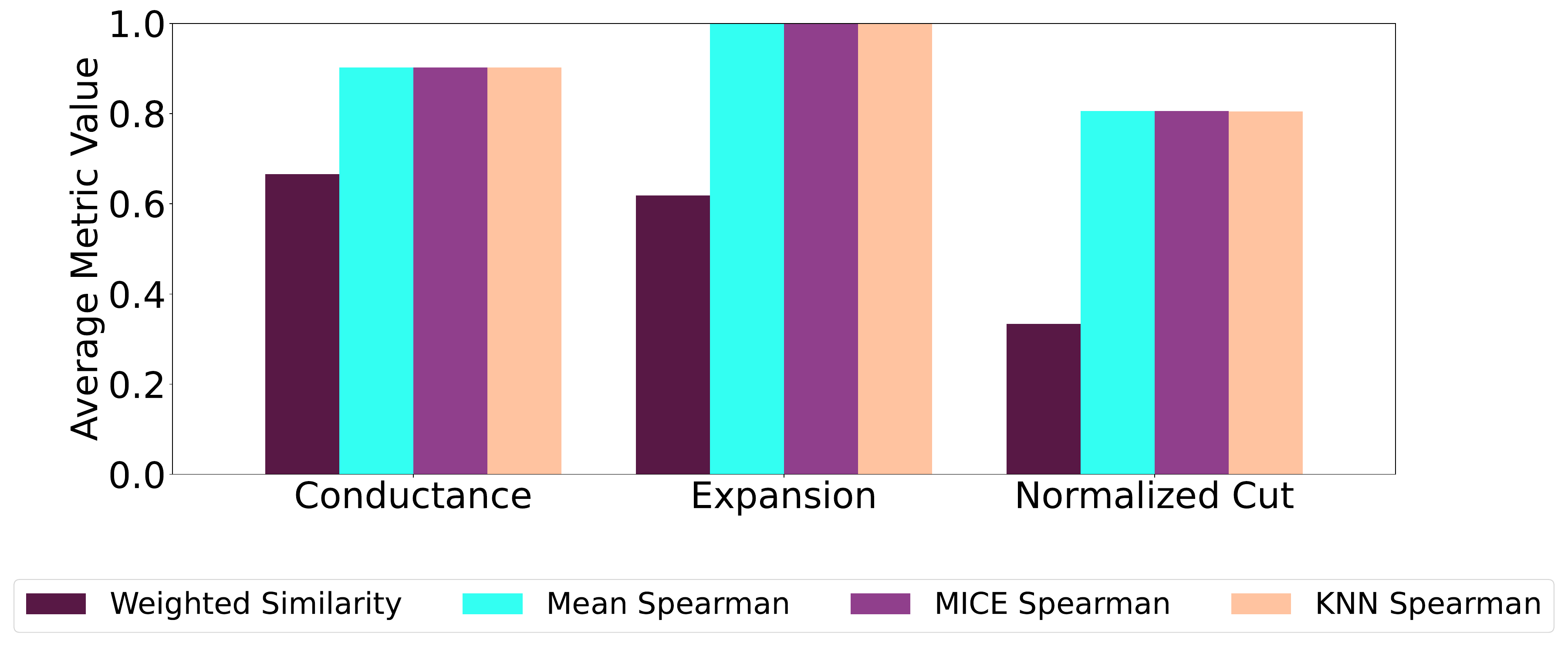}
        \label{fig:right}}
    \caption{Average community qualities with 1200 connected edges and 5 clusters. Shorter bar indicates better community.}
    \label{fig:performance1200E5C_aveComL}
\end{figure*}
%%%
\end{document}